\documentclass{aa} 
\usepackage[lofdepth,lotdepth]{subfig}
\usepackage{xspace}
\usepackage{graphicx}
\usepackage{lscape}
\usepackage{txfonts}
\usepackage{supertabular}
\usepackage{longtable}
\usepackage{amssymb}
\usepackage{caption}
\usepackage{color}
\usepackage{amsmath,amsfonts,amssymb}
\usepackage{comment}

\usepackage[titletoc,title]{appendix}
\usepackage[colorlinks=true,allcolors=blue]{hyperref}
\usepackage{natbib,twoopt}
\bibpunct{(}{)}{;}{a}{}{,}             
\makeatletter
  \newcommandtwoopt{\citeads}[3][][]{\href{http://adsabs.harvard.edu/abs/#3}%
    {\def\hyper@linkstart##1##2{}%
     \let\hyper@linkend\@empty\citealp[#1][#2]{#3}}}
  \newcommandtwoopt{\citepads}[3][][]{\href{http://adsabs.harvard.edu/abs/#3}%
    {\def\hyper@linkstart##1##2{}%
     \let\hyper@linkend\@empty\citep[#1][#2]{#3}}}
  \newcommandtwoopt{\citetads}[3][][]{\href{http://adsabs.harvard.edu/abs/#3}%
    {\def\hyper@linkstart##1##2{}%
     \let\hyper@linkend\@empty\citet[#1][#2]{#3}}}
  \newcommandtwoopt{\citeyearads}[3][][]%
    {\href{http://adsabs.harvard.edu/abs/#3}
    {\def\hyper@linkstart##1##2{}%
     \let\hyper@linkend\@empty\citeyear[#1][#2]{#3}}}
\makeatother

\renewcommand{\arraystretch}{1.3}

\begin{document} 

\def\HII{H\,{\sc{ii}}\,}
\def\mm{\,$\mu$m\,}
\def\spitzer{$\it{Spitzer}$\,}
\def\herschel{$\it{Herschel}$\,}
\def\cutex{$\it{Cutex}$\,}
\def\vialactea{$\it{Vialactea}$\,}
\defcitealias{zav07}{ZAV07}
\defcitealias{deh09}{DEH09}
   \title{Multiwavelength study of the G345.5+1.5 region\thanks{Tables~\ref{tab_param} \ref{ap:higal_clumps} and \ref{ap:co_clumps} are available in electronic form
at the CDS via anonymous ftp to cdsarc.u-strasbg.fr (130.79.128.5)
or via \url{http://cdsweb.u-strasbg.fr/cgi-bin/qcat?J/A+A/}}}

   \author{ M. Figueira\inst{1,2} 
            \and C. López-Calderón\inst{3}  
            \and L. Bronfman\inst{4}
            \and A. Zavagno\inst{1}              
            \and C. Hervías-Caimapo\inst{5}
            \and N. Duronea\inst{6}         
            \and L-\AA. Nyman\inst{3,7}      
}
          \offprints{Miguel.Figueira@ncbj.gov.pl }
          
   \institute{Aix Marseille Univ, CNRS, LAM, Laboratoire d'Astrophysique de Marseille, Marseille, France
   \and National Centre for Nuclear Research, ul. Ho\.za 69, 00-681, Warszawa, Poland
   \and Joint ALMA Observatory (JAO), Alonso de Córdova 3107, Vitacura, Santiago, Chile
   \and Departamento de Astronomía, Universidad de Chile, Casilla 36-D, Santiago, Chile
   \and Jodrell Bank Centre for Astrophysics, School of Physics \& Astronomy, University of Manchester, Oxford Road, Manchester M13 9PL, U.K.
   \and Instituto Argentino de Radioastronomía, CONICET, CCT-La Plata. C.C.5., 1894 Villa Elisa, Argentina
   \and European Southern Observatory, Alonso de Córdova 3107, Vitacura, Santiago, Chile
     }

   \date{Received 11 March 2018 ; accepted 26 January 2019}
 
  \abstract
  {The star formation process requires the dust and gas present in the Milky Way to self-assemble into dense reservoirs of neutral material where the new generation of stars will emerge. Star-forming regions are usually studied in the context of Galactic surveys, but dedicated observations are sometimes needed when the study reaches beyond the survey area.}
   {A better understanding of the star formation process in the Galaxy can be obtained by studying several regions. This allows increasing the sample of objects (clumps, cores, and stars) for further statistical works and deeper follow-up studies. Here, we studied the G345.5+1.5 region, which is located slightly above the Galactic plane, to understand its star formation properties.}
   {We combined the Large Apex BOlometer CAmera (LABOCA) and $^{12}$CO(4$-$3) transition line (NANTEN2) observations complemented with the Hi-GAL and $\it{Spitzer}$-GLIMPSE surveys to study the star formation toward this region. We used the \textit{Clumpfind} algorithm to extract the clumps from the 870~$\mu$m and $^{12}$CO(4$-$3) data. Radio emission at 36~cm was used to estimate the number of \HII regions and to remove the contamination from the free-free emission at 870~$\mu$m. We employed color-color diagrams and spectral energy distribution (SED) slopes to distinguish between prestellar and protostellar clumps. We studied the boundedness of the clumps through the virial parameter. Finally, we estimated the star formation efficiency (SFE) and star formation rate (SFR) of the region and used the Schmidt-Kennicutt diagram to compare its ability to form stars with other regions of the Galactic plane.}
 {Of the 13 radio sources that we found using the MGPS-2 catalog, 7 are found to be associated with \HII regions corresponding to late-B or early-O stars. We found 45 870~$\mu$m clumps with diameters between 0.4 and 1.2~pc and masses between 43~$M_{\sun}$ and 3923~$M_{\sun}$, and 107 $^{12}$CO clumps with diameters between 0.4~pc and 1.3~pc and masses between 28~$M_{\odot}$ and 9433~$M_{\odot}$. More than 50\% of the clumps are protostellar and bounded and are able to host (massive) star formation. High SFR and SFR density ($\Sigma _{SFR}$) values are associated with the region, with an SFE of a few percent.}
 {With submillimeter, CO transition, and short-wavelength infrared observations, our study reveals a population of massive stars, protostellar and bound starless clumps, toward G345.5+1.5. This region is therefore actively forming stars, and its location in the starburst quadrant of the Schmidt-Kennicutt diagram is comparable to other star-forming regions found within the Galactic plane.}


\maketitle
\titlerunning{Star formation around G345.5+1.5}
\authorrunning{Figueira et al.}

%

\section{Introduction}

Giant molecular clouds (GMCs) are huge and massive reservoirs of cold molecular gas. These objects are hierarchically structured into clumps ($\sim$0.2 to $\sim$1~pc), cores ($\sim$0.2pc), and filaments ($\sim$0.1~pc width, \citealt{arz11}) in which star formation is taking place. For instance, the Taurus B211 filament was found to undergo gravitational collapse with an associated supercritical mass per unit length, and is fragmented into several prestellar and protostellar cores \citep{pal13}. Toward the Aquila complex, up to 75\% of the prestellar core sample are found inside filaments \citep{kon15}. The large star-forming complexes of our Galaxy within $-2^{\circ}<b<1^{\circ}$ are found to host 75\% of ATLASGAL sources in which star formation activity has been detected \citep{cse14}. The abundance of gas and dust as well as the environmental conditions of these complexes are ideal for the formation of high-mass stars. This type of stars is very important because they can dictate the evolution of galaxies, and they dynamically and chemically regulate the interstellar medium (ISM) through powerful feedback \citep{kru14}. These mechanisms complicate our understanding of star formation since they affect the molecular cloud in which the stars are born. Stellar feedback such as outflows and jets is able to remove part of the material in the cluster, which slows down the accretion of the other stars \citep{wan10,mur18} or adds some turbulence that halts the contraction of the cloud \citep{mau09}. More powerful feedback, such as photoionization or supernovae (SN) explosions, can strongly modify the cloud structure, favoring or stopping the formation of stars in it \citep{dal13,dal15,gee16}. Detailed studies about \HII regions \citep{rive13,fig17,liu17,tig17} were obtained with the HOBYS program \citep{mot10}, which allowed us to better understand the properties of young stellar objects (YSOs) around these structures. In particular, massive dense cores (MDCS, $M>75$~$M_{\odot}$) are found at the edges or in the filamentary parts of these regions, and they are good candidates for high-resolution observations \citep{fig18,lou18}. 

While almost all the studies of star formation focus on regions in the Galactic plane that are covered by most of the surveys ($|b|<1^{\circ}$), other interesting star-forming regions are also found above and below.  
There are different advantages to extending a study to beyond the Galactic plane. First, the foreground and background dust emission that is unrelated to the region is less abundant\footnote{A map of the Galactic plane is available at \url{http://astro.phys.wvu.edu/wise/}} \citep{and14}, and the estimation of physical parameters, such as the mass or the column density, is more accurate. Second, the superimposition of sources that lie at a different distances along the line of sight cannot be easily treated in general, and this difficulty is less frequently encountered when observing regions away from the Galactic plane. Finally, since these regions are not covered by most of surveys, they need dedicated observations and are therefore less well studied. Until now, only \citet{lop11} and \citet{lop16} have conducted specifical studies of G345.5+1.5.\\
In this work, we present submillimeter and molecular observations of this region, which is located just above the Galactic plane, to better understand its star-forming properties and compare them to other known star-forming regions. Section~\ref{sect:region} presents the G345.5+1.5 region. In Section~\ref{sect:observations} we detail the observations and the reduction process for the 870~$\mu$m and $^{12}$CO(4$-$3) observations, Section~\ref{sect:analysis} presents the analysis made on the kinematic distance of the region, the H$\rm{\alpha}$ and radio continuum emission, and the properties of clumps that we extracted from the observations. Section~\ref{sect:result} discusses the star-forming properties of the G345.5+1.5 structure. Finally, Section~\ref{sect:conclusion} presents the conclusions of this work.

\section{G345.5+1.5 region}\label{sect:region}

\begin{figure}
\begin{center}
 \centering
\hspace*{-0.3cm}\includegraphics[scale=0.85]{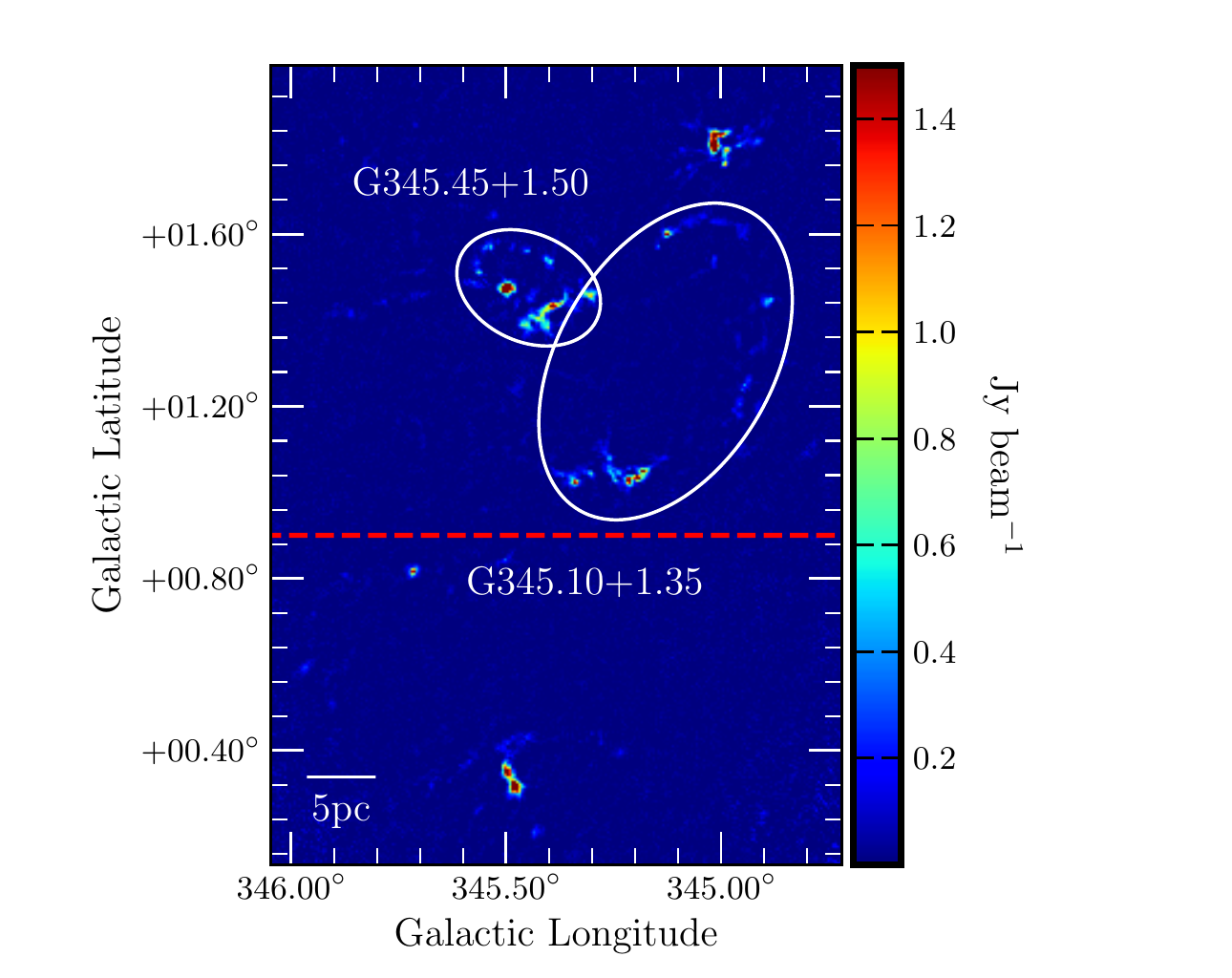}
  \caption{1.2~mm emission map of the GMC G345.5+1.0 \citep{lop11}. The white ellipses represent the two rings of the region: G345.45+1.50 and G345.10+1.35. The area above the red dashed line represents the region we studied here.}
             \label{Fig:G345_1200}
             \end{center}
\end{figure}

The G345.5+1.5 region is composed of two dusty ring-like structures referred to as G345.45+1.5 and G345.10+1.35 (see Fig.~\ref{Fig:G345_1200}) and classified as \HII regions \citep{cas87,and14}. They are part of the GMC G345.5+1.0 ($344.5^{\circ}<\ell<346.5^{\circ}$, $1^{\circ}<b<1.8^{\circ}$). The region is associated with molecular gas emission observed between a local standard of rest velocity ($V_{\rm{LSR}}$) of $-$33~km~s$^{-1}$ and $-$2~km~s$^{-1}$ with a peak at $-$13.6~km~s$^{-1}$, according to the Columbia University$-$Universidad de Chile $^{12}$CO(1$-$0) Survey of the Southern Galaxy \citep{bro89}. Using the rotation curve from \citet{alv90}, \citet{lop11} estimated the near kinematic distance of 1.8~kpc to be most probable since the region would be at 262~pc above the Galactic plane if the far distance were chosen, 4.4 times the half-width at half-maximum (HWHM) of the Galactic molecular disk \citep{bro00}. At 1.2~mm, the GMC is composed of 201 clumps with diameters ranging from 0.2 to 0.6~pc, masses ranging from 3~$M_{\sun}$ to 1.3$\times 10^{3}$~$M_{\sun}$ , and a star formation efficiency (SFE) of 2\% \citep{lop11}. The smallest ring, G345.45+1.5, mapped using APEX in the $^{13}$CO(3$-$2) emission line by \citet{lop16}, could have been created by a SN explosion represented by the 36.6~cm source J165920-400424 in its center and expanding at a velocity of 1~km~s$^{-1}.$ Because only 20\% of the clumps are associated with infrared (IR) counterparts \citep{lop11}, this GMC seems to have a low stellar activity. Several clump candidates are starless, but they only contain 28\% of the total mass forming clumps in the GMC.

\section{Observations and data reduction}\label{sect:observations}

\subsection{870\mm APEX-LABOCA}

The 870~$\mu$m continuum observations were carried out with the Large Apex BOlometer CAmera (LABOCA)\footnote{APEX is a collaboration between the Max-Planck-Institut f\"ur Radioastronomie, the European Southern Observatory, and the Onsala Space Observatory}. LABOCA is a 295-pixel bolometer array developed by the Max-Planck-Institut f\"ur Radioastronomie \citep{sir07}. Observations were carried out using the on-the-fly (OTF) mode to map four rectangular regions of $\delta \ell\times \delta$b $\sim$ 0.63$^{\circ} \times$2.66$^{\circ}$ in size, for a total area of $\delta \ell\times\delta$b $\sim$ 2.15$^{\circ}$ $\times$ 2.66$^{\circ}$ in size, centered on ($\ell$,$b$) = (345.4585$^{\circ}$, 1.30413$^{\circ}$). During the observations, the amount of precipitable water vapor (PWV) varied between 0.16 and 0.24~mm. Absolute flux calibrations were achieved through observations of Mars as primary calibrator and the stars N207\,11R and  VY\,CMa as secondary calibrators. The uncertainty due to flux calibration was estimated to be $\sim$20\%. The telescope focus and pointing were checked using the star $\eta$ Carinae. Observations were smoothed down to a beam of 21.8\arcsec. The data were reduced using the {\it \textup{Comprehensive Reduction Utility for SHARC-2 software package}} (CRUSH-2\footnote{\url{http://www.submm.caltech.edu/~sharc/crush/index.html}}; \citealt{kov08}) following the standard procedure, and this resulted in a rms of 0.2~Jy beam$^{-1}$.

\subsection{$^{12}$CO(4$-$3) NANTEN2}

Observations of the $^{12}$CO(4$-$3) transition line at 461.041~GHz were performed with the 4m NANTEN2 Telescope at Pampa La Bola between May and July 2012\footnote{NANTEN is a collaboration between Nagoya and Osaka Universities, Seoul National University, Universität zu Köln, Argelander-Institut Universität Bonn, ETH Zürich, University of New South Wales and Universidad de Chile.}. The KOSMA SMART receiver is a dual-frequency, 2x8 pixel array receiver operating between 460 and 880 GHz. The half-power beam width (HPBW) at 461.041~GHz is 40\arcsec~(0.35~pc at 1.8~kpc) and the velocity resolution is 0.7~km~s$^{-1}$. The smallest zones corresponding to 320\arcsec $\times$160\arcsec~in size were observed using the OTF mode with a total observed area of 0.37$^{\circ}\times$0.37$^{\circ}$ in size, centered on ($\ell$,$b$)=(345.2194$^{\circ}$, 1.30554$^{\circ}$) and a grid size of 8.5\arcsec$\times$8.5\arcsec. Each spectrum was observed for three seconds, giving a total observing time of 6.6~hours. The data reduction was made with the IRAM software GILDAS/CLASS\footnote{\url{https://www.iram.fr/IRAMFR/GILDAS}}. First, the data were filtered by opacity and noise temperature. Then, polynomial baselines of order three were subtracted from each spectrum and were then filtered by the RMS temperature of the fit. If needed, higher-order baselines were subtracted manually. After eliminating bad spectra (noisy and coupled signals, and holes), the data cube was produced by convolution with a Gaussian kernel with a grid of 20\arcsec$\times$20\arcsec, giving a final rms of 0.24~K.

\subsection{Ancillary data}

Our study is complemented with near-IR, mid-IR, and centimetric data. We used the 2MASS\footnote{\url{http://irsa.ipac.caltech.edu/Missions/2mass.html}} (1.25$-$2.15~$\mu$m, \citealt{skr06}) and $\it{Spitzer}$ GLIMPSE\footnote{\url{http://irsa.ipac.caltech.edu/Missions/spitzer.html}} (3.6$-$8~$\mu$m, \citealt{ben03}) observations to study the IR content of the region. The Supercosmos H$\rm{\alpha}$ Survey (SHS, \citealt{par05}) and the Molonglo Galactic Plane Survey at 36~cm (MGPS, \citealt{mur07}) are useful to study \HII regions (H$\rm{\alpha}$ and free-free emission), which are powered by massive stars as well as the free-free contamination, which can affect the 870~$\mu$m observations.  \herschel data from the Hi-GAL survey \citep{mol10a,eli17} between 70~$\mu$m (8\arcsec) and 500~$\mu$m (36.6\arcsec) were used to study the evolutionary stages of the clumps as well as  the large-scale temperature and column density in the region.

\section{Data analysis}\label{sect:analysis}

\subsection{Kinematic distance}

\begin{table*}
\caption{Masers detected toward G345.5+1.5}
\centering
\begin{tabular}{ccrrcc}
\hline
\hline
$\ell$ & $b$ & Masers  & Velocity & Kinematic distance & References\\
\cline{1-2}
\multicolumn{2}{c}{($^{\circ}$)} &  & (km~s$^{-1}$) & (kpc) & \\
\hline
\hline
345.013 & 1.798 & 1.665~GHz-OH & $-$23 & 2.5 & \citet{cas83a,cas95d}  \\
& & 22~GHz-H$_2$O & $-$18& 2.1 & \citet{cas83b,for89}  \\
& & 12~GHz-CH$_3$OH & $-$17 & 2.0 & \citet{cas83a,cas95c}  \\
& & 6.6~GHz-CH$_3$OH & $-$18 & 2.1 & \citet{men91,cas95d}  \\
& & 6.035~GHz-OH & $-$20 & 2.3 & \citet{cas95b}  \\
& & 4.7~GHz-OH & $-$25 & 2.7 & \citet{coh95}  \\
\hline
345.015 & 1.801 & 22~GHz-H$_2$O & $-$20 & 2.3 & \citet{cas83b} \\
& & 6.6~GHz-CH$_3$OH & $-$13 & 1.6 & \citet{cas95a,cas95d} \\
& & 12~GHz-CH$_3$OH & $-$13 & 1.6 & \citet{cas95a,cas95c} \\
\hline
345.123 & 1.592 & 1.7~GHz-OH & $-$17 & 2.0 & \citet{cas04} \\
\hline
345.464 & 1.460 & 4.7~GHz-OH & $-$15 & 1.9 & \citet{coh95}  \\
\hline
345.492 & 1.470 & 4.7~GHz-OH & $-$15 & 1.9 & \citet{coh95}  \\
& & 1.7~GHz-OH & $-$15 & 1.9 & \citet{cas04}  \\
\hline
345.210 & 1.035 & 4.7~GHz-OH & $-$9 & 1.2 & \citet{coh95}  \\
\hline
\hline
\end{tabular}
\tablefoot{
The kinematic distance is computed following the model of \citet{rei14}, which can be found at \url{http://bessel.vlbi-astrometry.org/revised_kd_2014}
}
\label{tab:masers}
\end{table*}

\begin{figure}
 \centering
 \includegraphics[angle=0,width=90mm]{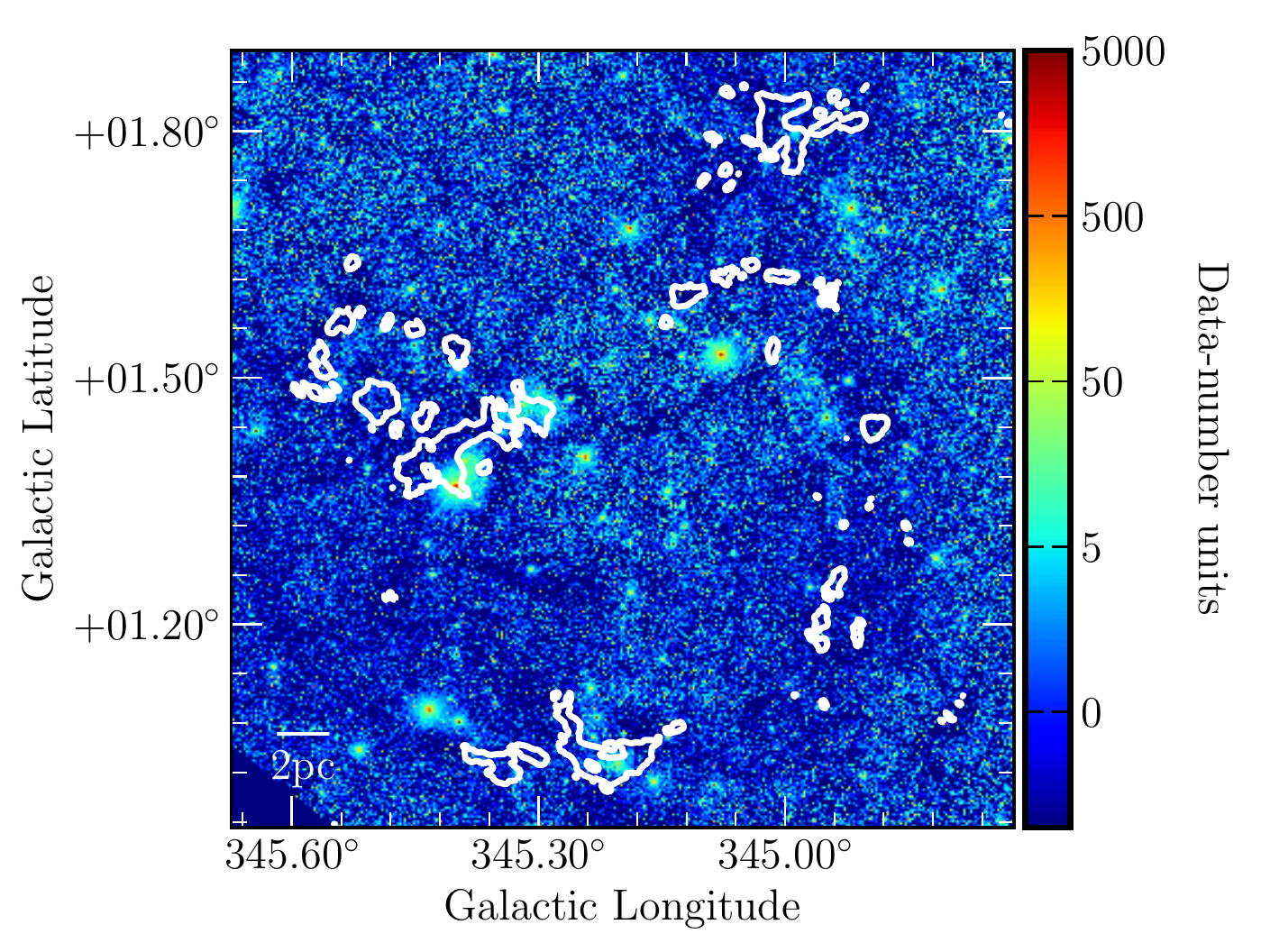}
  \caption{2MASS J-band image of G345.5+1.5 with the 1.2~mm emission superimposed as a white contour at 0.1~Jy~beam$^{-1}$.}
             \label{fig:Aj}
\end{figure}

\citet{lop11} rejected the far kinematic distance because it gives an elevation above the Galactic plane of 262~pc and a mass of 4.4$\times 10^{7}$~$M_{\odot}$. Using NH$_3$, N$_2$H$^+$ , and CS emission lines coupled with {H\sc{i}} absorption and self-absorption, \citet{wie15} resolved the kinematic distance ambiguity (KDA) for the ATLASGAL clumps. Kinematic distances from 1.86 to 1.95~kpc were assigned to the clumps of the G345.5+1.5 region, but only up to $b=1^{\circ}$ because the survey is limited in latitude. In Tab.~\ref{tab:masers} we summarize the different masers found toward the region and the different near kinematic distance computed from the central peak velocity using the Bayesian model of \citet{rei14}. The typical uncertainty for the distance is $\pm$0.6~kpc. Since the masers are distributed within different parts of the regions and the near kinematic distance agrees with the analysis of \citet{wie15}, the entire region should be located at approximately 1.8~kpc. Finally, in order to confirm whether the near kinematic distance is correct for G345.5+1.5, we show in Fig.~\ref{fig:Aj} the J-band image from 2MASS of G345.5+1.5. This figure shows that the 1.2~mm emission is well correlated with the extinction structures, suggesting that the region lies on the near side of the Galaxy \citep{rus11}.

\subsection{Large-scale H$\rm{\alpha}$ emission around G345.5+1.5}

\begin{figure}
 \centering
 \includegraphics[angle=0,width=90mm]{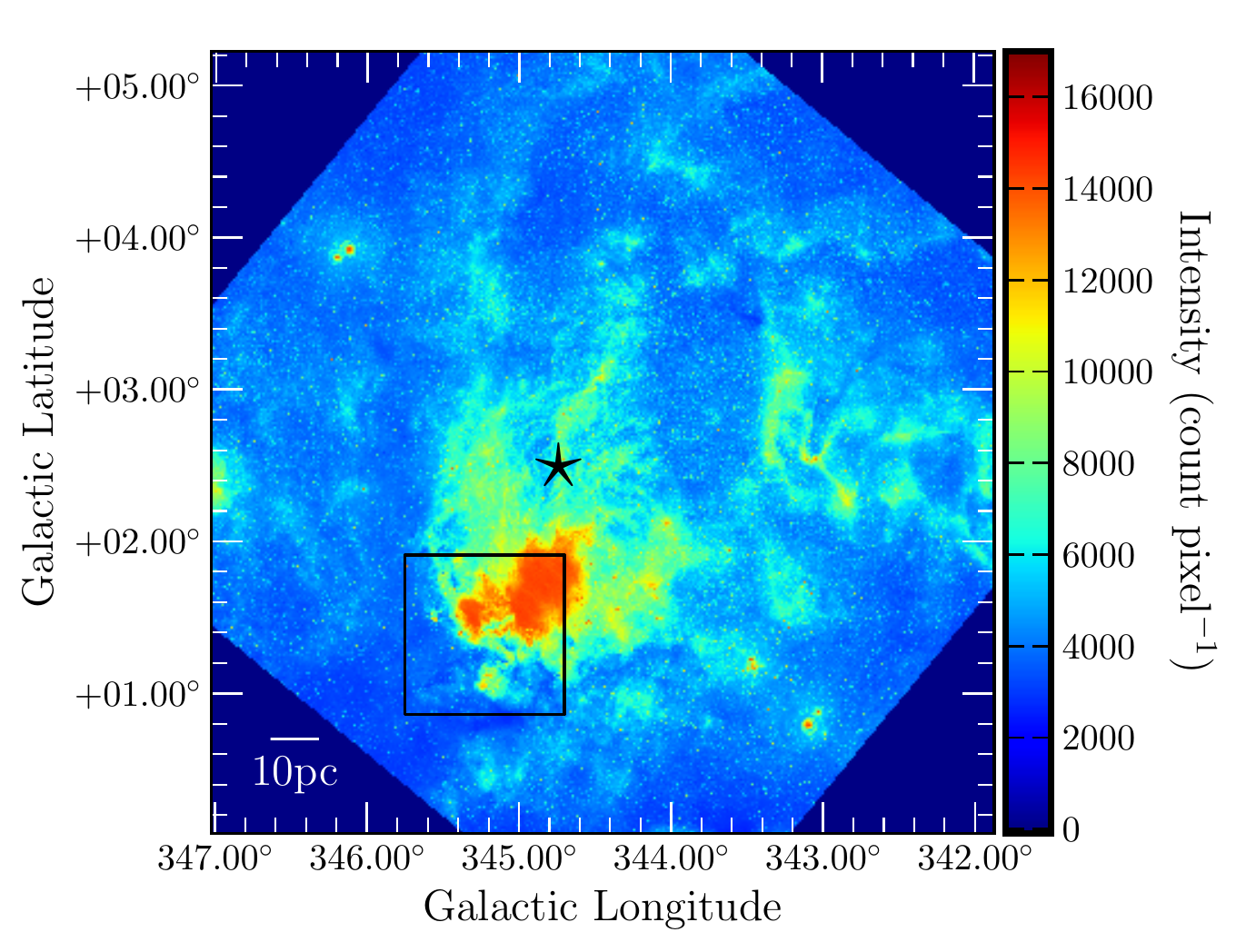}
  \caption{H$\rm{\alpha}$ image from the SHS. The black rectangle represents the location of G345.5+1.5 as shown in Fig.~\ref{Fig:G345_SuperCosmos}, and the black star represents the location of Nova Sco 1437.}
             \label{Fig:SuperCosmos_all}
\end{figure}

\begin{figure*}
\centering
\subfloat{
\includegraphics[angle=0,width=90mm]{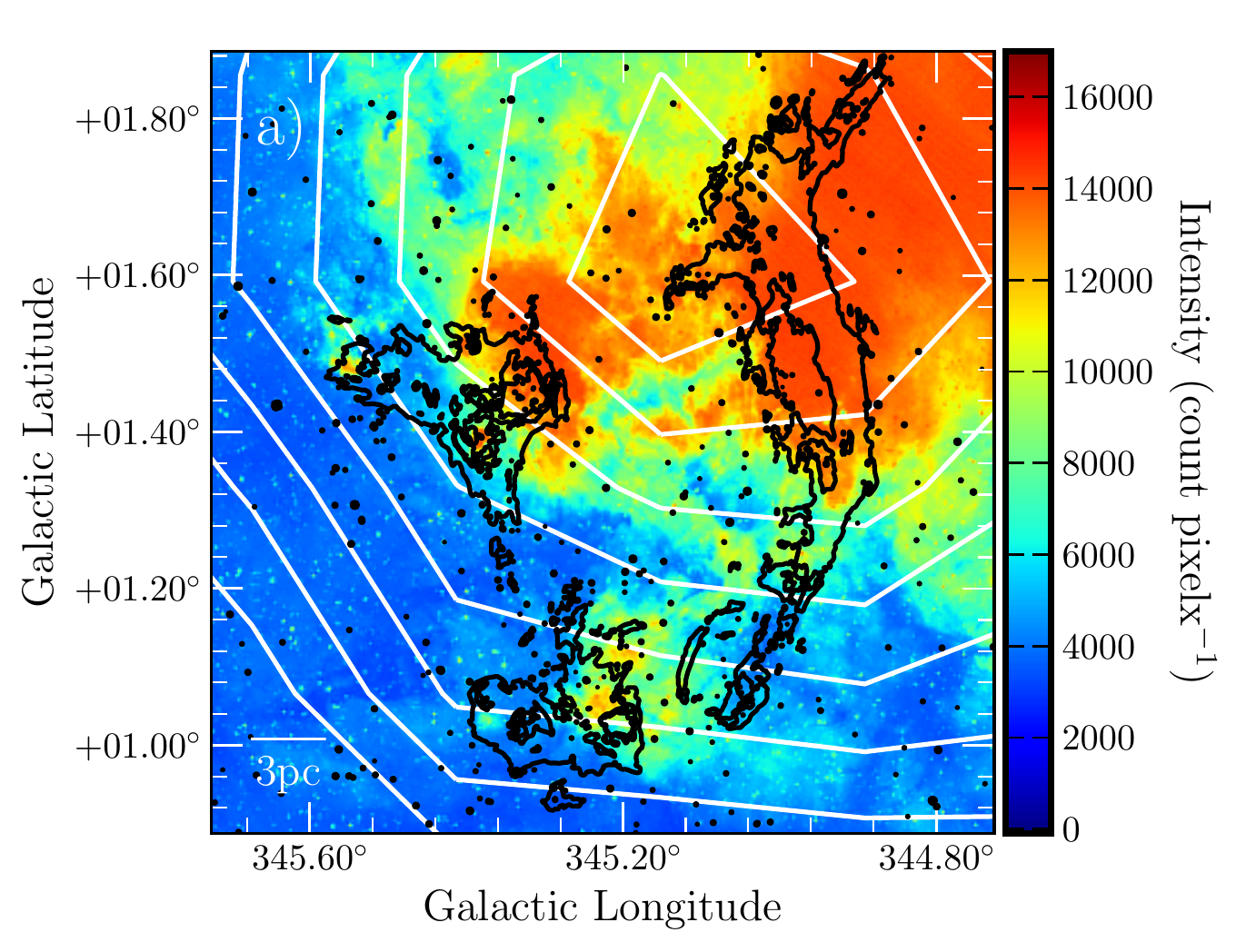}
\label{Fig:G345_SuperCosmos}}
\subfloat{
\includegraphics[angle=0,width=87mm]{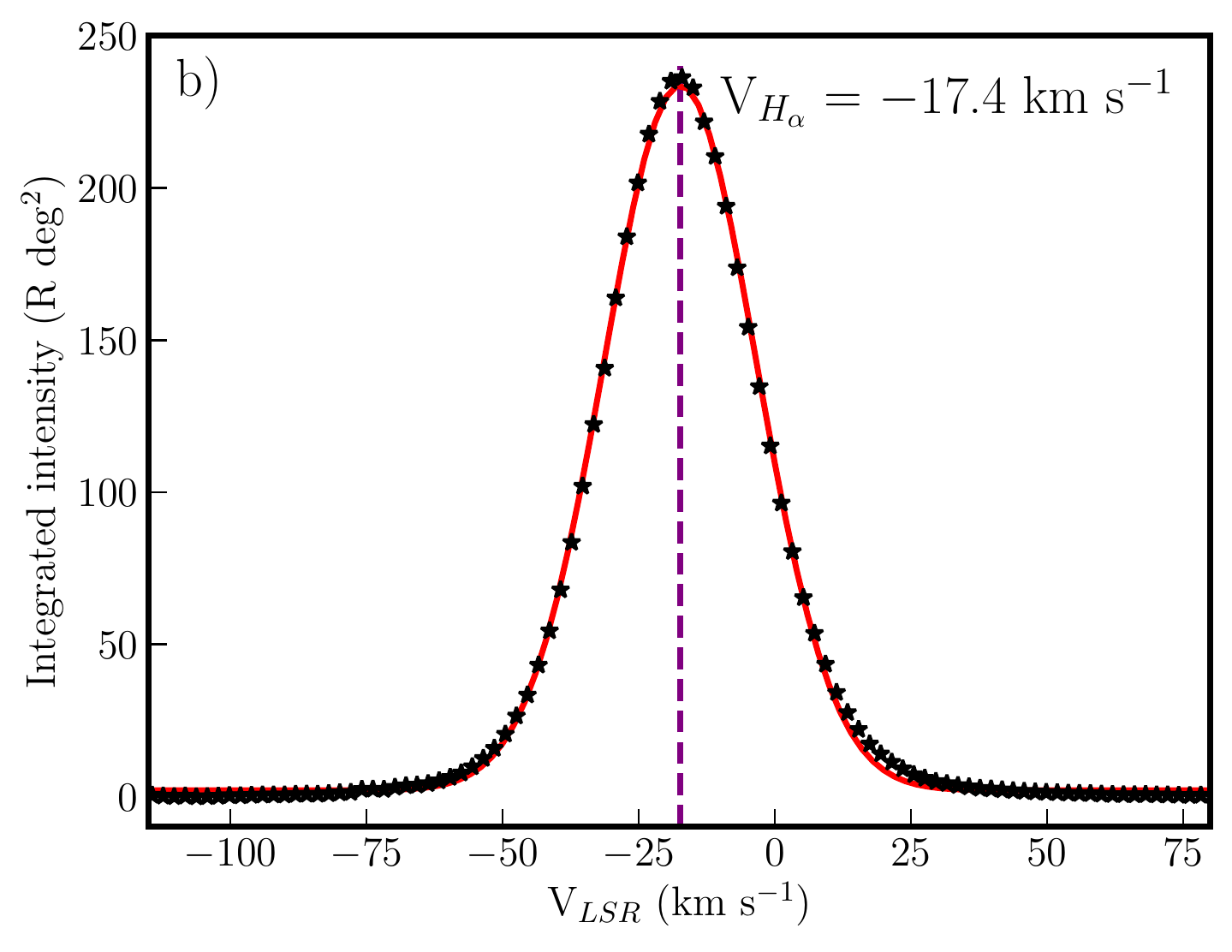}
\label{Fig:G345_average}}
\caption{a) H$\rm{\alpha}$ image from the SHS toward the G345.5+1.5 region. The black contours represent the 8~$\mu$m emission (100, 300, 500~MJy~sr$^{-1}$), and the white contours represent the integrated H$\rm{\alpha}$ emission (from 35 to 530~R~km~s$^{-1}$ with 55~R~km~s$^{-1}$ steps) from the WHAM-SS. b) Spatially integrated H$\rm{\alpha}$ emission where the star-points represent the data and the red curve represents the Gaussian fit to the data.}
\label{Fig:G345_WHAM+Fit}
\end{figure*}

 Figure~\ref{Fig:SuperCosmos_all} presents an image from SHS. This shows that G345.5+1.5 is observed toward a large H$\rm{\alpha}$ emission structure. To determine whether the large-scale emission of ionized gas is associated with the region, we used the Wisconsin H-Alpha Mapper Sky Survey (WHAM-SS, \citealt{haf03})\footnote{\url{http://www.astro.wisc.edu/wham-site/}} data with a spatial resolution of 1$^{\circ}$ and a velocity resolution of 2~km~s$^{-1}.$ Fig.~\ref{Fig:G345_SuperCosmos} presents the H$\rm{\alpha}$ emission from SHS with the 8~$\mu$m emission and the integrated WHAM-SS. The spectroscopic data agree with the continuum because the inner white contour is centered on the main H$\rm{\alpha}$ emission region from SHS. By spatially averaging the datacube, we plot in Fig.~\ref{Fig:G345_average} the integrated intensity versus V$_{\rm{LSR}}$ together with the Gaussian fit to the data. The peak is located at $-$17.4~km~s$^{-1}$ , which gives a near kinematic distance of 2.16$^{+0.53}_{-0.62}$~kpc. This is close to the distances found using the maser velocities (see Tab.~\ref{tab:masers}). If the H$\rm{\alpha}$ emission were clearly associated with G345.5+1.5, we would have observed a correlation between the H$\rm{\alpha}$ and dust emission through extinction features. Nonetheless,  Fig.~\ref{Fig:G345_SuperCosmos} shows that the dust emission at 8~$\mu$m seems to be mostly uncorrelated with the H$\rm{\alpha}$ cloud and should be located behind it, the uncertainty being $\sim$0.6~kpc for each kinematic distance. We searched in the public databases Vizier and SIMBAD for planetary nebulae, SN remnants, SN, white dwarfs, Wolf-Rayet stars, and novae to understand the presence of this H$\rm{\alpha}$ emission. We found Nova Sco 1437 (indicated by a black star in Fig.~\ref{Fig:SuperCosmos_all}). Unfortunately, we found no additional information on this nova, and its relation with the large H$\rm{\alpha}$ structure, if any, is beyond the scope of this paper. No particular H$\rm{\alpha}$ spatial distribution is found toward the G345.10+1.35 structure, whose ring morphology might be explained by the expansion of an \HII region powered by a cluster of OB stars toward its center.\\

\begin{figure}
 \centering
 \includegraphics[width=\linewidth]{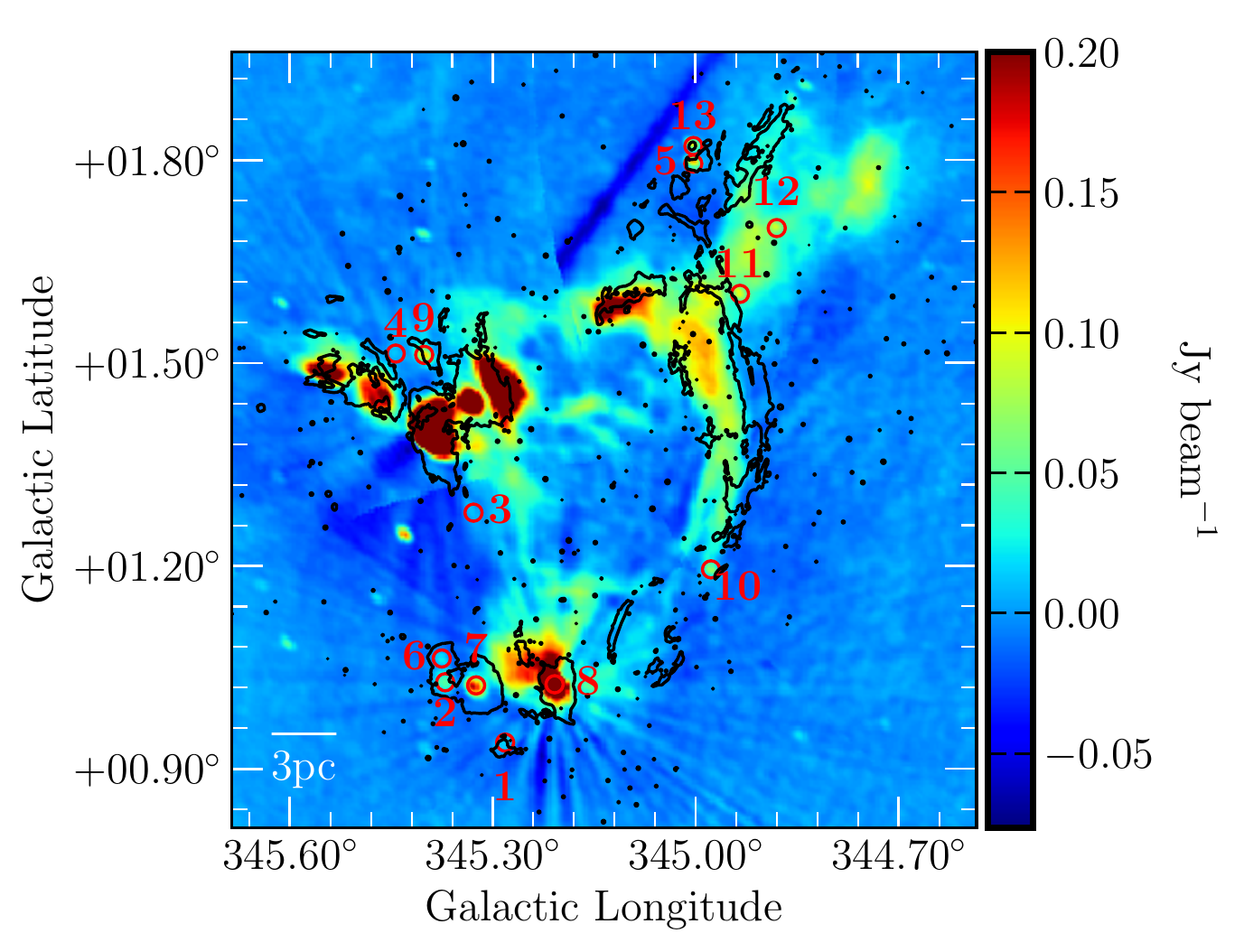}
  \caption{Radio continuum map at 843~MHz. The 8\mm contours at 40~Jy~beam$^{-1}$ are superimposed. Red circles represent the compact radio sources.}
\label{MGPS_radiosources}
\end{figure} 

\subsection{\HII regions and radio emission}\label{subsec:radio_compact}

To determine if \HII regions are associated with G345.5+1.5, we used the WISE catalog of \citet{and14}. This catalog contains 8399 objects divided into known \HII regions (detected radio recombination line, and/or H$\rm{\alpha}$ emission), \HII region candidates (coincident with radio continuum emission), group candidates (coincident with radio continuum emission and found within an \HII complex), and the radio-quiet sample (no radio continuum emission detected). Toward G345.5+1.5, 19 objects are found in the catalog, with five known \HII regions, eight group candidates, and six radio-quiet objects. To confirm possible \HII regions found in the candidate and radio-quiet samples, we used the 36~cm radio emission from the MGPS-2 survey \citep{mur07}, the H$\rm{\alpha}$ emission from SHS, and the 8~$\mu$m emission from \textit{Spitzer}. Emission at radio wavelength is less strongly affected by absorption than H$\rm{\alpha}$ emission, and G345.5+1.5 is covered by a huge H$\rm{\alpha}$ cloud. Moreover, it is a good tracer of \HII regions because it represents the free-free process that is caused by the bremsstrahlung. Fig.~\ref{MGPS_radiosources} presents the radio continuum emission at 843~MHz. In the catalog, we found five compact radio sources toward the region, but some very clearly visible compact radio sources were not included in the catalog. They were probably filtered out by the algorithm. We added them manually after checking that the peak brightness was higher than 5$\sigma_{\rm{MGPS}}$ ($\sim$10~mJy). Eight compact sources were added to the catalog, and they are presented in Fig.~\ref{MGPS_radiosources} and Tab.~\ref{tab:radio_sources_param}. The integrated flux of these radio sources was computed using aperture photometry and subtracting a background taken close to the radio source, but devoid of emission. In the following, we list the radio compact sources and estimate the spectral type of the star that is responsible for this radio emission if it has a clear effect on the surrounding medium seen at 8~$\mu$m.\\

\textbf{Radio source 1. }This source is listed as a radio-quiet object in the WISE catalog but presents a counterpart in MGPS-2 and at 8~$\mu$m. However, it could be a radial spike due to large brightness of radio source 8, and we did not consider it due to this uncertainty.

\textbf{Radio source 2. }This radio source is correlated with an \HII complex \citep{and11} and is part of the group sample because it is spatially coincident with two known \HII regions: G345.324+1.022 and G345.235+1.408, the first one being associated with radio source 7. Two layers of gas seen at 8~$\mu$m could represent dust pushed away by the expansion of the \HII region. This radio source is also correlated with H$\rm{\alpha}$ emission, whose central stars are clearly visible in the SHS image. To determine the spectral type of the ionizing source, we computed the number of Ly$\rm{\alpha}$ continuum photons from the MGPS-2 flux. We used the formula of \citet{mez67}:

\begin{equation}\label{eq:lyalpha}
\frac{N_{\rm{Ly \alpha}}}{8.9\times10^{46}~s^{-1}}=\left(\frac{S_{\nu}}{\rm{Jy}}\right)\times \left(\frac{\nu}{\rm{GHz}}\right)^{0.1}\times \left(\frac{T_e}{10^4~\rm{K}}\right)^{-0.45}\times \left(\frac{D}{\rm{kpc}}\right)^2
,\end{equation}

where $S_{\nu}$ is the integrated flux, $\nu$ is the frequency of the observation, $T_e$ is the electron temperature (of about 10$^4$~K, \citealt{kur99}), D is the distance to the region, and $N_{\rm{Ly \alpha}}$ is the ionization rate. The integrated flux at 843~MHz given in the MGPS-2 catalog for this source is 103.7~mJy. We found that $N_{\rm{Ly \alpha}}=2.9\times 10^{46}$~s$^{-1}$ , which corresponds to a B0.5V star according to \citet{pan73}. Because of the correlation with H$\rm{\alpha}$ and 8~$\mu$m, this group object is considered as an \HII region.

\textbf{Radio source 3. }The surrounding emission does not seem to be modified by this radio source, and no IR source is correlated with it. It is therefore possible that this source is located in the background, or might be an artifact because of the high ratio of its major to minor axis. Additionally, no particular emission in H$\rm{\alpha}$ is seen toward this source, and no counterpart is detected in the WISE catalog.

\textbf{Radio source 4. }This source presents no IR counterpart and has been proposed by \citet{lop16} to be the remnant of a SN explosion causing the G345.45+1.5 bubble to expand at a velocity of 1~km~s$^{-1}$.

\textbf{Radio source 5. }This source, together with radio source 13, is enclosed in a WISE radio-quiet object. Nonetheless, a 843~MHz counterpart is detected and could represent a very young \HII region, as identified by the QUaD Galactic Plane Survey \citep{cul11}. The H$\rm{\alpha}$ cloud in the foreground does not allow us to see the H$\rm{\alpha}$ emission associated with this compact source. A curved 8~$\mu$m emission is seen around it and is probably due to the ionization pressure acting on the molecular material. This radio source has an integrated flux of 119.7~mJy, which corresponds to a B0.5V star \citep{pan73}. 

\textbf{Radio source 6. }This source is listed in the group sample of the WISE catalog and has a counterpart at 843~MHz. Moreover, we clearly see a H$\rm{\alpha}$ counterpart in the form of a disk surrounded by 8~$\mu$m emission. Therefore this source is considered as an \HII region. From the integrated intensity of 326.8~mJy, the corresponding spectral type of the ionizing star is a B0.5III star.

\textbf{Radio source 7. }This radio source is correlated with a far-IR source whose colors are characteristic of an ultra-compact (UC) \HII region and CS(2$-$1) emission is detected toward it \citep{bro96}. It is part of a known \HII region from \citet{and14}, but no convincing H$\rm{\alpha}$ emission is seen toward this source, probably because of the contamination of the foreground H$\rm{\alpha}$ cloud. The integrated flux of the radio source is similar to that of radio source 6, giving a similar spectral type, B0.5III, for the ionizing star.

\textbf{Radio source 8. }This is the brightest radio source of the region with an integrated flux of 4.7~Jy and associated with an \HII region in the WISE catalog. We have to note that a smaller radio-quiet object in the WISE catalog is present at the same location, but the low resolution of MGPS-2 does not allow for its detection. The radio source indicates that an O9V star could be the ionizing source. Observations of \HII regions using atomic carbon C\,{\sc{i}}[$^3$P$_1\rightarrow^3$P$_0$] toward this region reveals a velocity of $-$15.7~km~s$^{-1}$ (2$^{+0.6}_{-0.7}$~kpc, \citealt{mao99}), which is associated with a cluster of stars and CS(2$-$1) emission at the same velocity \citep{bro96}. Spectroscopy in the K band with the VLT toward strongly reddened sources embedded in UC\HII regions reveals an early B, B0.5V, or O9V star (depending on the method used), which is part of a cluster with another O8V-BIV star \citep{bik06}, but at a distance of 1~kpc.

\textbf{Radio source 9. }This radio source is part of the WISE group sample. The integrated flux of 152.6~mJy corresponds to a B0.5V star and is correlated with H$\rm{\alpha}$ emission and an 8~$\mu$m curved emission arc. The star responsible for this radio source could be the $-$12.4~km~s$^{-1}$ IRAS source. A cluster of stars is also detected in this area, but at a velocity of $-$18.7~km~s$^{-1}$ \citep{kha13}. 

\textbf{Radio sources 10, 11, and 12. }These sources have no effect on their surrounding, and no H$\rm{\alpha}$ emission is associated with them. Since these sources were added to the catalog based on a visual inspection of the MGPS-2 map, they can be either artifacts, or foreground or background radio sources.

\textbf{Radio source 13. }This source is associated with a Red MSX Source (RMS) candidate \citep{lum13}, and the spectral type of the star responsible for this radio emission is B0.5V for a flux of 84.7~mJy. As discussed before, this radio source is close to radio source 5 and part of the WISE radio-quiet bubble sample. \citep{and14}.\\

\begin{table}
\tiny
\caption{Detected 36~cm radio sources toward G345.5+1.5. }
\centering
\begin{tabular}{c|ccrcc}
\hline
\hline
 Id & $\ell$ & $b$ & Flux & Spectral Type & Note \\
 \cline{2-3}
  & \multicolumn{2}{c}{($^{\circ}$)} & (mJy) & & \\
\hline
\hline
        1       &       345.282 &       0.939   &       22.7    &               &       No      \HII region  \\                              
        2       &       345.370 &       1.028   &       103.7   &       B0.5V   &       \HII region  \\                                              
        3       &       345.329 &       1.279   &       15.0    &               &       No      \HII    region  \\                              
        4       &       345.446 &       1.514   &       21.9    &               &       SN remnant ?       \\                                      
        5       &       345.003 &       1.795   &       119.7   &       B0.5V   &       \HII region  \\      
        \hline                                  
        6       &       344.375 &       1.063   &       326.8   &       B0.5III &       \HII region \\                                               
        7       &       345.325 &       1.023   &       323.9   &       B0.5III &       \HII region \\                                                       
        8       &       345.209 &       1.025   &       4718.1  &       O9V         &\HII region    \\                                              
        9       &       345.401 &       1.512   &       152.6   &       B0.5V   &       \HII region          \\                      
        10&     344.977 &       1.195   &       62.3    &               &       No \HII region     \\
        11      &       344.934 &       1.602   &       38.6    &               & No \HII region  \\                                                      
        12&     344.889 &       1.706   &       39.2    &               & No \HII region  \\                                                      
        13&     345.003 &       1.821   &       84.7    &       B0.5V   & \HII region     \\                                                      
\hline
\hline
\end{tabular}
\tablefoot{
Radio sources 1 to 5 were extracted from the MGPS-2 catalog, while radio sources 6 to 13 were added manually.
}
\label{tab:radio_sources_param}
\end{table}

To summarize, three \HII region candidates of the group sample (MGPS sources 2, 6, and 9) and one radio-quiet object (containing the MGPS sources 5 and 13) in the WISE catalog are considered as \HII regions because of the distortion of the dust emission seen around them and/or associated H$\rm{\alpha}$ emission. MGPS source 1, associated with a radio-quiet object, can be an \HII region or a radial spike artifact. The 5 remaining \HII candidates of the group sample, despite the high probability of being an \HII region \citep{and14}, could not be confirmed. Finally, 7 \HII regions (up to 12, considering the group sample) are found toward G345.5+1.5, five MGPS sources are not related to \HII regions, and it is possible that one of them represents a SN remnant \citep{lop16}. Mosaics of these sources at 8~$\mu$m, 22~$\mu$m, H$\rm{\alpha,}$ and 36~cm are presented in Fig.~\ref{fig:mosaic_radiosource}.

\subsection{870\mm clumps detection}\label{subsect:870clumpsdetect}

\begin{figure}
 \centering
\hspace*{-0.3cm}\includegraphics[scale=0.8]{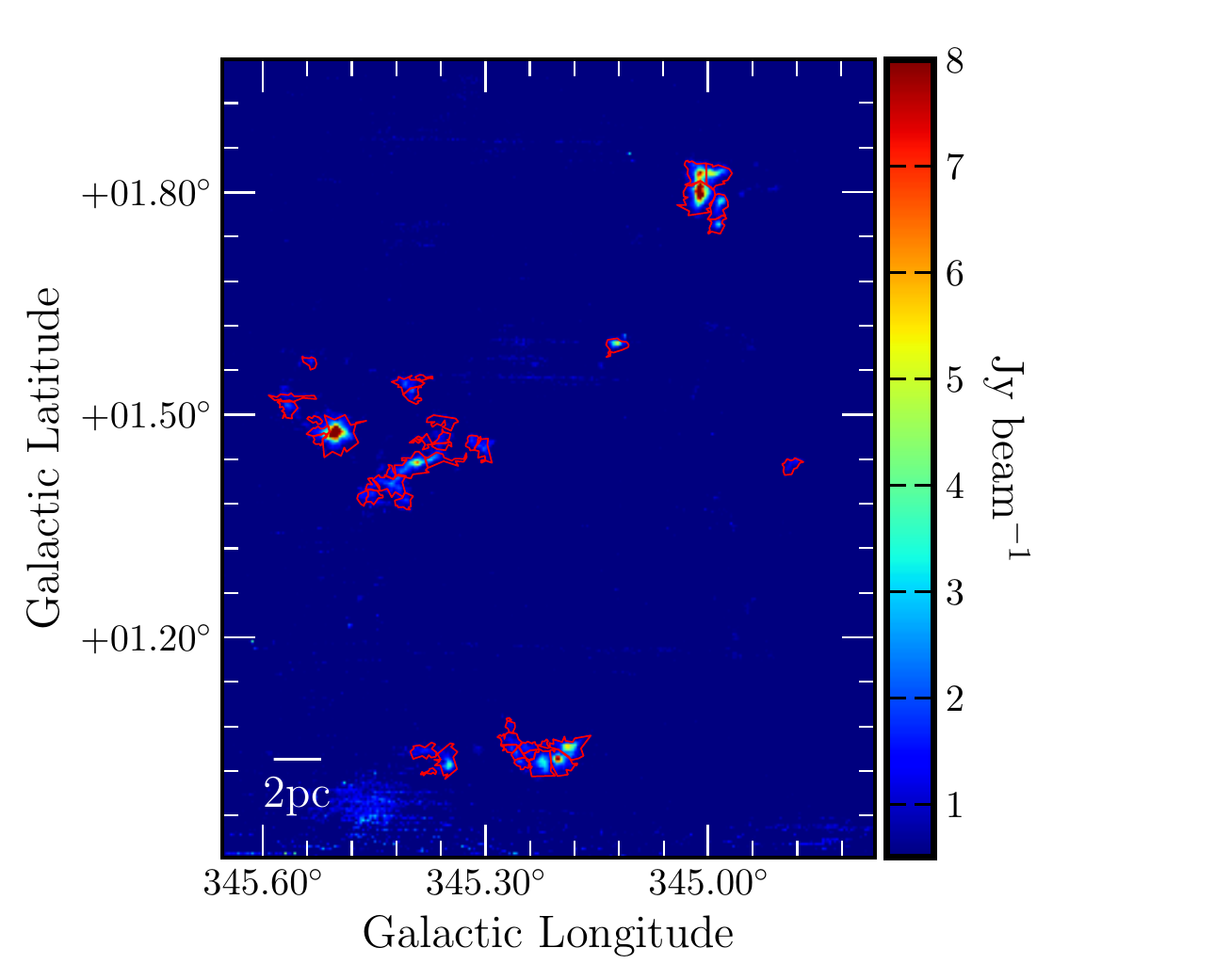}
  \caption{APEX-LABOCA 870\mm emission map of G345.5+1.5 at 3$\sigma$ ($\sim$0.6~Jy beam$^{-1}$). The red polygons represent the 45 clumps detected using the \textit{Clumpfind} algorithm.}
             \label{fig:G345_870}
\end{figure}

The G345.5+1.5 region is, as shown by \citet{lop11} with 1.2~mm data, fragmented into several clumps. Since the region extends up to a latitude of 1.9$^{\circ}$, the ATLASGAL catalogs \citep{con13,cse14} could not be used, and we had to extract the clumps independently. We used the \textit{CUPID} implementation of the \textit{Clumpfind} algorithm, fully described in \citet{wil94}, to extract the clumps in the 870\mm continuum emission map. \textit{Clumpfind} searches for emission peaks in the map and tracks them down until a contour level $C_{low}$ with a contour step of $\Delta_{inc}$, these parameters being provided by the user. In this method, a pixel can only belong to one clump. During the extraction, clumps whose size is smaller than the beam are rejected and the clump size is deconvolved in quadrature. For this study, we chose a $\Delta_{inc}$ of $2.5\sigma$ and a $C_{low}$ of $3\sigma$ and also excluded the stripes seen in the map, which could lead to false detections. With this method and these initial parameters, we found a sample of 45 clumps, whose contours are superimposed on the 870~$\mu$m map in Fig.~\ref{fig:G345_870}.

\subsection{870\mm clump properties}

\subsubsection{Contamination by radio emission sources}\label{subsect:contamination}

Submillimeter emission is usually dominated by dust, but in star-forming regions, free-free or synchrotron emission can significantly contaminate this wavelength range and affect the computation of the physical parameters. Because we found 7 to 12 \HII regions toward G345.5+1.5, we assumed that the contaminating radio emission comes entirely from the free-free process. To quantify this contamination, we compared the clump emission at 870~$\mu$m with the emission from MGPS-2 extrapolated at the same wavelength with a spectral index $\alpha _{ff}$ equal to $-0.1$, which is characteristic of the free-free emission in diffuse \HII regions \citep{mez67,rum16}. The 870~$\mu$m map was first convolved to the MGPS-2 resolution so that both observations could be compared. The percentage of contamination by the free-free emission is on average equal to 4\% in our sample of 45 870~$\mu$m clumps, with a peak at 25\%. While a close and strong radio source can highly contaminate a clump, low-mass clumps can be more affected than higher mass clumps even if the clump in question is not close to a radio source. For instance, six clumps present a contamination percentage higher than 10\%. Three of these (clumps 3, 14, and 43) are found toward the \HII region that is represented by the brightest radio source 8, and the three others (clumps 15, 22, and 24, associated with the contamination peak) are found toward the small ring G345.45+1.5, where most of the \HII candidates are observed.

\subsubsection{Physical parameters}
 
The \textit{Clumpfind} algorithm directly provides us physical parameters of the clumps, such as the standard deviation of the Gaussian, $\sigma _{\rm{c}}$, that was used to defined the clump, as well as the integrated flux. As stated before, an automatic quadratic deconvolution was performed, and the flux was scaled accordingly. The physical diameter $D_{\rm{c}}$ of the clumps was taken to be the FWHM computed as $\sqrt{8\rm{ln(}2\rm{)}}\times \sigma _.{\rm{c}}$. The integrated intensity was corrected for the free-free contamination to obtain better estimates of the mass, column density, and volume density values. The mass was computed using the formula of \citet{hil83} :

\begin{equation}\label{eq:hil}
M_{\rm{c}}=\frac{R\times S_{870\mu m}\times D^2}{B_{870\mu m}(T_{\rm{dust}})\times \kappa_{870\mu m}}
,\end{equation}

where $R$ is the gas-to-dust ratio, $S_{870\mu m}$ is the integrated flux, $D$ is the distance to the region, $B_{870\mu m}(T_{\rm{dust}})$ is the blackbody flux at 870\mm and at a temperature $T_{\rm{dust}}$ , and $\kappa_{870\mu m}$ is the opacity factor. We chose a value of $T_{\rm{dust}}$=20~K based on \citet{deh09} and \citet{cse17} at the same wavelength. We do not expect the temperature to be lower than this, and the resulting mass can be considered as an upper limit. The parameter $R$ is uncertain, but a value of 100 is commonly used \citep{mot10,pal13,cse17}. We followed the opacity law given in \citet{oss94}, ${\kappa_{\nu}=10\times (\nu/1\rm{THz})^{\beta}}$, with a spectral index $\beta$ of 2, and a value of 1.18~cm$^{2}$~g$^{-1}$ was adopted at $870$~$\mu$m. The value of the opacity factor is still debated because it depends on the dust environment and thus gives rise to an uncertainty of up to a factor of $\sim$ 2 for the clump mass \citep{deh12}. We also have to note that ground-based telescopes filter the large-scale emission during the data reduction, and consequently, a fraction of the mass is lost \citep{cse16}. The average column and volume density were computed following 

\begin{equation}\label{nh2_formula}
N({\rm{H_{2}}})=\frac{M_{\rm{c}}}{\mu m_{\rm{H}}\times\pi\times(\frac{D_{\rm{c}}}{2})^2}
\end{equation}

\begin{equation}\label{vnh2_formula}
n({\rm{H_{2}}})=\frac{M_{\rm{c}}}{\frac{4}{3}\pi\times \mu m_{\rm{H}}\times(\frac{D_{\rm{c}}}{2})^3}
,\end{equation}

where $\mu$ is the mean molecular weight per hydrogen molecule set to 2.8 \citep{kau08}, $m_{\rm{H}}$ is the hydrogen mass, $M_{\rm{c}}$ is the clump mass, and $D_{\rm{c}}$ is the clump diameter. Table~\ref{tab_param} lists all the parameters for each of the 45 870~$\mu$m clumps corrected for the free-free emission, together with early class YSOs (see Sect.~\ref{subsect:IR}).\\
The clump extraction was also performed by \citet{lop11} using the 1.2~mm SIMBA observations and the \textit{Clumpfind} algorithm. Because the observed region in their work is larger (see Fig.~\ref{Fig:G345_1200}), the sample of 201 1.2~mm clumps was reduced to 166 in order to match the area observed by APEX. Therefore, they detected between three to four times more clumps than at 870~$\mu$m. This difference cannot be fully explained by the different resolution and the extraction method because the resolution is approximately the same and the extraction was made in the same way. When we assume that the dust emission behaves as a modified blackbody (Eq.~\ref{eq:graybody-fd}), the emission at 870~$\mu$m should be higher than at 1.2.~mm and we should have recovered more clumps. However, the rms at 870~$\mu$m and 1.2~mm differs by a factor of 10 ($\sigma _{1.2mm}$=0.02~Jy~beam$^{-1}$), and the higher flux expected at 870~$\mu$m is not enough to compensate for this difference in the rms. Therefore, most of the weak emission clumps were completely removed during the extraction down to 3$\sigma$. Clumps whose emission peak was higher than 3$\sigma$ but whose size was reduced because of the threshold were removed if they were smaller than the beam. This can be seen by comparing Fig.~\ref{Fig:G345_1200} and Fig.~\ref{fig:G345_870}, where the ring part of G345.1+1.35 at $\ell=344.9^{\circ}$ is above 3$\sigma _{1.2mm}$ but below 3$\sigma _{870\mu m}$. In this area, 42 clumps at 1.2~mm are detected, but we only recovered 2 of them at 870~$\mu$m. The same statement can be made for the northern part of G345.45+1.5, where 15 clumps could not be recovered, and around the concentration of clumps seen toward ($b=1.04^{\circ}, 1.5^{\circ}, 1.8^{\circ}$), where low-emission clumps are lost. When we use the integrated emission at 1.2~mm (see Tab.~5 of \citealt{lop11}) and Eq.~\ref{eq:hil}, the total mass of the 1.2~mm clumps that are also found in this work is 2.1$\times 10^{4}$~$M_{\odot}$ , and this agrees with the total dust mass of 2.2$\times 10^{4}$~$M_{\odot}$ at 870~$\mu$m. The total mass of clumps thatare not recovered in our observation is equal to 4.1$\times 10^{3}$~$M_{\odot}$. This is about 17\% of the total mass. The average clump mass that we do not detect is about 48~$M_{\odot}$. 

\subsection{Spatial distribution and properties of \herschel sources}\label{subsect:herschel}

To analyze the star formation around G345.5+1.5, we also used the catalog from the Hi-GAL survey \citep{mol10b,eli17}. The extraction, made with the \cutex algorithm \citep{mol11} and the 70, 160, 250, 350, 500~$\mu$m bands from PACS \citep{pog10} and SPIRE \citep{gri10}, is limited in latitude since PACS observations only reach $\sim$1.2$^{\circ}$. Toward the southern part of G345.5+1.5, 79 sources were extracted from the catalog together with their associated physical parameters (see Fig.~\ref{fig:preproto}). However, the distance chosen to compute the mass and the bolometric luminosity was not necessarily accurate \citep{tra18b} and not given for 25 sources (set to an arbitrary value of 1~kpc in that case). To be consistent, the distance-dependent parameters were scaled to the G345.5+1.5 distance of 1.8~kpc. The diameter of the sources is equal to the deconvolved FWHM of the source footprints at 250~$\mu$m, and following the scheme of \citet{ber07}, 21 sources are classified as cores ($D<0.2$~pc) and 58 as clumps ($0.2$~pc $\le D \le 3$~pc). If a source was associated with an IR counterpart (see Sect.~{\ref{subsect:IR}), it was included in the computation of the bolometric luminosity. Explanations about the spectral energy distribution (SED) fitting method and the computation of the physical parameters can be found in \citet{eli17}. The uncertainty on the temperature and the envelope mass was derived from the SED fitting, and the uncertainty on the bolometric luminosity followed from them. The parameters of the Hi-GAL clumps are listed in Tab.~\ref{ap:higal_clumps}.\\

\begin{figure}
 \centering
 \includegraphics[angle=0,width=90mm]{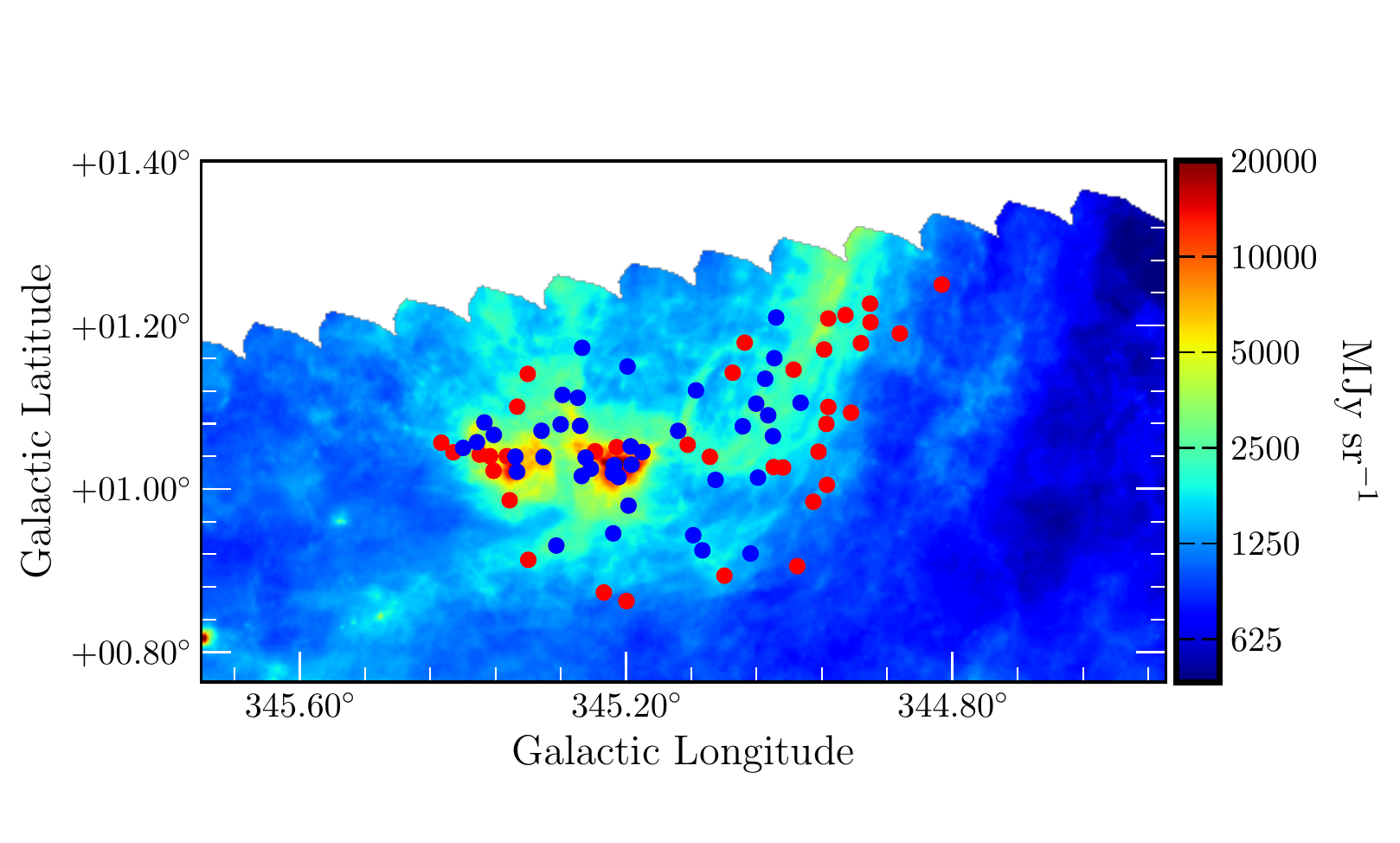} 
 \caption{\herschel 160~$\mu$m map with protostellar (blue) and prestellar sources (red) superimposed.}
 \label{fig:preproto}
\end{figure}

\subsection{Association with IR sources}\label{subsect:IR}

To characterize the star formation occurring in G345.5+1.5, we searched for IR counterparts in the GLIMPSE 3D \citep{chu09} merged with the 2MASS Point Sources Catalog (PSC, \citealt{skr06}) for each of the clumps detected at 870~$\mu$m and applied different criteria to the sources in order to exclude stars and keep the YSOs. We only kept sources with a close source flag of 0 (no sources within 3\arcsec) to prevent contamination by other sources during the photometry step. Depending on their evolution, these sources can be grouped into different classes \citep{and00}: main accretion phase (Class~0), late accretion phase (Class~I), pre-main sequence (PMS) stars with protoplanetary disks (Class~II), and PMS stars with debris disks (Class~III). We used three different indicators: the $J-H$ versus $H-K$, the $[3.6]-[4.5]$ versus $[5.8]-[8.0]$ color-color diagrams where YSOs and stars are expected to be at different locations, and the value of the SED slope from 3.6~$\mu$m to 8~$\mu$m, which can also be linked to the evolutionary class of the source \citep{lad87}.

\begin{figure*}
\centering
\begin{minipage}{0.325\textwidth}
\subfloat{
\includegraphics[width=\textwidth]{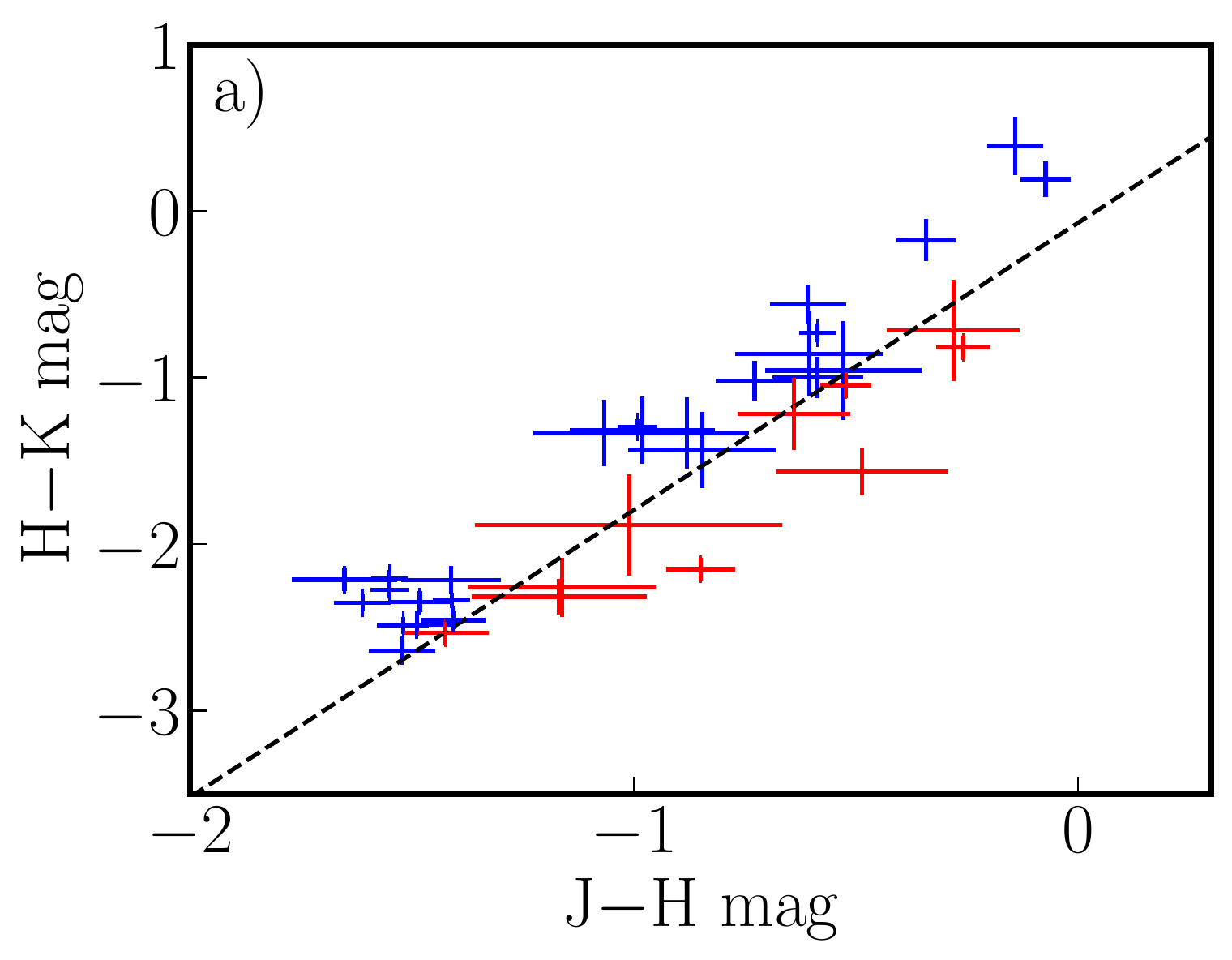}
\label{fig:Mass_cc}}
\end{minipage}
\begin{minipage}{0.325\textwidth}
\subfloat{
\vspace*{0.1cm}\includegraphics[width=\textwidth]{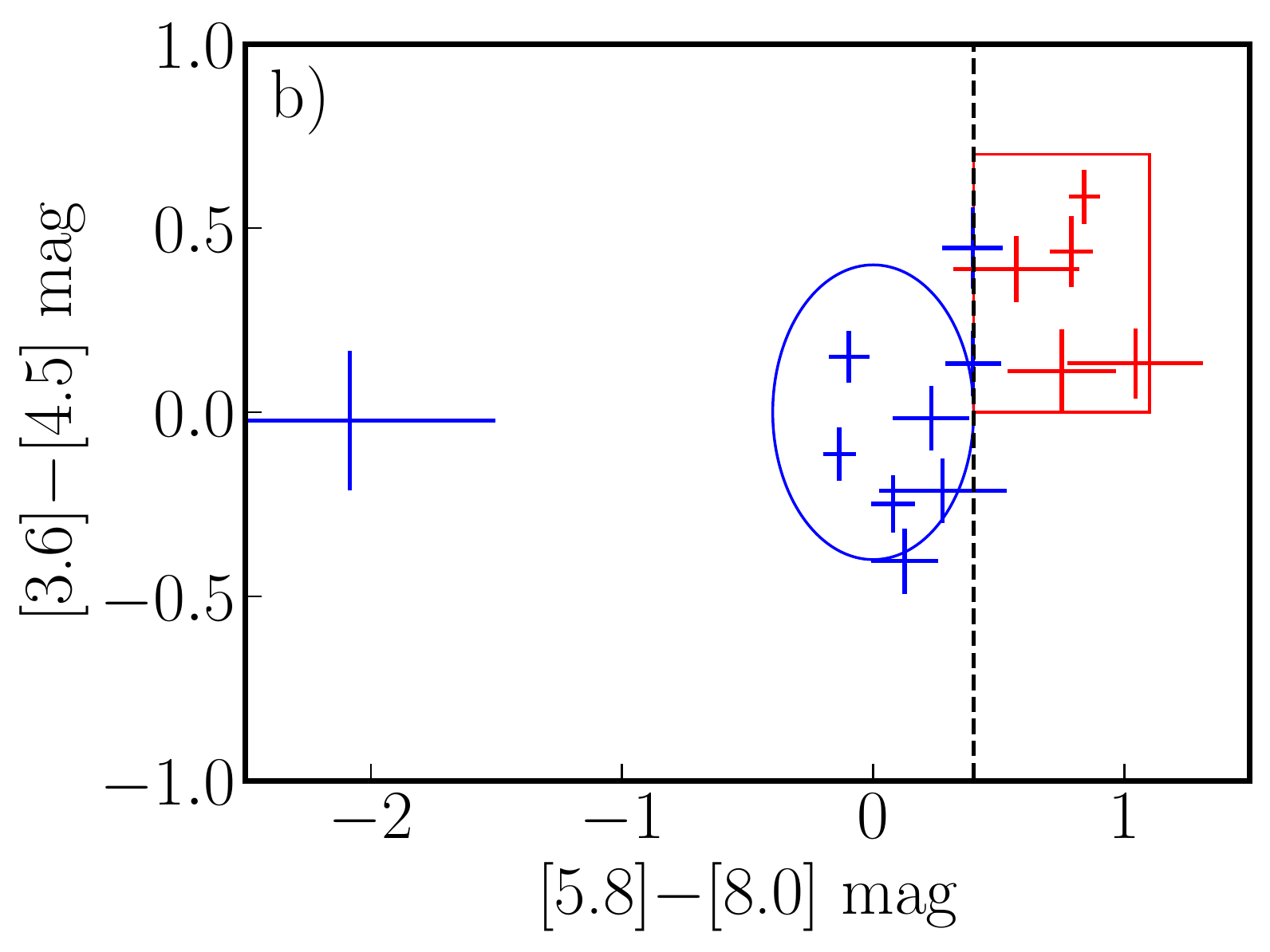}
\label{fig:Glimpse_cc}}
\end{minipage}
\begin{minipage}{0.34\textwidth}
\subfloat{
\includegraphics[width=\textwidth]{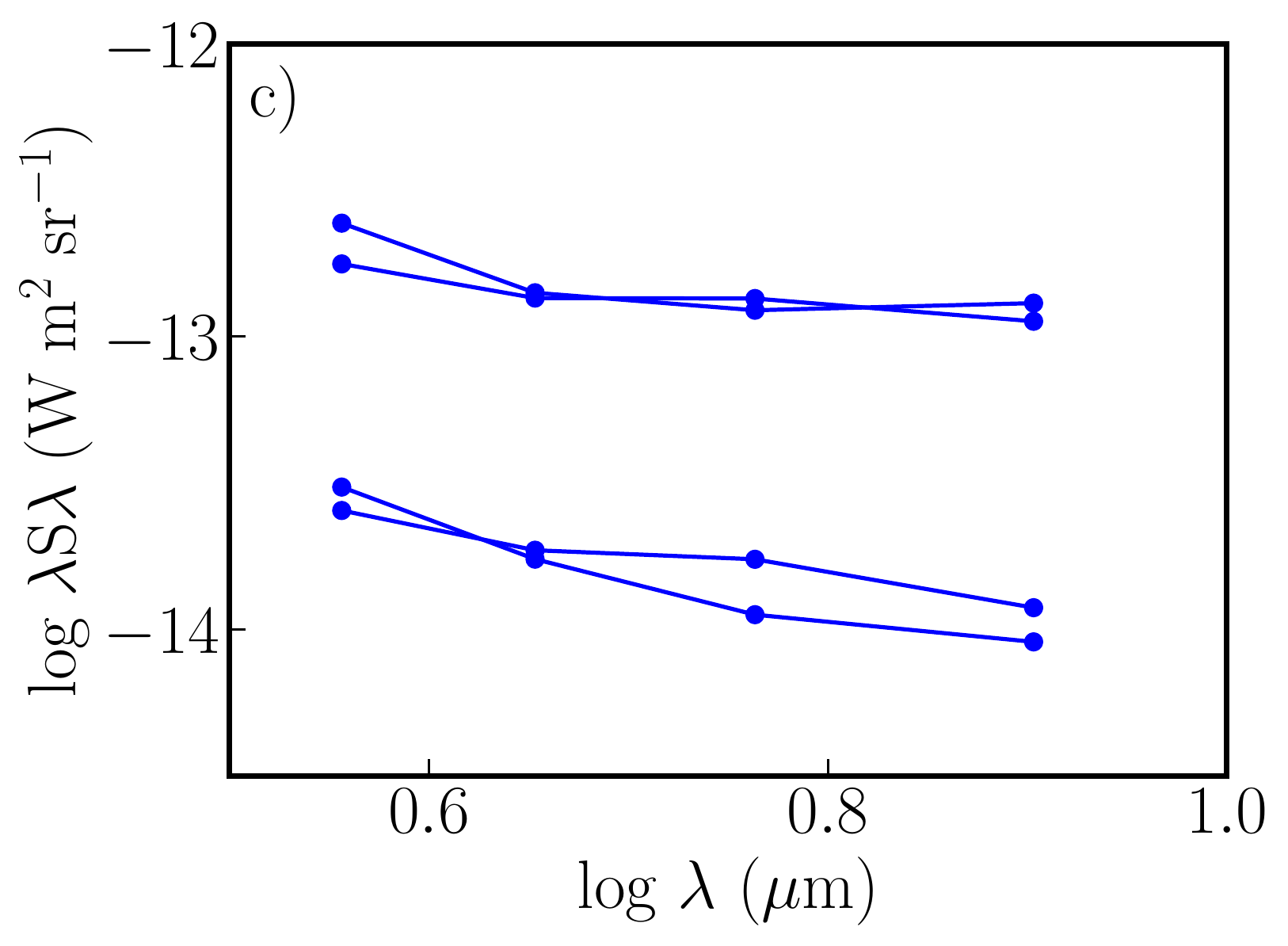}
\label{fig:Glimpse_sed}}
\end{minipage}
\caption{a) J$-$H vs. H$-$K color-color diagram. The line represents the location of an O9V star depending on the reddening \citep{tok00}. Blue and red crosses represent the IR counterparts that are considered Class~III YSO or stars and early YSOs (from Class~0 to Class~II). b) $[3.6]-[4.5]$ vs. $ [5.8]-[8.0]$ color-color diagram. The blue ellipse and red box represent the location of Class III YSO or stars and YSOs following the models of \citet{all04}. Red crosses represent Class~II (or earlier) YSOs located right of the blue dashed line that delineates the Class~II and Class~III stars. c) SED from 3.6\mm to 8~$\mu$m.}
\label{color-color}
\end{figure*}

Fig.~\ref{color-color} shows an example of the three indicators that we used for one clump. By assuming $A_V=7.5A_K$ with $R_V=5$ \citep{car89}, where $A_V$, $A_K$ , and $R_V$ are the visual extinction, the K-band extinction, and the visual extinction to color excess ratio, respectively, all the sources were dereddened with a visual extinction corresponding to the column density of the hosting clump that was computed with the relation of \citet{boh78}. Following the extinction law of \citet{ind05}, we plotted on the J$-$H versus H$-$K color-color diagram the location of an O9V star \citep{tok00} along a line representing the extinction due to the clump (Fig.~\ref{fig:Mass_cc}). Sources above this line are expected to represent mostly stars, while YSOs (Class~0-I-II) populate the bottom right corner of the plot as a result of the IR excess that originated in the circumstellar dusty layers.
Following the disk and envelope models of \citet{all04}, the location of sources in the $[3.6]-[4.5]$ versus $[5.8]-[8.0]$ color-color diagram (Fig.~\ref{fig:Glimpse_cc}) is also representative of their class. Stars and Class~III YSOs should be located within the blue ellipse centered on the origin, while Class~II and Class~I objects are found toward the red rectangle and above it, respectively. We selected the early YSOs as all the sources located to the right of the dashed blue line that defines the separation between Class~III stars and Class~0-I-II objects. To avoid doubtful classifications, we excluded the YSOs that were classified differently following color-color diagrams. To confirm the class of the selected YSOs, we computed the slope of the SED between 3.6\mm and 8~$\mu$m (Fig.~\ref{fig:Glimpse_sed}). Each of the sources classified as Class~0 to II has a slope higher than $-$1.5, the limit for Class~II objects \citep{lad87,and00}. Class~0 to II YSOs were found toward 26 of the 45 clumps ($\sim$58\%) of the sample. However, it is possible that some sources are located in the foreground and not associated with the region, while others remain undetected if they are deeply embedded. We note that a high free-free contamination does not necessarily mean that the clump hosts IR counterparts because we searched for early class YSOs, while the high-mass stars emitting the ionizing flux are late class YSOs or stars. As explained in Sect.~\ref{subsect:contamination}, a high free-free contamination can also be due to the low mass of a clump.\\
For the Hi-GAL sources, IR counterparts at 21~$\mu$m (MSX), 22~$\mu$m (WISE), and/or 24~$\mu$m (MIPSGAL) were considered only if a 70~$\mu$m source was found in the 250~$\mu$m emission footprint. To ensure the presence or absence of it, \citet{eli17} performed another extraction using a lower threshold than was used for the general extraction. If no 70~$\mu$m detection was found in this new extraction, no IR counterparts at 21, 22, or 24~$\mu$m were associated with the source. This method is fully described in \citet{eli17}. In the sample of 79 Hi-GAL sources, 37 are prestellar and 42 are protostellar. Fig.~\ref{fig:preproto} shows the prestellar and protostellar sources superimposed on the 160~$\mu$m image. The densest parts and the inner parts of the region host more protostellar than prestellar sources, but a few sources contradict this tendency. The temperature of these protostellar clumps given by their SED is found to be higher than that of the prestellar clumps \citep{rag12,liu17} because of the ongoing star-formation that occurs inside, even if the temperature given by the SED is an average over the clump and does not indicate the temperature of the inner parts. The distribution is shown in Fig.~\ref{fig:histo_temp}, where the temperature of the prestellar clumps ranges from 8 to 18~K and peaks at (14$\pm$0.9)~K, while for protostellar clumps, it ranges from 10 to 30~K and peaks at (16$\pm$1.2)~K \citep{oha16,olm16,eli17}. When we compared the 870~$\mu$m clumps and Hi-GAL sources below $b=1.2^{\circ}$ with respect to their IR counterpart, we observed a good agreement between the distribution of star-forming (protostellar) and non-star-forming (prestellar) clumps, where only one of them is classified as protostellar with the Hi-GAL criteria and as prestellar with 870~$\mu$m. 

\begin{figure}
 \centering
 \includegraphics[angle=0,width=90mm]{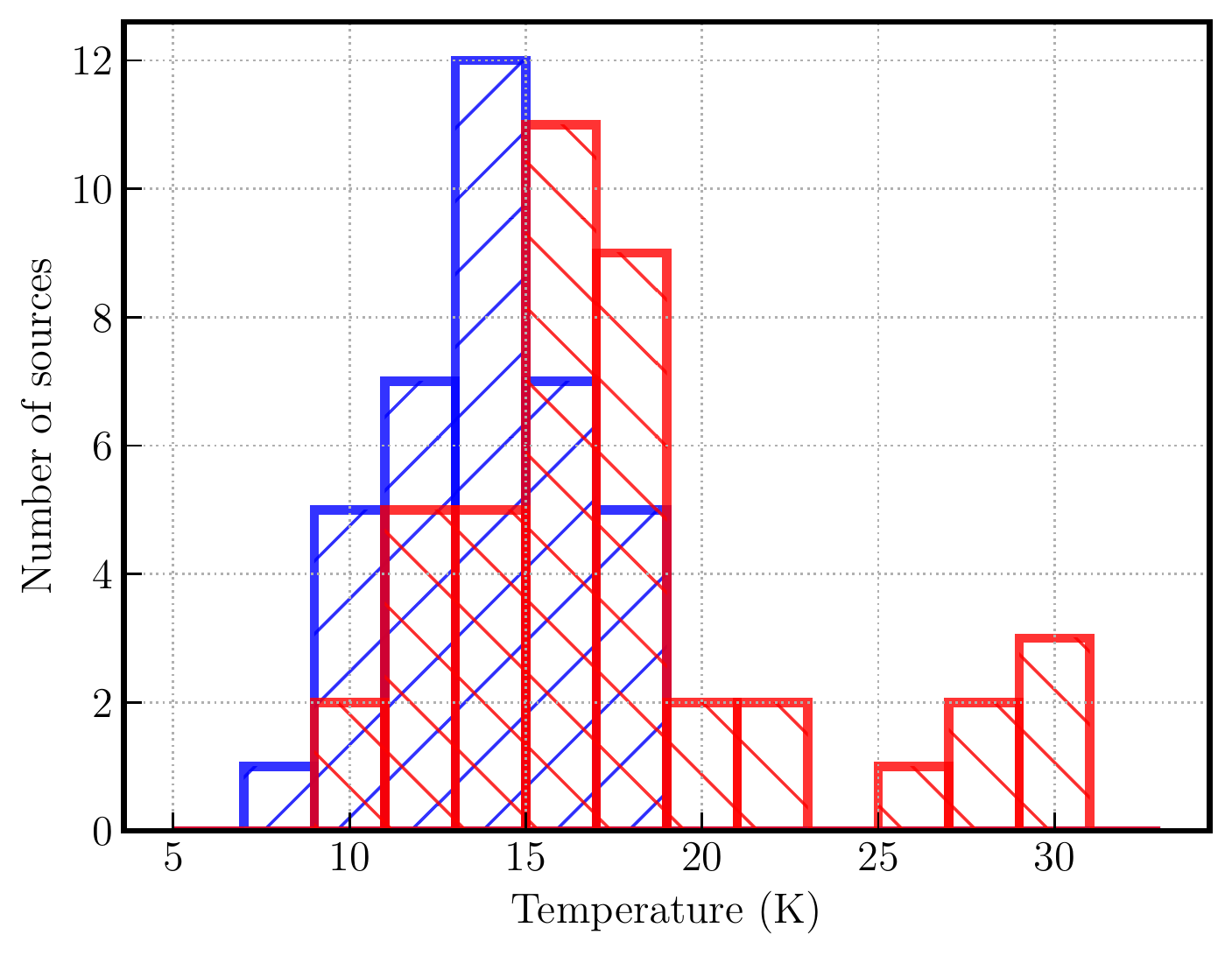}
\caption{Histogram of the source temperatures derived from the SED. Prestellar sources are plotted in blue and protostellar sources in red.}
\label{fig:histo_temp}
\end{figure}

\subsection{Temperature and column density maps}\label{subsect:temp_columndensity}

Using the Hi-GAL data, we created the temperature and column density maps for the G345.5+1.5 region using a pixel-to-pixel SED fitting from 160~$\mu$m onward. The observations were convolved to the resolution of the 500~$\mu$m map (36.6\arcsec) and resampled to 14\arcsec~pix$^{-1}$. The pixel-to-pixel fitting was performed using a modified blackbody model:

\begin{equation}\label{eq:graybody-fd}
F_{\nu}=B_{\nu}(T) \times \kappa_{\nu} \times \mu m_{\rm{H}}  N(\rm{H_2})\times \Omega
,\end{equation}

where the parameters are the same as for Eq.~\ref{eq:hil}. The resulting temperature and column density maps are shown in Figs.~\ref{fig:tempP} and \ref{fig:columnP}. However, the spatial area observed with PACS extends to $b=1^{\circ}$ , which limits the spatial coverage of the temperature and column density maps. To fully exploit the spatial coverage of \herschel, we also constructed them using SPIRE bands alone, which extends the spatial coverage to $b=1.5^{\circ}$. These new maps give more spatial information at the expense of loosing accuracy in the obtained SEDs, and they are shown in Figs.~\ref{fig:tempPS} and \ref{fig:columnPS} with the same color scale. To understand how these maps differ according to the number of bands used, we plot in Fig.~\ref{fig:HiGAL_diff} the temperature and column density obtained using SPIRE bands only versus those obtained using PACS and SPIRE bands, respectively. The obtained Pearson correlation coefficient for the linear least-squares regression is 0.92 and 0.98 for the temperature and the column density, respectively, which shows that the two sets of maps are in good agreement. When using SPIRE bands alone, the temperature is higher by less than 5~K. The peak of the SED occurs generally around 160~$\mu$m in our case, and using this band allows us to better constrain its peak and therefore the temperature. When only SPIRE bands are used, the peak will tend to be less well constrained as the temperature increases. Consequently, there is a significant deviation by 10$-$15~K for a few pixels when $T_{\rm{PACS}}>32$~K. Concerning the column density maps, the differences are lower than a factor of 2. Three areas of high temperature ($>27$~K) are observed: one toward the small ring G345.5+1.45 (difficult to analyze because of its proximity to the edge of the survey), corresponding to bright radio emission and \HII candidates, and the other two around $b=1^{\circ}$, corresponding to known \HII regions. The temperature of 25$-$30~K is in agreement with other temperatures found around \HII regions \citep{and12,liu17}. The other part of the ring has a temperature of 22~K, while the temperature outside the region is 18~K. Some very cold areas are found at 16~K and are correlated with high-column density ($>5\times$10$^{22}$~cm$^{-2}$) areas and with emission observed at 870~$\mu$m and 1.2~mm. 

\begin{figure*}[ht!]
\centering
\subfloat{
\includegraphics[width=0.5\textwidth]{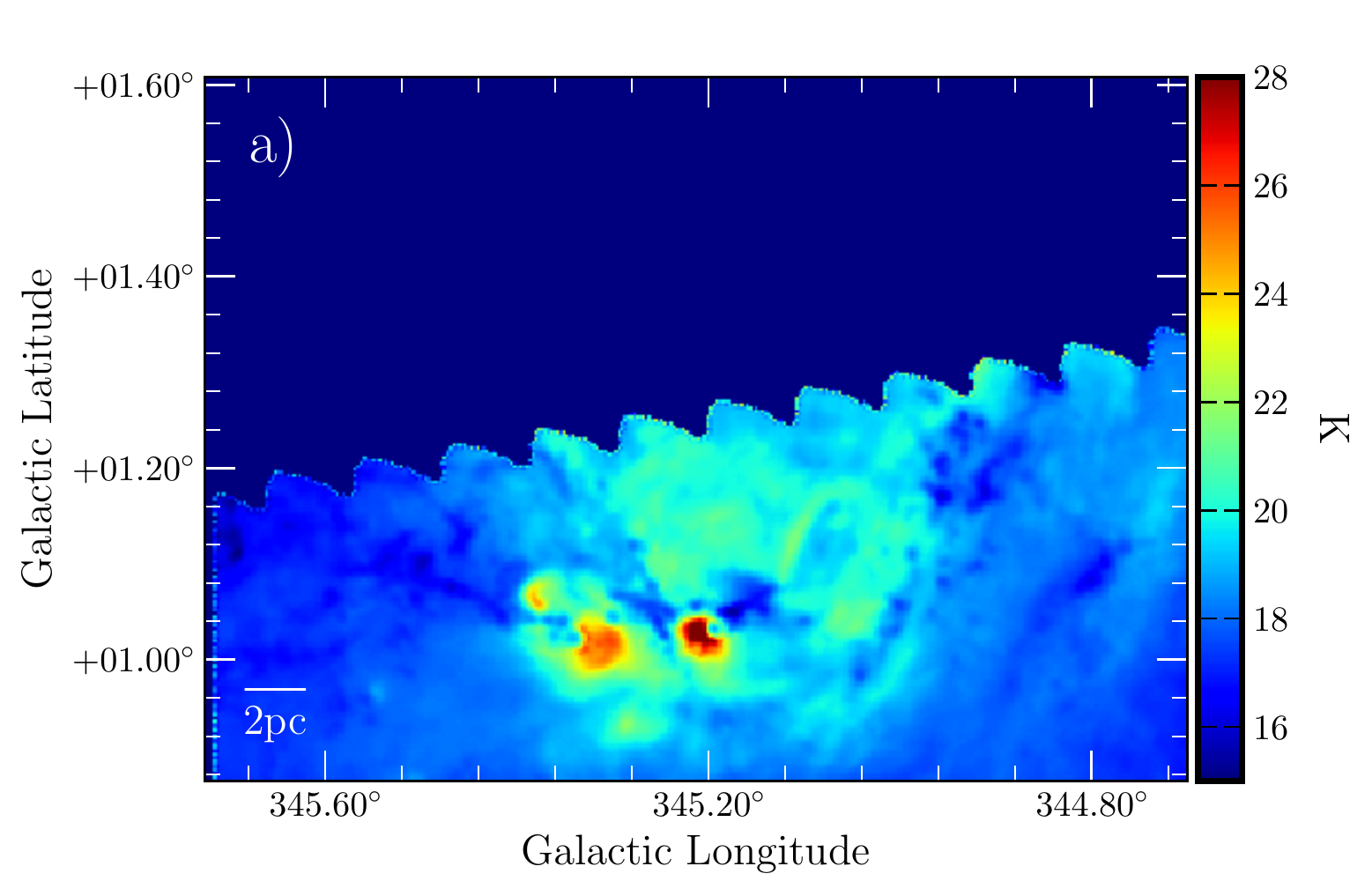}
\label{fig:tempP}}
\subfloat{
\includegraphics[width=0.5\textwidth]{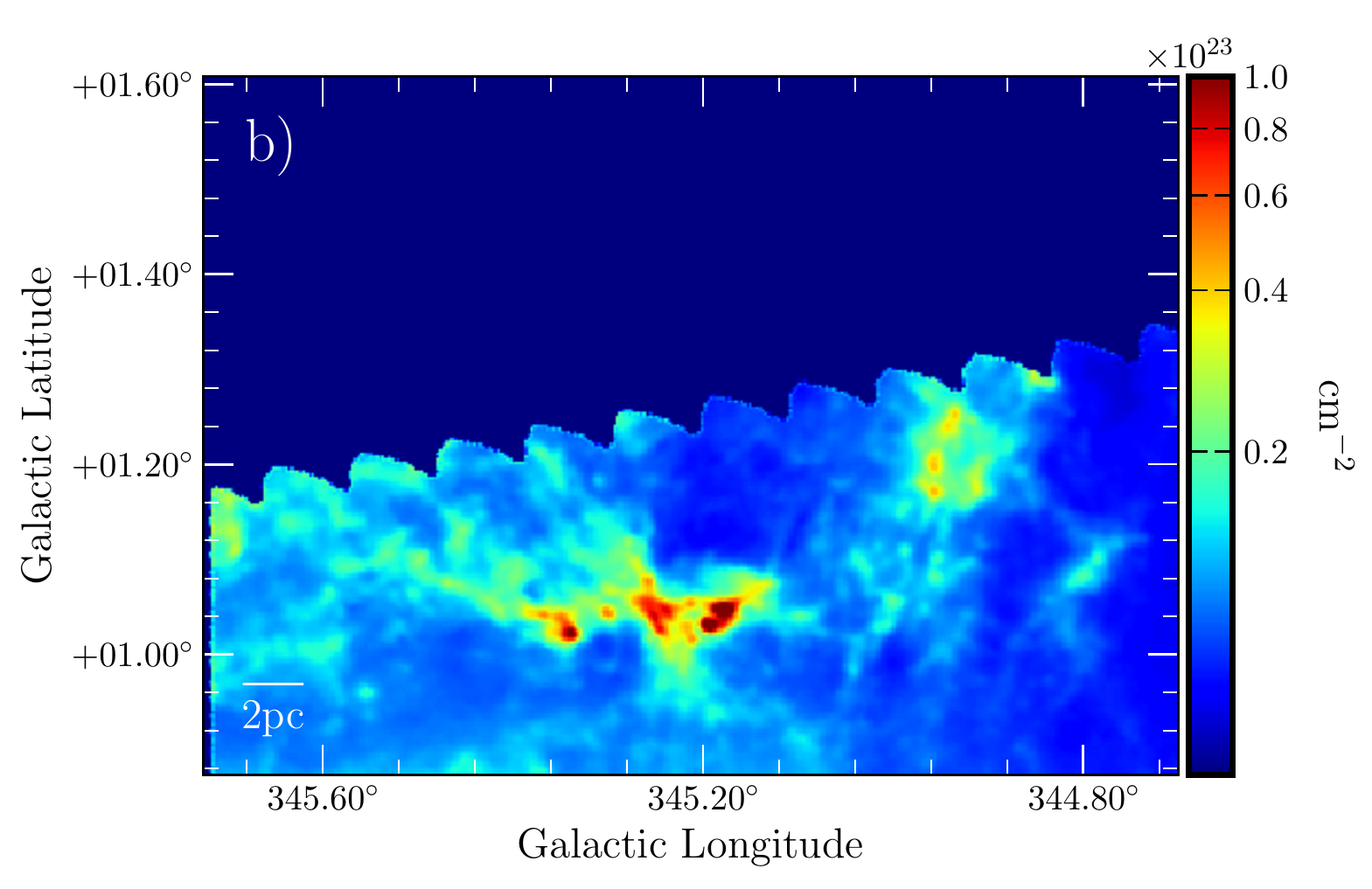}
\label{fig:columnP}}
\quad
\subfloat{
\includegraphics[width=0.5\textwidth]{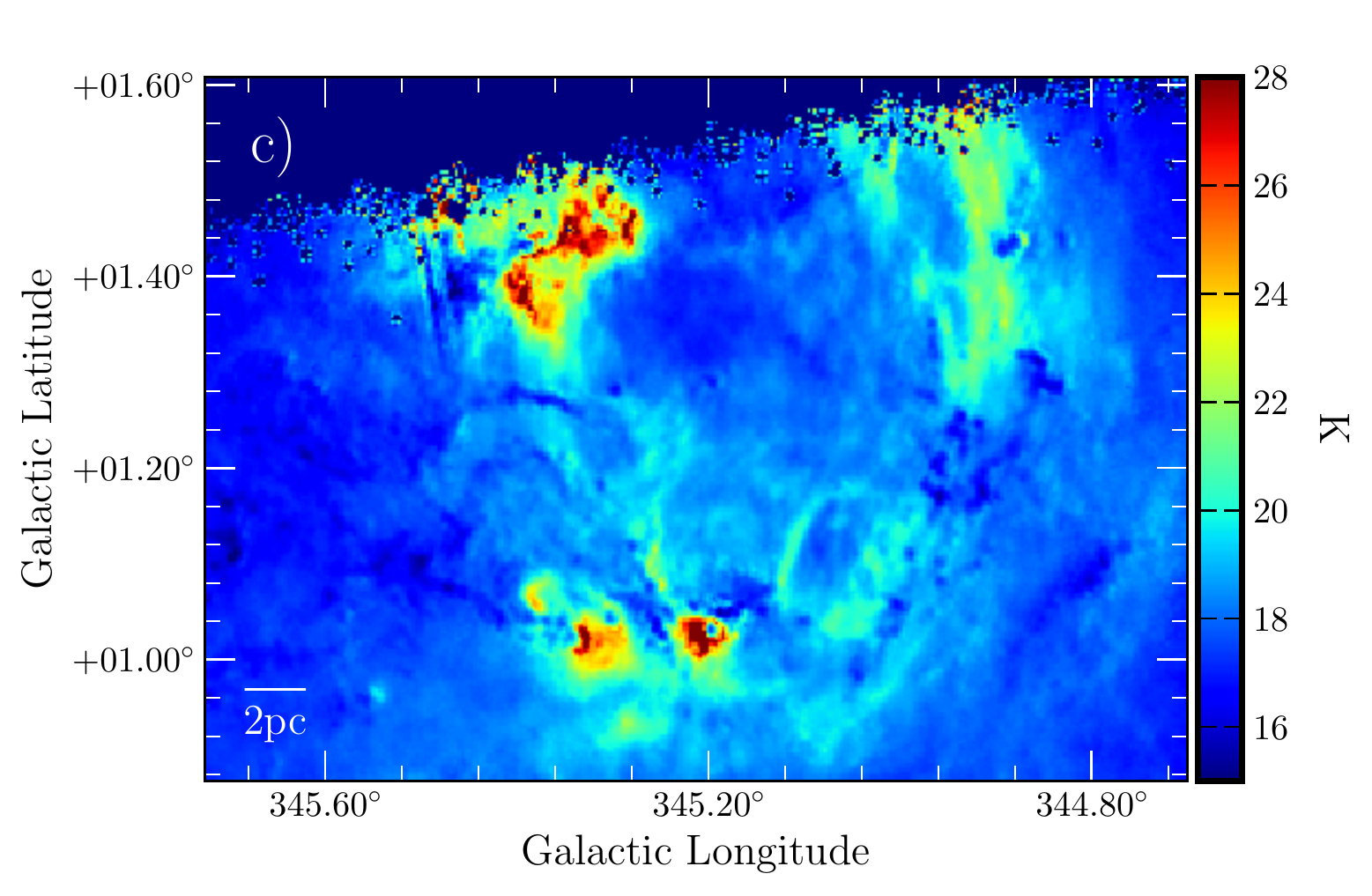}
\label{fig:tempPS}}
\subfloat{
\includegraphics[width=0.5\textwidth]{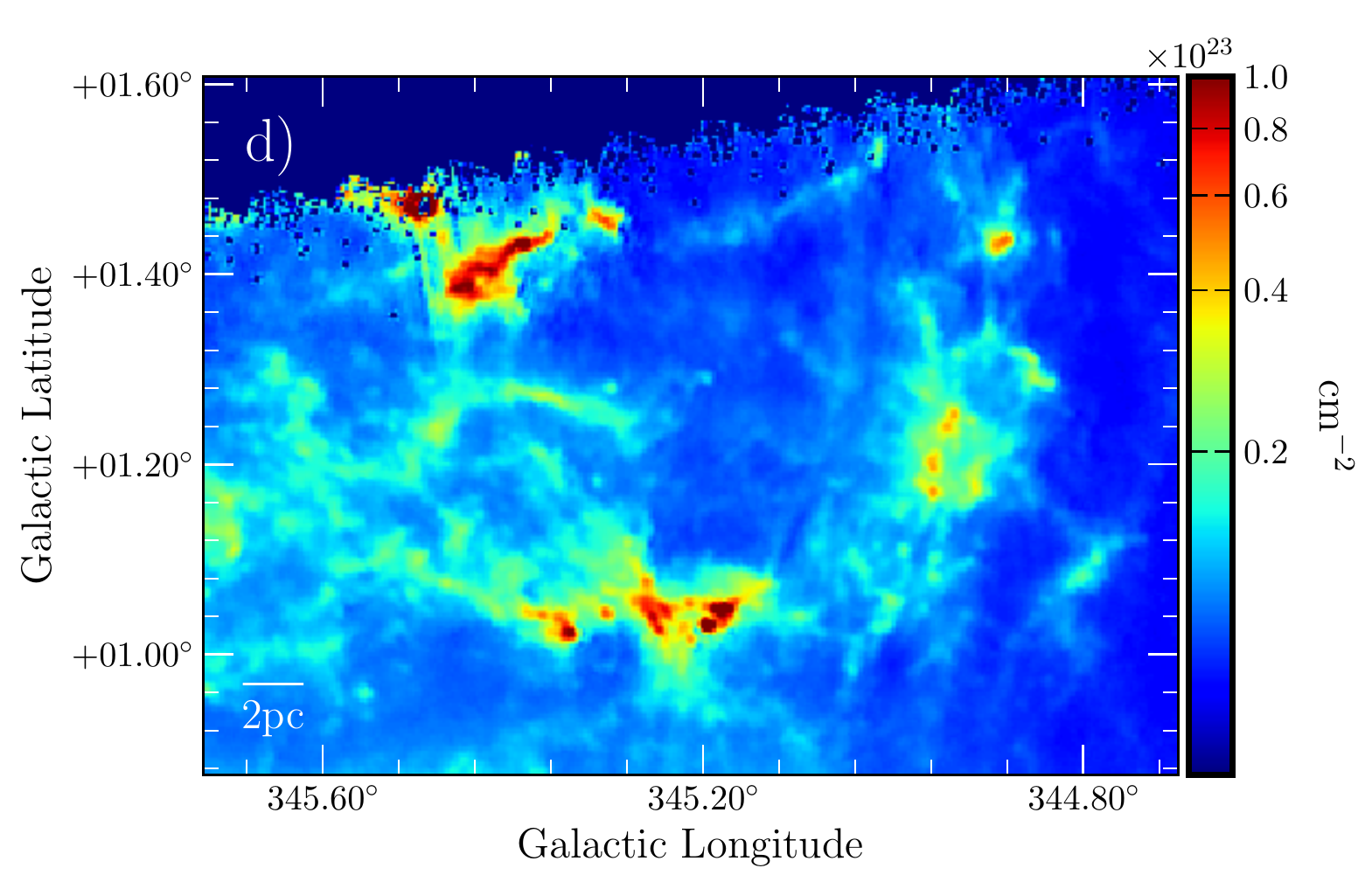}
\label{fig:columnPS}}
\caption{Temperature and column density maps using the 160~$\mu$m and SPIRE bands (a and b) and SPIRE bands alone (c and d).}
\label{fig:HiGAL_TC}
\end{figure*}

\begin{figure*}[t!]
\centering
\subfloat{
\includegraphics[width=0.5\linewidth]{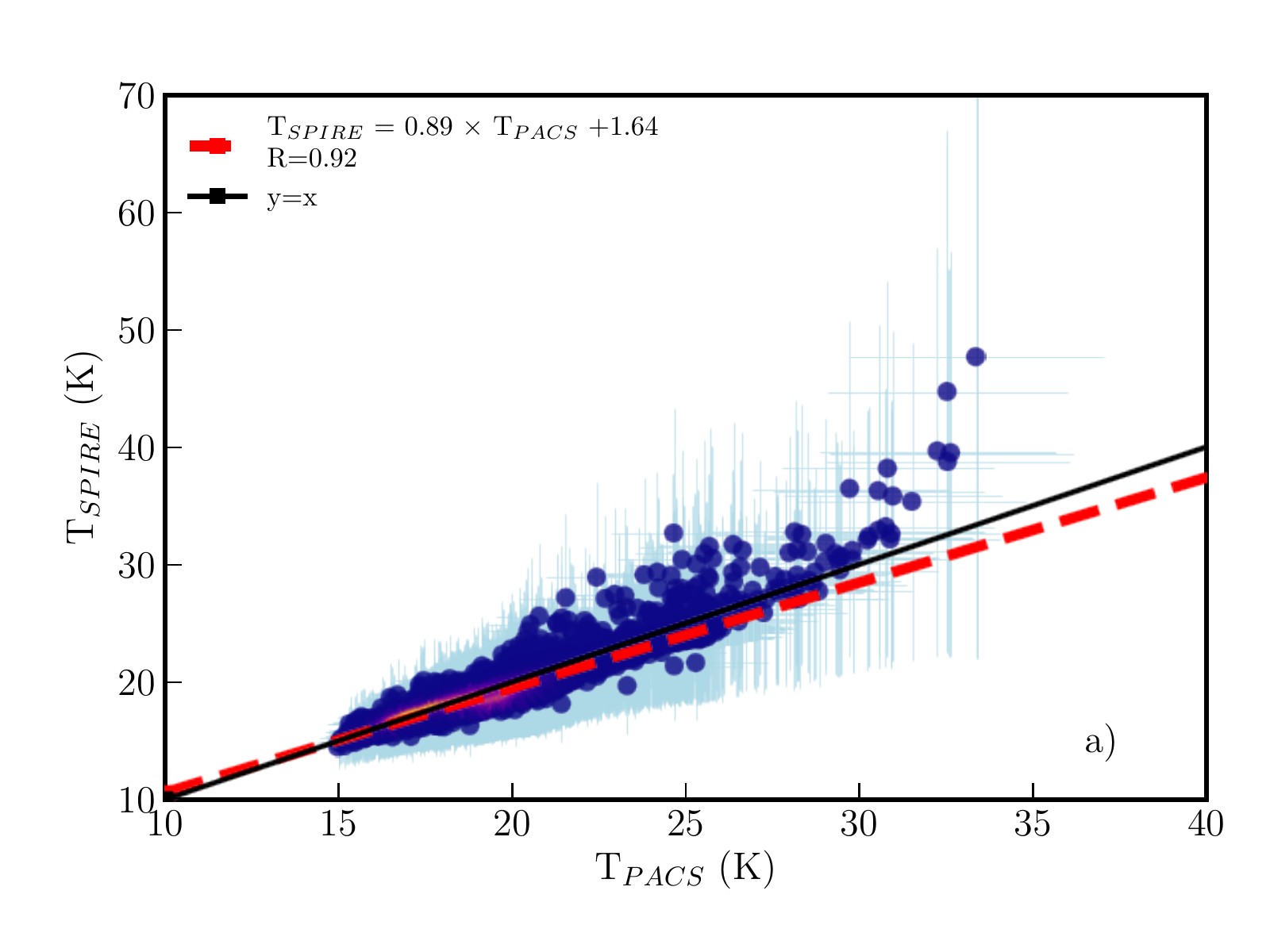}
\label{fig:temp_diff}}
\subfloat{
\includegraphics[width=0.5\linewidth]{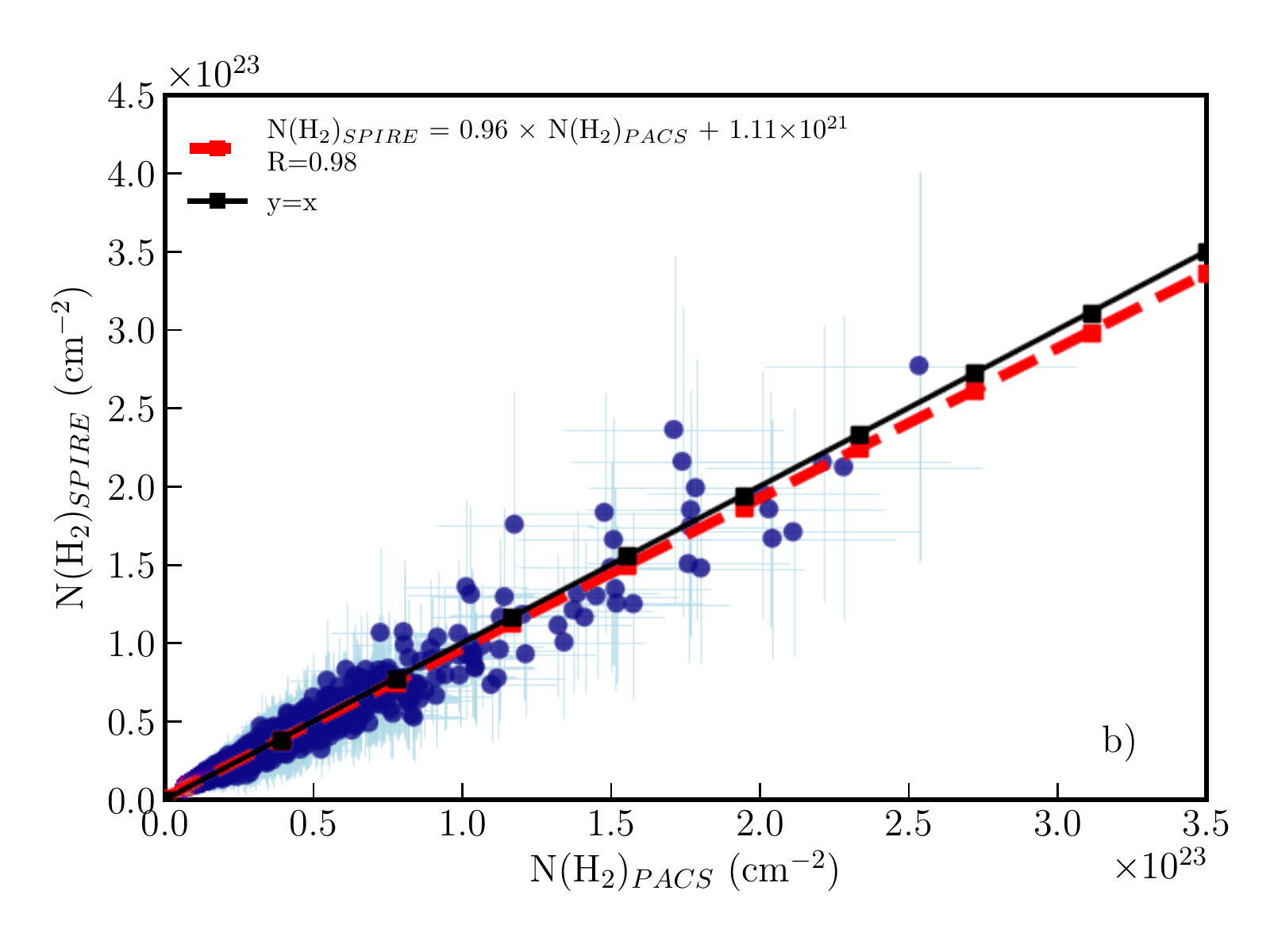}
\label{fig:column_diff}}
\caption{a) Temperature obtained with SPIRE bands alone vs. PACS and SPIRE bands. b) Column density obtained with SPIRE bands alone vs. PACS and SPIRE bands. The black continuous line is the one-to-one relation, and the red dashed line is the linear fit to the data.}
\label{fig:HiGAL_diff}
\end{figure*}

\subsection{$^{12}$CO emission spectrum}

In Fig.~\ref{histo_coclumps} we show the spatially averaged main-beam temperature ($T_{\rm{MB}}$) of the $^{12}$CO(4$-$3) emission line over the entire observed area with respect to V$_{\rm{LSR}}$. We observed the two-component velocity structure along the GMC, first seen in the $^{12}$CO(1$-$0) line by \citet{bro89}, in the CS(2$-$1) line (main component only) by \citet{bro96} and more recently in the $^{12}$CO(3$-$2) and $^{13}$CO(3$-$2) for G345.45+1.5 by \citet{lop16}. Using a Gaussian fitting, we found the main and second peak to be located at $-$12.8~km~s$^{-1}$ and $-$24.1~km~s$^{-1}$. Fig.~\ref{fig:G345_intz_main}) represents the integrated emission around the main peak, between $-$21~km~s$^{-1}$ and 1.4~km~s$^{-1}$ , where the two rings, G345.45+1.5 and G345.10+1.35, are clearly visible. Fig.~\ref{fig:G345_intz_weak} displays the integrated emission around the secondary peak, between $-$32.2~km~s$^{-1}$ and $-$21.7~km~s$^{-1}$. The velocity averaged $T_{\rm{MB}}$ around the secondary peak is so weak that the emission over the entire range of velocity is similar to that of the main peak. This weak $^{12}$CO emission mainly arises around G345.45+1.5 \citep{lop16}, and no counterparts in the dust emission seen by \herschel as well as extinction features in the near-IR are observed. When the model of \citet{rei14} is assumed, this gas is located either at a distance of 2.6$\pm$0.5~kpc or 14.1$\pm$0.5~kpc and is therefore not associated with G435.5+1.5. The emission peak of the map is located at ($\ell$,$b$)=(345.19$^{\circ}$,1.03$^{\circ}$) with an intensity of 26.9~K, a line width of 4.9~km~s$^{-1}$ , and an integrated intensity of 220~K~km~s$^{-1}$ corresponding to a known \HII region of the WISE catalog associated with the brightest radio source 8. Another peak is also seen toward the \HII regions and radio sources 2 and 7. A bright area is seen towards G345.45+1.5, corresponding to the maximum line width of 12.8~km~s$^{-1}$ and is associated with a young high-mass star \citep{guz16}. Both rings are composed of several clumps where the largest concentration is found for G345.45+1.5 \citep{lop16} and to the south of G345.10+1.35.

\begin{figure}
 \centering
 \includegraphics[angle=0,width=90mm]{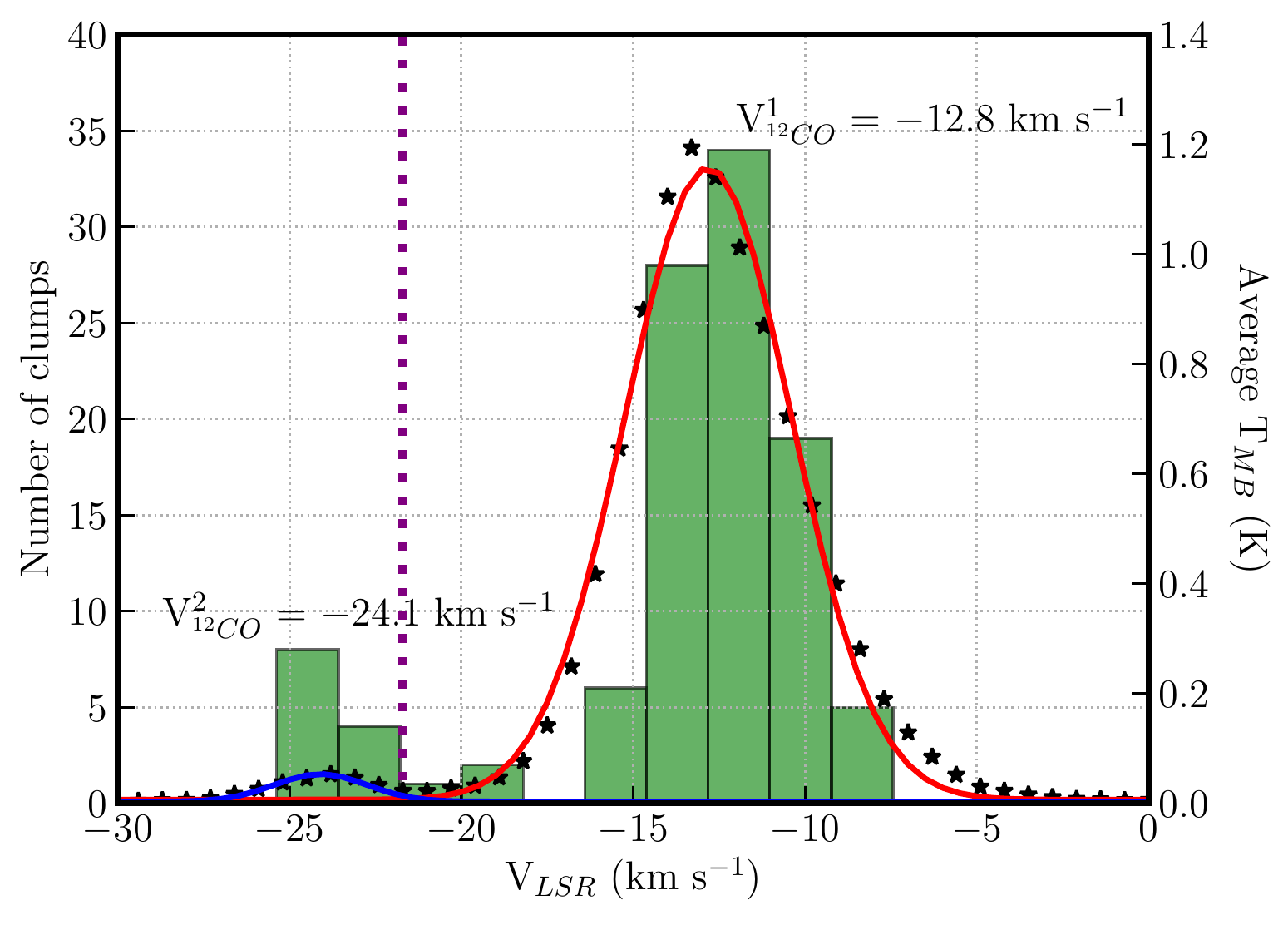}
  \caption{Averaged spectra of the $^{12}$CO(4$-$3) data. The red and blue curves correspond to the best Gaussian fitting for the main and weakest component, respectively. Histogram of the 107 CO-clump velocities, where the blue line at $-$21.7~km~s$^{-1}$\,represents the edge between the main peak and the secondary peak.}
\label{histo_coclumps}
\end{figure}

\begin{figure*}
\centering
\subfloat{
\includegraphics[width=0.5\textwidth]{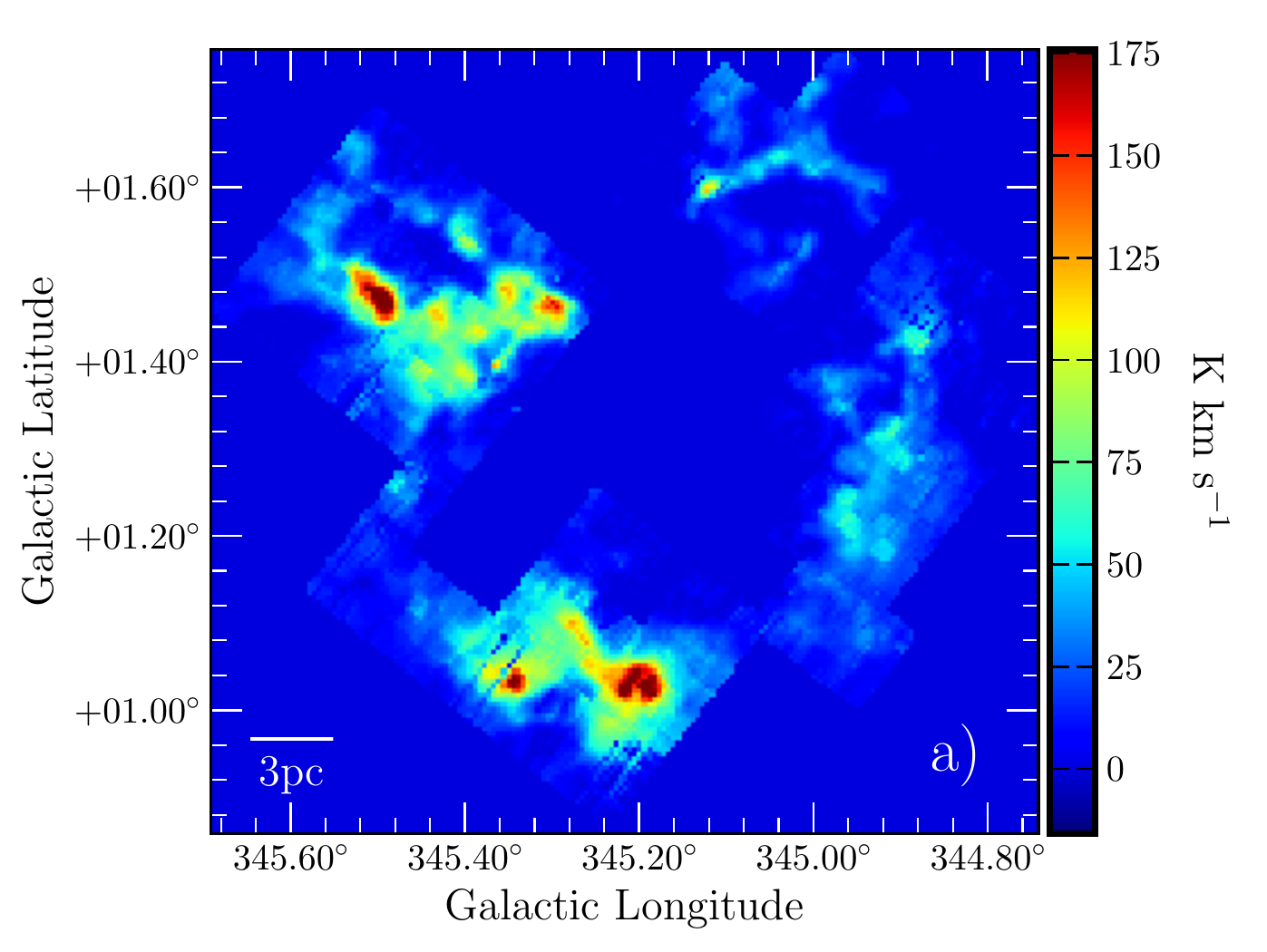}
\label{fig:G345_intz_main}}
\subfloat{
\includegraphics[width=0.5\textwidth]{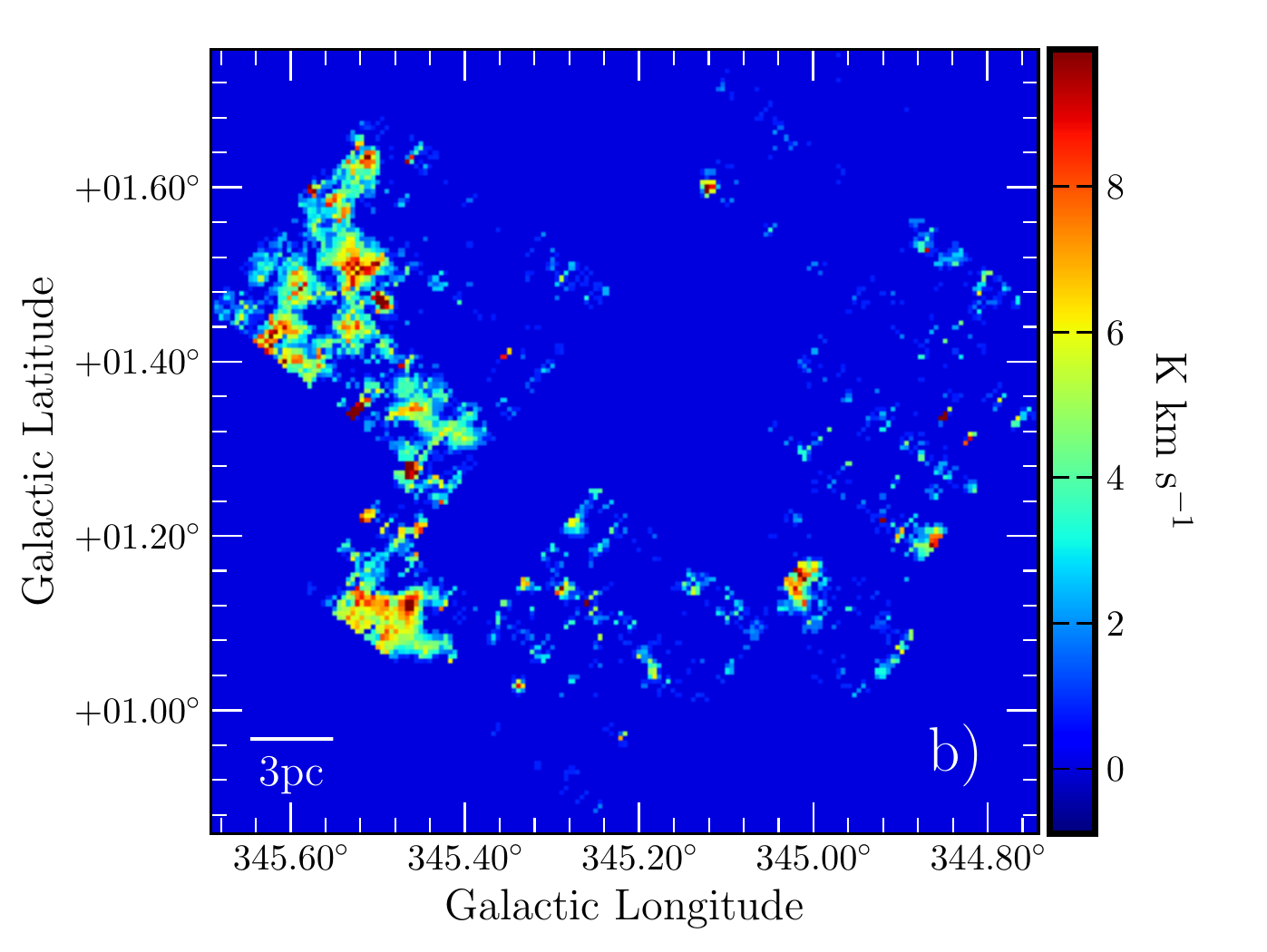}
\label{fig:G345_intz_weak}}
  \caption{Integrated emission over a) $-$21~km~s$^{-1}$ and 1.4~km~s$^{-1}$ and b) $-$32.2~km~s$^{-1}$ and $-$21.7~km~s$^{-1}$}
\label{Mosaique_int_z}
\end{figure*}

\subsection{$^{12}$CO clump detection and properties}\label{subsect:12CO_detect}

To extract the clumps detected in the $^{12}$CO observations, we used the same algorithm as for the 870~$\mu$m clump extraction with a lowest level of detection at $3\sigma$ and an increment step of $2\sigma$. From the 416 CO clumps extracted from the whole data cube, we removed those touching the edges of the map and whose temperature peak was lower than $5\sigma$. We ended up with a sample of 107 $^{12}$CO clumps, whose histogram is presented in Fig.~\ref{histo_coclumps} together with the spectra integrated over the entire cube. As expected, we found the same behavior between the spectrum and the histogram. The blue line divides the histogram at $-$21.7~km s$^{-1}$ , where 12 clumps belongs to the second peak range ($-$32.2~km~s$^{-1}$ to $-$21.7~km~s$^{-1}$) and 95 clumps to the main peak range ($-$21~km~s$^{-1}$ to 1.4~km~s$^{-1}$). Only this latter group of clumps, associated with G345.5+1.5, was considered. As for the previous extraction, the \textit{Clumpfind} output catalog gives us the standard deviation of the Gaussian representing the clump as well as the integrated flux. In order to compute the mass of the clumps, we scaled their emission to that of the $^{12}$CO(1$-$0) and used a CO-to-H$_2$ conversion factor ($X_{\rm{CO}}$) of 2$\times$10$^{20}$~cm$^{-2}$~(K~km~s$^{-1}$)$^{-1}$, the value recommended by \citet{bol13}, with an uncertainty of $\pm$30\%. To compute this scale factor, the $^{12}$CO(1$-$0) and $^{12}$CO(4$-$3) data were integrated over the spatial area and main peak velocity range. The ratio between the two transition levels is equal to 0.15, which is comparable to the Orion-A GMC (\citealt{ish16}, see their Fig.~11a). We have to note that the mass of the region is different from the mass given in \citet{lop11} because of the different $X_{\rm{CO}}$ value used and the larger spatial coverage, which corresponds to the G345.5+1.0 GMC, while only a fraction of this region is studied in the present work. At 3$\sigma$, we estimate the completeness limit to be equal to $\sim 7~M_{\odot}$. The parameters of the 95 $^{12}$CO(4$-$3) clumps are listed in Tab.~\ref{ap:co_clumps}. The mass and the diameter of the clumps extend from 28 to 9432~$M_{\odot}$ , with a median of 795~$M_{\odot}$ , and from 0.35 to 1.34~pc, with a median of 0.7~pc.
 
\section{Discussion}\label{sect:result}

\subsection{Star-forming clumps in G345.5+1.5}

In Sects.~\ref{subsect:870clumpsdetect} and \ref{subsect:herschel} we showed that the dust emission is distributed into several cores and clumps where star formation is taking place depending on the association with IR counterparts. Because the agreement with IR counterparts found for the 870~$\mu$m clumps and Hi-GAL sources in the same area ($b<1.2^{\circ}$) is good, we are confident about the IR counterparts found for the 870~$\mu$m clumps, where no \herschel data are available. The percentage of star-forming clumps for the 870~$\mu$m clumps or Hi-GAL sources (58\% and 53\%, respectively) shows that G345.5+1.5 is an active star-forming region, in agreement with the numerous known and candidate \HII regions found toward it. Part of the remaining starless clumps where no IR counterparts are found might be future sites of star formation under the condition that they are gravitationally bound. As in \citet{eli17} or \citet{pal17}, we considered a clump as bound when its mass was higher than $M_{\rm{Bal}}>460$~$M_{\odot}$ $\times$ r$^{1.9}$ , with r in parsec \citep{lar81}. We found 34 bound and three unbound prestellar clumps in the Hi-GAL sample, while all 870~$\mu$m clumps without an IR counterpart are bound. This result is in agreement with \citet{ward16}, who found that SCUBA-2 (850~$\mu$m) prestellar cores are mostly located in the region of the mass versus FWHM diagram where bound cores are found, while \textit{Herschel} cores can also be found below, in the unbound core region. This is related to the fact that SCUBA-2 mostly detects high-surface brightness sources that tend to be bound, while \textit{Herschel} can detect lower brightness structures. Therefore, many sources seem to be able to form stars in the future. Because G345.5+1.5 appears to have several \HII regions, we also used the criteria of \citet{bal17} to predict whether high-mass star formation can occur in the future. According to their work, the mass limit at which a clump is able to form high-mass stars is given by $M_{\rm{lim}}>1282$~$M_{\odot}$~$\times$ $r^{1.42}$ , with r in parsec. We found that 15 of the 45 870~$\mu$m clumps and 32 of the \herschel bound sources are good candidates. These objects are located toward parts of the region where the high-density and low temperature conditions are ideal for the formation of massive stars.\\
Based on the submillimetric to bolometric luminosity ratio ($L_{\rm{sub}}$/$L_{\rm{bol}}$ where $L_{\rm{sub}}=L_{\rm{bol}}^{\rm{\lambda>350~\mu m}}$), which is linked to the evolutionary stage, we found that this ratio is lower than 1\% in only five \textit{Herschel} sources \citep{and00,and93}, corresponding to Class~I objects at least. One corresponds to the YSO represented by radio source 6 (and to an \HII bubble), two correspond to sources found toward radio source 8, and the two last ones are found in the south of the region. In total, 94\% are Class~0 sources toward G345.5+1.5 (up to $b=1.2^{\circ}$), showing that a new generation of stars is currently under formation. We also plotted the sample of Hi-GAL sources in the $L_{\rm{bol}}$ versus $M_{\rm{env}}$ diagram together with the evolutionary paths of \citet{sar96} and \citet{mol08}. In this model, the paths are composed of an accelerating accretion phase and an envelope clean-up phase based on the turbulent core model of \citet{mck03}. Fig.~\ref{fig:G345_evol_diag} shows the Hi-GAL sources superimposed on the evolutionary diagram according to their protostellar and prestellar nature. As expected, the majority of the sources are found in the accelerating accretion phase, and the sources having an IR counterpart are found in a later evolutionary stage than starless sources. This is expected because \herschel wavelengths characterize the early stages of star formation. 
 
\begin{figure}
 \centering
\includegraphics[angle=0,width=\linewidth]{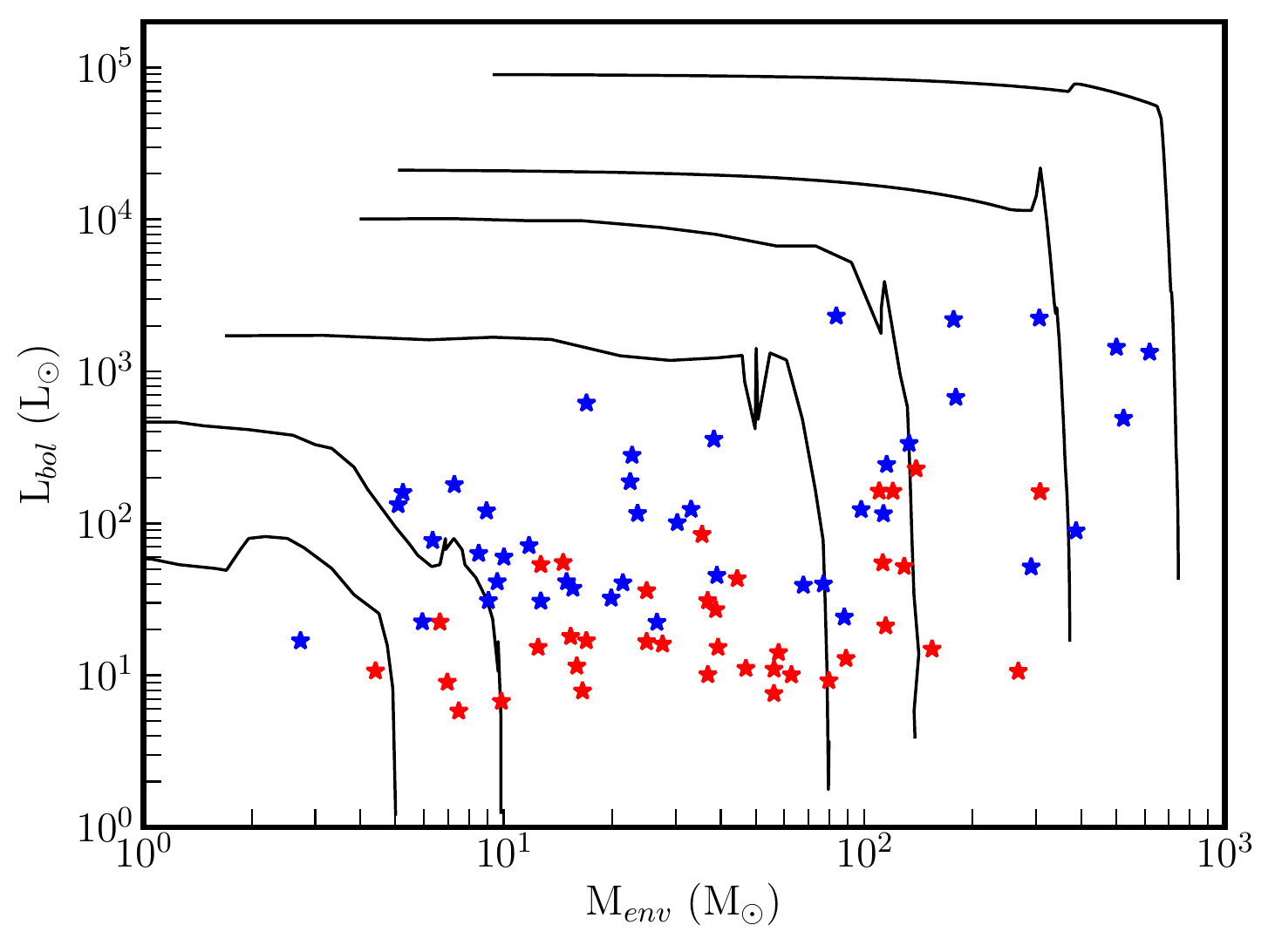}
\caption{$L_{\rm{bol}}$-$M_{\rm{env}}$ diagram with the evolutionary tracks of \citet{mol08} and the sample of sources superimposed on it. Protostellar sources (with a 70~$\mu$m counterpart) are shown in blue and prestellar sources (without a 70~$\mu$m counterpart) in red.}
\label{fig:G345_evol_diag}
\end{figure}

\subsection{Gravitational stability of the clumps}

In order to determine whether the $^{12}$CO clumps we extracted in Sect.~\ref{subsect:12CO_detect} can form stars, we studied their gravitational stability through their virial parameter $\rm{\alpha}$ \citep{ber92}. This is defined as the kinetic (E$_{\rm{K}}$) to the gravitational energy (E$_{\rm{G}}$) ratio, which also translates as

\begin{equation}
\alpha=\frac{5\sigma _{\rm{v}}^2R}{GM}
,\end{equation}

where $\sigma _{\rm{v}}$ is the line width of the clump spectrum, $R$ is the radius of the clump, and $G$ is the Newton constant. Depending on the value of $\rm{\alpha}$, clumps are either supercritical ($\rm{\alpha}<1$, collapsing), critical ($\rm{\alpha}=1$, in equilibrium), or subcritical ($\rm{\alpha}>1$, unbound). Fig.~\ref{fig:virial} shows the virial parameter for the sample of clouds with respect to their mass. The virial ratio of more than two-thirds of the clumps (76\%) is below the critical value, and they are localized in the south of G345.45+1.5 (where several \HII region candidates are found) and along the G345.5+1.35 ring, while the unbound clumps are found around these structures where the density is lower. This indicates that roughly half of the clumps are collapsing/will collapse in the future if no external effects counteracting the gravitation are at work. The magnetic field could play this role, as proposed by \citet{kau13}. Several low $\rm{\alpha}$ clumps were found toward the ring at $b=344.9^{\circ}$ , but no \HII regions (known, candidates, or radio-quiet regions) were found in the WISE catalog \citep{and14}, suggesting that this part of the G435.5+1.5 region is not actively forming stars at the moment but is expected to do so in the future. The median value of $\rm{\alpha}$ in the region is 0.94 and is on the same order of magnitude as the value found for the southern Galactic plane sample of ATLASGAL clumps by \citet{wie18}. Additionally, a decreasing virial parameter with increasing clump mass was observed and follows a power law of the form $\rm{\alpha}=\gamma \times M^{\beta}$. This behavior of the virial parameter was also reported by \citet{lad08} and \citet{fos09}, which implies that the massive clumps tend to have a low virial parameter. It has also been reported that the value of the exponent in this power law is enclosed between $-$1 and 0 in different studies \citep{pil11,lad08,wie18,tan13}. For our sample of clumps, the fit gives $\gamma=9.9\pm0.1$ and $\beta=-0.38\pm0.04$. Therefore, the exponent found for G345.5+1.5 is in the common range of exponent values observed toward low- and high-mass star-forming regions. As explained by \citet{kau13}, this implies that the mass-size and line width-size relations here are also similar to that of their samples.

\begin{figure}
 \centering
 \includegraphics[angle=0,width=90mm]{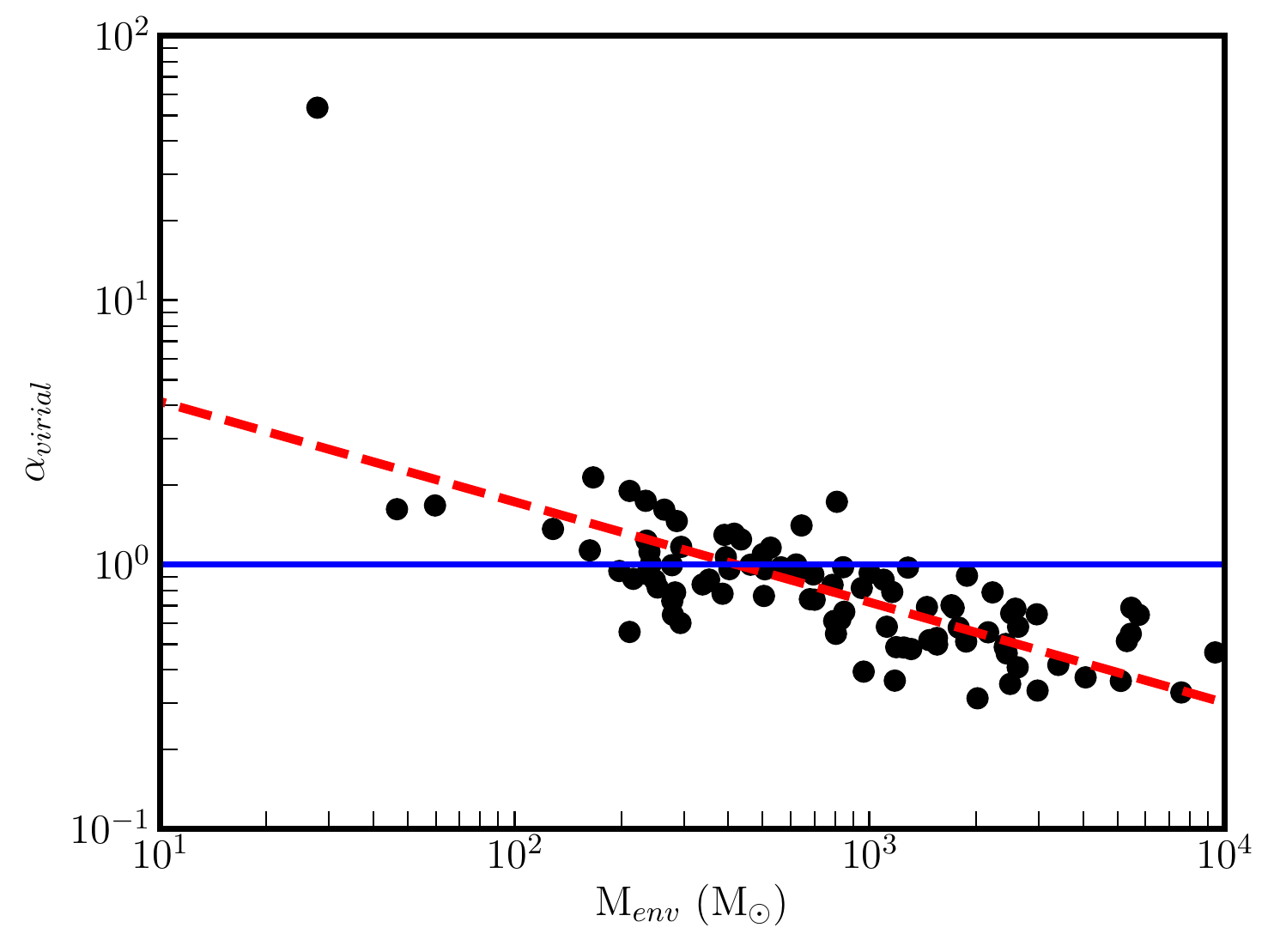}
  \caption{Virial parameter of the $^{12}$CO clumps with respect to their mass. The blue line represents the value for a cloud in equilibrium.}
\label{fig:virial}
\end{figure}

\subsection{Star formation rate and star formation efficiency of G345.5+1.5}\label{substc:sfr}

Based on the number of compact radio sources identified at 843~MHz and on the analysis of their surroundings, G345.5+1.5 contains at least seven high-mass stars. This number is a lower limit because we considered that the resulting \HII regions come from a single star, and we did not consider candidate \HII regions either. Moreover, the absorption of the Ly$\rm{\alpha}$ continuum radiation by the surrounding dust can also contribute to lower the true ionizing rate and therefore the number of massive stars in this region. We made another computation to determine the approximate number of high-mass stars by taking into account all the radio continuum emission present within the G345.5+1.5. Using aperture photometry, we found a total integrated intensity of 39~Jy with an average background emission of 0.001~Jy~beam$^{-1}$ estimated in an empty region. Because the \HII regions are powered by O9/B0.5 stars (see Tab.~\ref{tab:radio_sources_param}), we assumed an average spectral type O9.5 for all the ionizing stars. Based on the calibrations of \citet{mar05}, their mass is approximately equal to 20~$M_{\odot}$ with $N_{\rm{Ly \alpha}}=10^{48.5}$~s$^{-1}$. We estimate that 27 O9.5 stars are responsible for the free-free emission in the region. To compute the star formation properties of this region, namely the star formation rate (SFR) and the SFE, we assumed that the distribution of stellar mass followed the Kroupa initial mass function (IMF, \citealt{kro02}):

\begin{equation}\label{eq:kroupa}
\Phi(M)= \left\{
    \begin{array}{cc}
        k_1\left(\frac{M}{0.08}\right)^{-1.3} & 0.08~M_{\odot}\le M \le  0.5~M_{\odot}  \\
        & \\
        k_2\left(\frac{M}{0.5}\right)^{-2.3} & 0.5~M_{\odot}\le M  \\
    \end{array}
\right.
\end{equation}
\begin{equation}\label{eq:kroupa}
    \begin{array}{cc}
k_2=k_1\left(\frac{0.5}{0.08}\right)^{-1.3} & \\
  \end{array}
.\end{equation}

\noindent The normalization constant $k_2$ was computed assuming that the number of stars with a mass from 8 to 20~$M_{\odot}$ corresponds to the number of O9.5V stars found. The total stellar mass is obtained from 0.08 to 20~$M_{\odot}$. We assumed a typical massive protostellar lifetime, $t_{*}=0.2$~Myr \citep{dua13}, the SFE as SFE=$M_*$/$(M_{G345}+M_*),$ and the SFR as SFR=$M_*$/$t_{*}$, where $M_*$ is the expected total stellar mass equal to $3024^{+587}_{-1036}~M_{\odot}$ and $M_{G345}$ is the mass of the region equal to 1.2$\times$10$^{5}$~$M_{\odot}$. The values are reported in Tab.~\ref{tab:starforming_regions}. Compared to W43-MM1 and RCW~106 \citep{lou14,ngu15}, where the resulting SFE was computed with the Kroupa (from 0.08 to 150~$M_{\odot}$, second approach) and Salpeter IMF (from 0.08 to 50~$M_{\odot}$), respectively, the SFE of G345.5+1.5 is found to lie in between. The SFR, computed in each case with the same $t_*$, is equal to 0.25 and 0.006~$M_{\odot}$~yr$^{-1}$ for RCW~106 and W46~MM1, respectively. The SFR of G345.5+1.5 is in between the value of these two regions, which have the potential to become starburst regions. In agreement with the \HII regions and protostellar sources found within it, G345.5+1.5 will continue to actively form stars in the future. 

A better way to characterize the star formation occurring in a GMC is by computing the SFR density ($\Sigma _{SFR}$), which is independent of the area. However, $\Sigma _{SFR}$ is dependent on the way the surface is chosen. In the case of a ring structure, we could choose either to enclose the whole cloud or to enclose the ring alone because the interior is mostly devoid of emission (see Fig.~\ref{Fig:G345_1200}, \ref{fig:G345_870}, \ref{fig:columnPS}). We chose the latter method, and the area corresponding to the emission ring is 265~pc$^2$. The $\Sigma _{SFR}$ is six times higher than the value of RCW~106, but 200 times lower than for W43~MM1. This latter region could be in a process of collision, which would drastically increase the high-mass star formation \citep{mot03,ngu13} and could explain this high value. Despite this difference, G345.5+1.5 also has a high $\Sigma _{SFR}$ for a star-forming region. Another way to compute the SFR is by using a scaling relation, as was done in \citet{ngu16}, who compared differnt star-forming regions with regard to different parameters. They made use of radio continuum emission at 1.42~GHz to determine the SFR of these regions extracted from the $^{12}$CO(1$-$0) survey \citep{bro89} through the relation of \cite{mur10}. In that case, the SFR of G345.5+1.5 drops to 3.5$\times 10^{-4}$~$M_{\odot}$~yr$^{-1}$ using their method and is lower by an order of magnitude at least compared to the massive molecular cloud complexes (MCC) listed in their work, such as W~51 or Cygnus-X. However, this value is comparable to the SFR values computed by \citet{ven17} using Monte Carlo procedures with Hi-GAL and $^{12}$CO observations toward four GMCs of the fourth quadrant for approximately the same mass of gas (see their Tab.~4). The values for $\Sigma _{SFR}$ range from approximately $10^{-2}$ to 10~$M_{\odot}$~yr$^{-1}$, G345.5+1.5 being comparable to Cygnus-X or or W~48. 
Finally, to determine whether the region is really efficient at forming stars, the SFR of the region needs to be compared to the available quantity of gas. In the Schmidt-Kennicutt diagram, $log(\Sigma _{SFR})$ versus $log(\Sigma _{gas})$, the region is located in the starburst quadrant (see Fig.~6 of \citealt{ngu16}), which confirms that it is an active star-forming region, as is also clear from the several \HII regions and the \HII region candidates \citep{and14} as well as from the protostellar and bound sources found in the Hi-GAL survey.

\begin{table*}
\tiny
\caption{Properties of different star-forming regions}
\centering
\begin{tabular}{c|cccccr}
\hline
\hline
Region & Mass & Area & SFE & $\Sigma _{gas}$ & SFR & $\Sigma _{SFR}$ \\
       & ($M_{\odot}$) & (pc$^2$) & (\%) & ($M_{\odot}$~pc$^{-2}$) & ($M_{\odot}$~yr$^{-1}$) & ($M_{\odot}$~yr$^{-1}$~kpc$^{-2}$) \\
\hline
\hline
G345.5+1.5 (IMF) & 1.2$\times$10$^{5}$ & 265 & 2.5 & 452 & 1.5$\times$10$^{-2}$ & 56.6 \\
RCW~106\tablefootmark{a} & 5.9$\times$10$^{6}$ & 2.6$\times$10$^{4}$ & 0.8 & 227 & 2.5$\times$10$^{-1}$ & 9.6 \\
W43~MM1\tablefootmark{b} & 2.0$\times$10$^{4}$ & 5.75 & 5.9 & 3.5$\times$10$^{3}$ & 6.0$\times$10$^{-3}$ & 1043.5 \\
\hline
G345.5+1.5 (Scaled) & 1.2$\times$10$^{5}$ & 265 & 2.5 & 452.1 & 3.5$\times$10$^{-4}$ & 1.3 \\
W33\tablefootmark{c} & 4.7$\times 10^{6}$ & 1.2$\times$10$^{4}$ &  & 393.2 & 1.8$\times$10$^{-3}$ & 0.149 \\
Cygnus~X\tablefootmark{c} & 2.2$\times 10^{6}$ & 3.7$\times$10$^{4}$ &  & 59.9 & 6.2$\times$10$^{-2}$ & 1.68 \\
W51\tablefootmark{c} & 2.5$\times 10^{6}$ & 1.8$\times$10$^{4}$ &  & 139.9 & 1.8$\times$10$^{-1}$ & 10.2 \\
R5\tablefootmark{d} & 1.3$\times 10^{5}$ & 4.2$\times$10$^{3}$ &  & 30.9 & 1.3$\times$10$^{-4}$ & 3.1$\times$10$^{-2}$ \\
\hline
\hline
\end{tabular}
\tablefoot{
\tablefoottext{a}{Values taken from \citet{ngu15} from the immediate past (see their Sect.~7.1).}
\tablefoottext{b}{Values taken from \citet{lou14} following their second approach (see their Sect.~5.2).}
\tablefoottext{c}{Values taken from \citet{ngu16} (see their Tab.~1).}
\tablefoottext{d}{Values taken from \citet{ven17} (see their Sect.~4.1).}
}
\label{tab:starforming_regions}
\end{table*}

\section{Conclusions}\label{sect:conclusion}

We used the APEX-LABOCA 870~$\mu$m and NANTEN2 $^{12}$CO(4$-$3) molecular line, complemented with 2MASS, \textit{Spitzer,} and \herschel Hi-GAL observations, to study the G345.5+1.5 region. The large-scale distribution of H$\rm{\alpha}$ emission that is found spatially close to the region is also close in distance and probably mainly in front because very few extinction features are observed in H$\rm{\alpha}$ despite the kinematic distance that is assigned. An analysis of the radio-compact sources at 36~cm in association with the WISE catalog of \HII regions confirmed the presence of seven late-B and early-O high-mass stars, which could increase to 12. The region is composed of 45 870~$\mu$m clumps, and 79 Hi-GAL sources are found in the south ($b<1.2^{\circ}$). The high percentage, more than 50\%, of IR counterparts found toward these clumps indicates that a new generation of stars will emerge. In particular, high-mass stars are able to form in most of the clumps because their mass is above the limit proposed in the literature. The boundedness of the prestellar clumps also indicates that most of them will be able to host star formation in the future. This is in agreement with the high percentage of bound $^{12}$CO clumps and its SFR and $\Sigma _{SFR}$, which are comparable to other well-known star-forming regions of the Galaxy. Located in the starburst quadrant of the Schmitt-Kennicutt diagram, G345.5+1.5 is currently actively forming stars and is therefore an interesting region that merits follow-up studies to better understand star formation.

\begin{acknowledgements}
L.B. acknowledges support from CONICYT project Basal AFB-170002.
\end{acknowledgements}

\bibliographystyle{aa}
\bibliography{/home/mfigueira/Documents/Articles/biblio.bib}


\renewcommand{\arraystretch}{1.1}

\begin{appendix}
\begin{table*}
\section{Properties of the 870~$\mu$m, Hi-GAL, and $^{12}$CO clumps}
\centering
\begin{tabular}{|ccccrrrrcc|}
\hline
Id & $\ell$ & $b$ & $D_{\rm{c}}$ &  $S_{870\mu m}$ & FF & $M_{\rm{c}}$ & $N(\rm{H_2})$ & $n{_{\rm{H_2}}}$ & IR-cp \\
\cline{2-3}
   &  \multicolumn{2}{c}{($^{\circ}$)} & (pc) & (Jy) & (\%) & ($M_{\sun}$) & (cm$^{-2}$) & (cm$^{-3}$)  &  \\
   \hline
   \hline
   &&&&&&&&& \\
1       &       345.500 &       1.474   &       1.1     &       141.5   &       1       &       3923    &       1.7$\times      10^{23}$        &       7.1$\times      10^{4}$ &       Y       \\
2       &       345.011 &       1.796   &       1.1     &       115.2   &       0       &       3212    &       1.5$\times      10^{23}$        &       6.7$\times      10^{4}$ &       Y       \\
3       &       345.200 &       1.035   &       0.8     &       41.1    &       10      &       1035    &       9.1$\times      10^{22}$        &       5.5$\times      10^{4}$ &       Y       \\
4       &       345.012 &       1.823   &       0.9     &       43.2    &       0       &       1204    &       9.3$\times      10^{22}$        &       5.3$\times      10^{4}$ &       Y       \\
5       &       345.397 &       1.432   &       1.1     &       42.0    &       7       &       1096    &       4.7$\times      10^{22}$        &       2.0$\times      10^{4}$ &       Y       \\
6       &       345.187 &       1.051   &       1.0     &       38.6    &       1       &       1064    &       6.2$\times      10^{22}$        &       3.1$\times      10^{4}$ &       Y       \\
7       &       345.123 &       1.594   &       0.6     &       16.1    &       6       &       423     &       6.3$\times      10^{22}$        &       4.9$\times      10^{4}$ &       Y       \\
8       &       344.992 &       1.822   &       0.9     &       35.2    &       0       &       982     &       6.3$\times      10^{22}$        &       3.2$\times      10^{4}$ &       Y       \\
9       &       345.519 &       1.478   &       0.8     &       10.2    &       4       &       274     &       2.7$\times      10^{22}$        &       1.7$\times      10^{4}$ &       N       \\
10      &       345.350 &       1.031   &       0.9     &       25.1    &       0       &       699     &       4.6$\times      10^{22}$        &       2.4$\times      10^{4}$ &       Y       \\
11      &       345.520 &       1.471   &       0.5     &       6.8     &       2       &       187     &       3.8$\times      10^{22}$        &       3.5$\times      10^{4}$ &       Y       \\
12      &       344.985 &       1.782   &       0.8     &       20.7    &       0       &       579     &       4.8$\times      10^{22}$        &       2.8$\times      10^{4}$ &       Y       \\
13      &       344.987 &       1.754   &       0.6     &       10.8    &       0       &       301     &       5.2$\times      10^{22}$        &       4.4$\times      10^{4}$ &       Y       \\
14      &       345.223 &       1.030   &       1.0     &       34.2    &       13      &       835     &       4.8$\times      10^{22}$        &       2.3$\times      10^{4}$ &       Y       \\
15      &       345.368 &       1.440   &       1.0     &       15.1    &       11      &       377     &       2.0$\times      10^{22}$        &       9.3$\times      10^{3}$ &       Y       \\
16      &       345.424 &       1.407   &       1.2     &       27.5    &       6       &       724     &       2.9$\times      10^{22}$        &       1.2$\times      10^{4}$ &       N       \\
17      &       345.398 &       1.529   &       0.6     &       8.5     &       1       &       234     &       3.4$\times      10^{22}$        &       2.6$\times      10^{4}$ &       Y       \\
18      &       345.257 &       1.043   &       0.5     &       6.0     &       4       &       163     &       3.7$\times      10^{22}$        &       3.6$\times      10^{4}$ &       Y       \\
19      &       345.252 &       1.032   &       0.5     &       5.8     &       5       &       154     &       3.5$\times      10^{22}$        &       3.5$\times      10^{4}$ &       Y       \\
20      &       345.564 &       1.510   &       0.7     &       11.0    &       3       &       298     &       3.9$\times      10^{22}$        &       2.9$\times      10^{4}$ &       Y       \\
21      &       345.405 &       1.541   &       0.8     &       12.4    &       0       &       345     &       2.8$\times      10^{22}$        &       1.7$\times      10^{4}$ &       Y       \\
22      &       345.303 &       1.454   &       0.8     &       13.3    &       11      &       333     &       3.2$\times      10^{22}$        &       2.0$\times      10^{4}$ &       Y       \\
23      &       345.45  &       1.404   &       0.5     &       4.8     &       1       &       133     &       2.8$\times      10^{22}$        &       2.6$\times      10^{4}$ &       N       \\
24      &       345.408 &       1.384   &       0.7     &       8.4     &       25      &       176     &       2.3$\times      10^{22}$        &       1.7$\times      10^{4}$ &       N       \\
25      &       345.360 &       1.459   &       0.5     &       4.3     &       2       &       117     &       2.6$\times      10^{22}$        &       2.5$\times      10^{4}$ &       Y       \\
26      &       345.265 &       1.052   &       0.5     &       4.4     &       6       &       115     &       3.0$\times      10^{22}$        &       3.1$\times      10^{4}$ &       Y       \\
27      &       345.317 &       1.464   &       0.6     &       6.8     &       5       &       182     &       3.4$\times      10^{22}$        &       3.0$\times      10^{4}$ &       Y       \\
28      &       345.453 &       1.391   &       0.6     &       7.8     &       0       &       216     &       3.1$\times      10^{22}$        &       2.4$\times      10^{4}$ &       Y       \\
29      &       345.464 &       1.390   &       0.5     &       5.2     &       0       &       145     &       3.5$\times      10^{22}$        &       3.4$\times      10^{4}$ &       Y       \\
30      &       345.244 &       1.052   &       0.6     &       7.5     &       6       &       198     &       2.7$\times      10^{22}$        &       2.1$\times      10^{4}$ &       N       \\
31      &       345.535 &       1.571   &       0.5     &       4.7     &       0       &       132     &       2.9$\times      10^{22}$        &       2.8$\times      10^{4}$ &       N       \\
32      &       345.426 &       1.422   &       0.5     &       3.4     &       2       &       93      &       2.5$\times      10^{22}$        &       2.6$\times      10^{4}$ &       Y       \\
33      &       345.384 &       1.047   &       1.0     &       12.0    &       0       &       336     &       2.0$\times      10^{22}$        &       1.0$\times      10^{4}$ &       Y       \\
34      &       345.240 &       1.044   &       0.5     &       3.9     &       6       &       102     &       2.6$\times      10^{22}$        &       2.6$\times      10^{4}$ &       Y       \\
35      &       344.887 &       1.433   &       0.7     &       8.0     &       0       &       224     &       2.5$\times      10^{22}$        &       1.7$\times      10^{4}$ &       N       \\
36      &       345.358 &       1.468   &       0.6     &       5.2     &       1       &       144     &       2.6$\times      10^{22}$        &       2.3$\times      10^{4}$ &       Y       \\
37      &       345.564 &       1.524   &       1.2     &       7.4     &       3       &       202     &       7.7$\times      10^{21}$        &       3.0$\times      10^{3}$ &       Y       \\
38      &       345.265 &       1.062   &       0.8     &       8.2     &       7       &       214     &       2.1$\times      10^{22}$        &       1.3$\times      10^{4}$ &       N       \\
39      &       345.266 &       1.082   &       0.5     &       3.6     &       7       &       93      &       2.2$\times      10^{22}$        &       2.2$\times      10^{4}$ &       N       \\
40      &       345.387 &       1.551   &       0.5     &       1.9     &       2       &       53      &       1.1$\times      10^{22}$        &       9.9$\times      10^{3}$ &       Y       \\
41      &       345.373 &       1.020   &       0.5     &       1.7     &       4       &       45      &       1.1$\times      10^{22}$        &       1.1$\times      10^{4}$ &       Y       \\
42      &       345.381 &       1.465   &       0.7     &       4.3     &       1       &       118     &       1.3$\times      10^{22}$        &       8.3$\times      10^{3}$ &       Y       \\
43      &       345.221 &       1.057   &       0.4     &       1.9     &       17      &       43      &       1.5$\times      10^{22}$        &       1.9$\times      10^{4}$ &       N       \\
44      &       345.527 &       1.494   &       0.5     &       2.7     &       7       &       69      &       1.3$\times      10^{22}$        &       1.2$\times      10^{4}$ &       Y       \\
45      &       345.357 &       1.492   &       1.2     &       7.1     &       2       &       195     &       8.2$\times      10^{21}$        &       3.4$\times      10^{3}$ &       Y       \\              \hline
\end{tabular}
\caption{Properties of the 45 870~$\mu$m clumps. Clump id (Col. 1), galactic coordinates (Cols.  2 and 3), diameter (Col. 4), integrated flux (Col. 5), free-free contamination (Col. 6), clump mass (Col. 7) column density (Col. 8), volume density (Col. 9), and early YSOs between 1.25~$\mu$m and 8~$\mu$m (Col. 10).}
\label{tab_param}
\end{table*}

\begin{sidewaystable*}
        \centering
        \small
        \begin{tabular}{|cccrrrrl|cccrrrrl|}
\hline
Id & $\ell$ & $b$ & $T$ & $M_{\rm{env}}$ & $L_{\rm{bol}}$ & $L_{\rm{sub}}$/$L_{\rm{bol}}$ & IR & Id & $\ell$ & $b$ & $T$ & $M_{\rm{env}}$ & $L_{\rm{bol}}$ & $L_{\rm{sub}}$/$L_{\rm{bol}}$ & IR  \\
\cline{2-3}
\cline{10-11}
 & \multicolumn{2}{c}{$^{\circ}$} & (K) & $M_{\odot}$ & $L_{\odot}$ & (\%) & &  & \multicolumn{2}{c}{$^{\circ}$} & (K) & $M_{\odot}$ & $L_{\odot}$ & (\%) &  \\
\hline
\hline
1       &       344.813 &       1.250   &       17.1    $\pm$   1.3     &       4       $\pm$   1       &       11      $\pm$   7       &       4.1     &       Prestellar      &       41      &       345.193 &       1.030   &       28.7    $\pm$   0.9     &       306     $\pm$   19      &       2245    $\pm$   541     &       2.8     &       Protostellar    \\
2       &       344.865 &       1.190   &       10.3    $\pm$   0.5     &       80      $\pm$   23      &       9       $\pm$   5       &       22.4    &       Prestellar      &       42      &       345.194 &       1.052   &       20.5    $\pm$   1.7     &       22      $\pm$   7       &       188     $\pm$   148     &       1.7     &       Protostellar    \\
3       &       344.900 &       1.204   &       8.7     $\pm$   0.4     &       268     $\pm$   111     &       11      $\pm$   7       &       35.8    &       Prestellar      &       43      &       345.197 &       0.980   &       15.2    $\pm$   1.2     &       38      $\pm$   31      &       358     $\pm$   449     &       0.8     &       Protostellar    \\
4       &       344.901 &       1.227   &       14.0    $\pm$   1.3     &       16      $\pm$   7       &       12      $\pm$   11      &       8.7     &       Prestellar      &       44      &       345.198 &       1.150   &       21.5    $\pm$   1.9     &       6       $\pm$   2       &       77      $\pm$   60      &       1.4     &       Protostellar    \\
5       &       344.912 &       1.178   &       13.8    $\pm$   0.9     &       28      $\pm$   7       &       16      $\pm$   10      &       10.1    &       Prestellar      &       45      &       345.199 &       0.863   &       15.4    $\pm$   1.5     &       7       $\pm$   6       &       9       $\pm$   13      &       6.1     &       Prestellar      \\
6       &       344.924 &       1.093   &       12.6    $\pm$   0.5     &       39      $\pm$   31      &       15      $\pm$   16      &       12.2    &       Prestellar      &       46      &       345.209 &       1.015   &       27.7    $\pm$   1.9     &       17      $\pm$   14      &       619     $\pm$   759     &       0.7     &       Protostellar    \\
7       &       344.931 &       1.213   &       11.3    $\pm$   0.7     &       56      $\pm$   19      &       11      $\pm$   8       &       17.4    &       Prestellar      &       47      &       345.211 &       1.051   &       16.5    $\pm$   1.9     &       139     $\pm$   53      &       229     $\pm$   248     &       5.6     &       Prestellar      \\
8       &       344.952 &       1.100   &       14.4    $\pm$   1.3     &       37      $\pm$   10      &       31      $\pm$   26      &       7.8     &       Prestellar      &       48      &       345.215 &       1.029   &       29.2    $\pm$   2.2     &       84      $\pm$   22      &       2315    $\pm$   1683    &       1.0     &       Protostellar    \\
9       &       344.952 &       1.209   &       14.9    $\pm$   1.3     &       44      $\pm$   16      &       43      $\pm$   38      &       7.4     &       Prestellar      &       49      &       345.216 &       0.946   &       14.8    $\pm$   1.0     &       20      $\pm$   16      &       32      $\pm$   40      &       4.4     &       Protostellar    \\
10      &       344.954 &       1.005   &       11.6    $\pm$   0.6     &       47      $\pm$   38      &       11      $\pm$   12      &       15.7    &       Prestellar      &       50      &       345.216 &       1.019   &       18.7    $\pm$   1.8     &       177     $\pm$   87      &       2196    $\pm$   2352    &       1.0     &       Protostellar    \\
11      &       344.954 &       1.080   &       10.7    $\pm$   0.5     &       56      $\pm$   16      &       8       $\pm$   4       &       21.1    &       Prestellar      &       51      &       345.227 &       0.873   &       13.9    $\pm$   1.3     &       10      $\pm$   8       &       7       $\pm$   10      &       8.9     &       Prestellar      \\
12      &       344.957 &       1.171   &       12.9    $\pm$   0.8     &       129     $\pm$   42      &       52      $\pm$   36      &       12.2    &       Prestellar      &       52      &       345.238 &       1.046   &       13.7    $\pm$   0.5     &       307     $\pm$   37      &       161     $\pm$   58      &       10.9    &       Prestellar      \\
13      &       344.960 &       1.195   &       14.3    $\pm$   0.7     &       98      $\pm$   21      &       123     $\pm$   63      &       5.2     &       Protostellar    &       53      &       345.243 &       1.025   &       17      $\pm$   0.2     &       115     $\pm$   3       &       244     $\pm$   21      &       4.6     &       Protostellar    \\
14      &       344.964 &       1.046   &       15.2    $\pm$   0.7     &       15      $\pm$   12      &       18      $\pm$   19      &       6.5     &       Prestellar      &       54      &       345.250 &       1.038   &       14.9    $\pm$   0.6     &       113     $\pm$   20      &       116     $\pm$   48      &       6.6     &       Protostellar    \\
15      &       344.970 &       0.984   &       14.8    $\pm$   1.5     &       17      $\pm$   14      &       17      $\pm$   24      &       7.0     &       Prestellar      &       55      &       345.254 &       1.173   &       15.4    $\pm$   1.2     &       16      $\pm$   5       &       37      $\pm$   28      &       3.3     &       Protostellar    \\
16      &       344.986 &       1.105   &       11.1    $\pm$   0.4     &       88      $\pm$   14      &       24      $\pm$   9       &       11.7    &       Protostellar    &       56      &       345.254 &       1.016   &       25.1    $\pm$   2.5     &       9       $\pm$   7       &       121     $\pm$   166     &       1.6     &       Protostellar    \\
17      &       344.990 &       0.905   &       14.2    $\pm$   1.0     &       8       $\pm$   6       &       6       $\pm$   7       &       8.3     &       Prestellar      &       57      &       345.256 &       1.077   &       15.4    $\pm$   0.4     &       133     $\pm$   16      &       336     $\pm$   95      &       3.0     &       Protostellar    \\
18      &       344.995 &       1.146   &       11.1    $\pm$   0.5     &       115     $\pm$   31      &       21      $\pm$   11      &       18.1    &       Prestellar      &       58      &       345.259 &       1.112   &       17.1    $\pm$   1.3     &       33      $\pm$   10      &       124     $\pm$   92      &       2.7     &       Protostellar    \\
19      &       345.008 &       1.026   &       10.8    $\pm$   0.7     &       89      $\pm$   30      &       13      $\pm$   10      &       20.3    &       Prestellar      &       59      &       345.278 &       1.115   &       15      $\pm$   0.8     &       39      $\pm$   9       &       45      $\pm$   25      &       6.3     &       Protostellar    \\
20      &       345.016 &       1.210   &       18.6    $\pm$   1.4     &       10      $\pm$   2       &       41      $\pm$   28      &       2.8     &       Protostellar    &       60      &       345.280 &       1.079   &       30.6    $\pm$   2.8     &       5       $\pm$   4       &       159     $\pm$   213     &       1.0     &       Protostellar    \\
21      &       345.018 &       1.160   &       18.3    $\pm$   1.8     &       9       $\pm$   3       &       63      $\pm$   57      &       1.6     &       Protostellar    &       61      &       345.285 &       0.931   &       16.8    $\pm$   0.7     &       23      $\pm$   19      &       280     $\pm$   295     &       0.8     &       Protostellar    \\
22      &       345.019 &       1.027   &       11.7    $\pm$   0.7     &       58      $\pm$   47      &       14      $\pm$   17      &       15.3    &       Prestellar      &       62      &       345.301 &       1.039   &       16      $\pm$   0.5     &       180     $\pm$   19      &       675     $\pm$   208     &       2.3     &       Protostellar    \\
23      &       345.020 &       1.065   &       16.9    $\pm$   0.2     &       13      $\pm$   10      &       31      $\pm$   26      &       4.1     &       Protostellar    &       63      &       345.304 &       1.071   &       17.2    $\pm$   2.1     &       30      $\pm$   25      &       101     $\pm$   157     &       3.0     &       Protostellar    \\
24      &       345.026 &       1.090   &       11.8    $\pm$   0.7     &       68      $\pm$   24      &       39      $\pm$   27      &       6.7     &       Protostellar    &       64      &       345.320 &       0.913   &       15.3    $\pm$   1.0     &       12      $\pm$   10      &       15      $\pm$   17      &       6.4     &       Prestellar      \\
25      &       345.029 &       1.135   &       18.4    $\pm$   1.3     &       10      $\pm$   2       &       60      $\pm$   40      &       2.0     &       Protostellar    &       65      &       345.320 &       1.141   &       18.8    $\pm$   2.1     &       13      $\pm$   4       &       54      $\pm$   55      &       3.0     &       Prestellar      \\
26      &       345.038 &       1.014   &       12.2    $\pm$   0.2     &       27      $\pm$   20      &       22      $\pm$   19      &       5.1     &       Protostellar    &       66      &       345.333 &       1.101   &       11      $\pm$   0.6     &       63      $\pm$   20      &       10      $\pm$   7       &       19.3    &       Prestellar      \\
27      &       345.040 &       1.104   &       12.2    $\pm$   1.1     &       77      $\pm$   33      &       40      $\pm$   39      &       8.3     &       Protostellar    &       67      &       345.334 &       1.021   &       16      $\pm$   0.3     &       619     $\pm$   53      &       1338    $\pm$   265     &       3.6     &       Protostellar    \\
28      &       345.047 &       0.921   &       20.4    $\pm$   1.8     &       3       $\pm$   2       &       17      $\pm$   23      &       2.4     &       Protostellar    &       68      &       345.336 &       1.040   &       10.3    $\pm$   0.4     &       290     $\pm$   44      &       52      $\pm$   19      &       13.3    &       Protostellar    \\
29      &       345.055 &       1.179   &       15.7    $\pm$   1.1     &       25      $\pm$   6       &       36      $\pm$   23      &       5.7     &       Prestellar      &       69      &       345.343 &       0.986   &       18.1    $\pm$   1.8     &       7       $\pm$   5       &       22      $\pm$   32      &       3.6     &       Prestellar      \\
30      &       345.057 &       1.077   &       16.4    $\pm$   1.3     &       24      $\pm$   19      &       116     $\pm$   149     &       1.8     &       Protostellar    &       70      &       345.346 &       1.040   &       13.7    $\pm$   0.5     &       113     $\pm$   19      &       55      $\pm$   21      &       11.5    &       Prestellar      \\
31      &       345.069 &       1.142   &       18.4    $\pm$   1.6     &       15      $\pm$   7       &       55      $\pm$   57      &       3.2     &       Prestellar      &       71      &       345.362 &       1.066   &       22.1    $\pm$   1.9     &       7       $\pm$   6       &       180     $\pm$   234     &       0.7     &       Protostellar    \\
32      &       345.079 &       0.894   &       11.9    $\pm$   1.0     &       37      $\pm$   30      &       10      $\pm$   14      &       14.6    &       Prestellar      &       72      &       345.362 &       1.022   &       17.3    $\pm$   1.9     &       35      $\pm$   17      &       85      $\pm$   98      &       4.3     &       Prestellar      \\
33      &       345.090 &       1.011   &       16.5    $\pm$   1.6     &       12      $\pm$   10      &       71      $\pm$   100     &       1.5     &       Protostellar    &       73      &       345.367 &       1.040   &       15.9    $\pm$   0.5     &       120     $\pm$   15      &       163     $\pm$   54      &       6.2     &       Prestellar      \\
34      &       345.097 &       1.039   &       13.3    $\pm$   1.0     &       17      $\pm$   5       &       8       $\pm$   6       &       11.2    &       Prestellar      &       74      &       345.374 &       1.081   &       30.9    $\pm$   3.3     &       5       $\pm$   4       &       133     $\pm$   186     &       1.2     &       Protostellar    \\
35      &       345.106 &       0.925   &       15.1    $\pm$   1.1     &       21      $\pm$   17      &       41      $\pm$   51      &       3.9     &       Protostellar    &       75      &       345.379 &       1.042   &       15.8    $\pm$   0.9     &       110     $\pm$   86      &       163     $\pm$   184     &       5.6     &       Prestellar      \\
36      &       345.114 &       1.121   &       14.6    $\pm$   1.3     &       15      $\pm$   8       &       41      $\pm$   45      &       2.5     &       Protostellar    &       76      &       345.383 &       1.057   &       11.7    $\pm$   0.5     &       501     $\pm$   115     &       1441    $\pm$   685     &       1.3     &       Protostellar    \\
37      &       345.118 &       0.943   &       17.8    $\pm$   1.7     &       6       $\pm$   5       &       23      $\pm$   31      &       2.9     &       Protostellar    &       77      &       345.400 &       1.050   &       17.6    $\pm$   1.3     &       9       $\pm$   3       &       31      $\pm$   23      &       3.1     &       Protostellar    \\
38      &       345.124 &       1.054   &       13.8    $\pm$   0.9     &       25      $\pm$   20      &       17      $\pm$   20      &       9.2     &       Prestellar      &       78      &       345.412 &       1.045   &       14.1    $\pm$   0.8     &       39      $\pm$   13      &       27      $\pm$   18      &       8.9     &       Prestellar      \\
39      &       345.136 &       1.071   &       10.7    $\pm$   0.5     &       387     $\pm$   84      &       89      $\pm$   43      &       12.4    &       Protostellar    &       79      &       345.426 &       1.057   &       10.1    $\pm$   0.5     &       154     $\pm$   45      &       15      $\pm$   9       &       24.5    &       Prestellar      \\
40      &       345.180 &       1.045   &       16.6    $\pm$   0.4     &       524     $\pm$   30      &       491     $\pm$   4       &       8.6     &       Protostellar    &               &               &                               &                               &                               &               &               &               \\
\hline
        \end{tabular}
    \caption{Properties of the Hi-GAL sources toward G345.5+1.5. Source id (Col. 1), galactic coordinates (Cols 2 and 3) envelope temperature (Col. 4), envelope mass (Col. 5), bolometric luminosity (Col. 6), submillimetric to bolometric luminosity ratio (Col. 7), nature of the source (pre/protostellar, Col. 8). }\label{ap:higal_clumps}
\end{sidewaystable*}

\clearpage

\begin{table*}
\centering
        \begin{tabular}{|cccrrr|cccrrr|}
        \hline
        Id & $\ell$ & $b$  & $D_{\rm{c}}$ & $M_{\rm{c}}$ & $\Delta$v$_{\rm{c}}$ & Id & $\ell$ & $b$  & $D_{\rm{c}}$ & $M_{\rm{c}}$ & $\Delta$v$_{\rm{c}}$  \\
\cline{2-3}
\cline{8-9}
&\multicolumn{2}{c}{($^{\circ}$)}&(pc)&($M_{\sun}$)&(km~s$^{-1}$) & &\multicolumn{2}{c}{($^{\circ}$)}&(pc)&($M_{\sun}$)&(km~s$^{-1}$) \\
\hline
1       &       345.182 &       1.030   &       1.0     &       7571    &       4.8     &       49      &       344.960 &       1.618   &       0.7     &       829     &       2.7     \\
2       &       345.206 &       1.054   &       0.8     &       4068    &       4.2     &       50      &       344.874 &       1.249   &       0.4     &       278     &       2.5     \\
3       &       345.222 &       1.023   &       0.8     &       5112    &       4.7     &       51      &       344.867 &       1.452   &       0.5     &       463     &       3.1     \\
4       &       345.259 &       1.080   &       0.7     &       2618    &       3.9     &       52      &       344.879 &       1.350   &       0.7     &       695     &       3.0     \\
5       &       345.308 &       1.040   &       0.9     &       2978    &       3.3     &       53      &       344.886 &       1.372   &       0.4     &       293     &       2.1     \\
6       &       345.493 &       1.470   &       1.3     &       9433    &       5.6     &       54      &       344.920 &       1.224   &       0.7     &       952     &       3.3     \\
7       &       345.504 &       1.484   &       1.2     &       5317    &       4.6     &       55      &       345.375 &       1.367   &       0.6     &       501     &       2.9     \\
8       &       345.361 &       1.397   &       0.4     &       701     &       3.5     &       56      &       344.947 &       1.087   &       0.8     &       787     &       2.8     \\
9       &       345.299 &       1.068   &       0.8     &       2493    &       3.3     &       57      &       345.301 &       0.998   &       0.7     &       622     &       2.9     \\
10      &       345.434 &       1.450   &       1.2     &       5746    &       5.4     &       58      &       344.860 &       1.267   &       0.4     &       339     &       2.6     \\
11      &       345.251 &       1.053   &       0.8     &       3407    &       4.1     &       59      &       344.887 &       1.269   &       0.5     &       563     &       3.2     \\
12      &       345.120 &       1.602   &       0.7     &       1884    &       4.9     &       60      &       344.870 &       1.471   &       0.5     &       353     &       2.4     \\
13      &       345.402 &       1.400   &       0.6     &       1874    &       3.9     &       61      &       345.133 &       1.051   &       0.6     &       278     &       1.9     \\
14      &       345.390 &       1.375   &       0.5     &       1120    &       3.5     &       62      &       344.870 &       1.300   &       0.8     &       1002    &       3.3     \\
15      &       345.454 &       1.391   &       1.3     &       5454    &       4.8     &       63      &       344.868 &       1.497   &       0.5     &       279     &       1.8     \\
16      &       345.416 &       1.365   &       1.0     &       2407    &       3.4     &       64      &       344.897 &       1.246   &       0.5     &       527     &       3.4     \\
17      &       344.883 &       1.429   &       0.9     &       1786    &       3.3     &       65      &       344.886 &       1.396   &       0.4     &       283     &       2.2     \\
18      &       345.348 &       1.480   &       0.8     &       2963    &       4.9     &       66      &       345.080 &       1.547   &       1.1     &       796     &       2.1     \\
19      &       345.318 &       1.094   &       0.9     &       2438    &       3.5     &       67      &       344.890 &       1.485   &       0.4     &       253     &       2.3     \\
20      &       344.996 &       1.638   &       0.9     &       2017    &       2.5     &       68      &       345.488 &       1.601   &       0.6     &       508     &       2.7     \\
21      &       345.331 &       1.500   &       0.9     &       2511    &       4.3     &       69      &       344.904 &       1.460   &       0.7     &       435     &       2.8     \\
22      &       345.352 &       1.446   &       0.7     &       2161    &       4.0     &       70      &       345.459 &       1.346   &       1.0     &       1284    &       3.4     \\
23      &       345.066 &       1.623   &       0.5     &       964     &       2.6     &       71      &       344.872 &       1.394   &       0.5     &       403     &       2.7     \\
24      &       345.234 &       0.975   &       1.2     &       5472    &       5.5     &       72      &       345.521 &       1.404   &       0.6     &       504     &       2.5     \\
25      &       345.034 &       1.641   &       0.8     &       1475    &       2.9     &       73      &       345.497 &       1.417   &       0.9     &       810     &       3.9     \\
26      &       345.261 &       1.010   &       0.8     &       1704    &       3.7     &       74      &       344.878 &       1.511   &       0.5     &       216     &       2.0     \\
27      &       345.094 &       1.611   &       0.6     &       850     &       3.0     &       75      &       344.847 &       1.448   &       0.5     &       234     &       2.9     \\
28      &       344.929 &       1.318   &       0.6     &       805     &       2.6     &       76      &       345.356 &       1.402   &       0.6     &       394     &       2.6     \\
29      &       345.386 &       1.435   &       0.7     &       2424    &       4.0     &       77      &       344.853 &       1.429   &       0.4     &       264     &       3.2     \\
30      &       344.954 &       1.185   &       0.8     &       1250    &       2.8     &       78      &       344.834 &       1.303   &       0.7     &       249     &       1.8     \\
31      &       344.952 &       1.207   &       0.5     &       680     &       3.0     &       79      &       344.844 &       1.293   &       0.4     &       236     &       2.6     \\
32      &       345.390 &       1.469   &       1.0     &       2625    &       3.8     &       80      &       345.222 &       1.175   &       0.4     &       242     &       2.3     \\
33      &       344.900 &       1.324   &       0.8     &       1550    &       3.2     &       81      &       345.024 &       1.537   &       0.4     &       236     &       2.3     \\
34      &       344.910 &       1.295   &       0.8     &       1453    &       3.5     &       82      &       345.544 &       1.395   &       0.6     &       351     &       2.2     \\
35      &       344.926 &       1.265   &       0.8     &       1552    &       3.1     &       83      &       345.335 &       1.529   &       0.6     &       390     &       2.8     \\
36      &       344.955 &       1.288   &       0.8     &       1179    &       2.3     &       84      &       344.863 &       1.365   &       0.5     &       287     &       2.8     \\
37      &       344.963 &       1.351   &       0.8     &       1191    &       2.6     &       85      &       345.259 &       1.129   &       0.6     &       167     &       2.3     \\
38      &       345.551 &       1.523   &       1.1     &       2224    &       3.9     &       86      &       344.979 &       1.332   &       0.5     &       163     &       1.8     \\
39      &       345.144 &       1.034   &       0.7     &       1097    &       3.6     &       87      &       345.456 &       1.112   &       0.7     &       416     &       2.8     \\
40      &       345.557 &       1.484   &       1.1     &       2580    &       3.9     &       88      &       345.486 &       1.556   &       0.6     &       211     &       2.6     \\
41      &       345.422 &       1.420   &       0.9     &       1728    &       3.5     &       89      &       345.434 &       1.104   &       0.8     &       295     &       2.0     \\
42      &       345.090 &       1.674   &       0.5     &       211     &       1.6     &       90      &       345.556 &       1.412   &       0.5     &       197     &       1.8     \\
43      &       345.151 &       1.083   &       0.8     &       1160    &       3.2     &       91      &       345.524 &       1.439   &       0.8     &       240     &       1.8     \\
44      &       345.097 &       1.685   &       0.6     &       386     &       2.1     &       92      &       344.910 &       1.405   &       0.4     &       60      &       1.6     \\
45      &       344.972 &       1.147   &       1.0     &       1312    &       2.5     &       93      &       345.254 &       1.178   &       0.5     &       128     &       1.8     \\
46      &       345.125 &       1.073   &       0.8     &       843     &       3.2     &       94      &       345.261 &       1.179   &       0.4     &       47      &       1.4     \\
47      &       345.094 &       1.657   &       0.7     &       644     &       3.5     &       95      &       345.354 &       1.406   &       0.4     &       28      &       6.3     \\
48      &       344.912 &       1.422   &       0.7     &       628     &       2.9     &    &            &               &               &       &          \\   
\hline
        \end{tabular}
\caption{Properties of the 95 $^{12}$CO(4$-$3) clumps extracted from the cube that belong to the main peak. Clump id (Col. 1), galactic position (Cols. 2 and 3), clump diameter (Col. 4), clump mass (Col. 5), and clump line width (Col. 6).}\label{ap:co_clumps}        
\end{table*}

\clearpage

\begin{figure*}

\section{\HII region images}

\centering
\subfloat{
\includegraphics[width=0.33\textwidth]{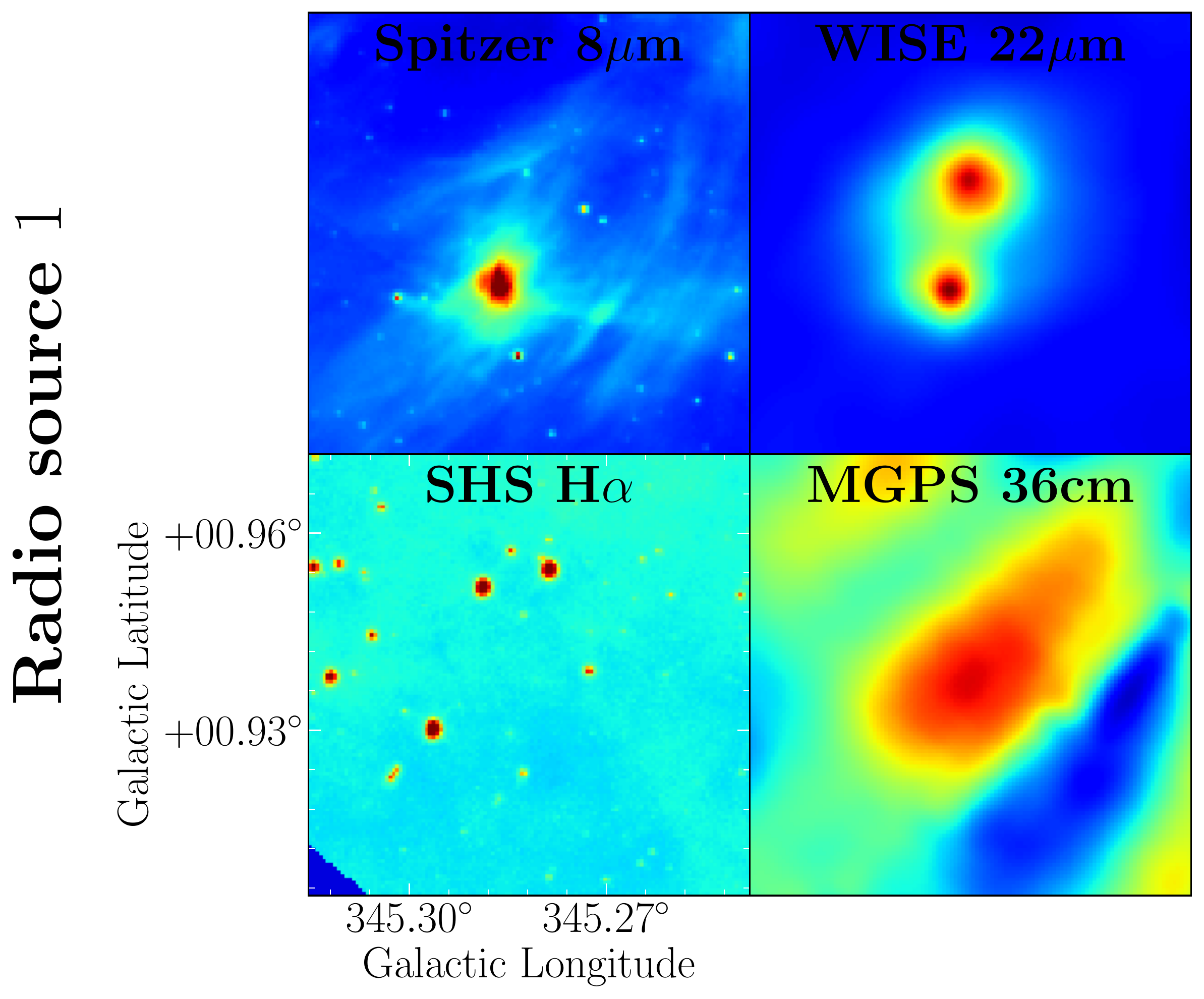}}
\subfloat{
\includegraphics[width=0.33\textwidth]{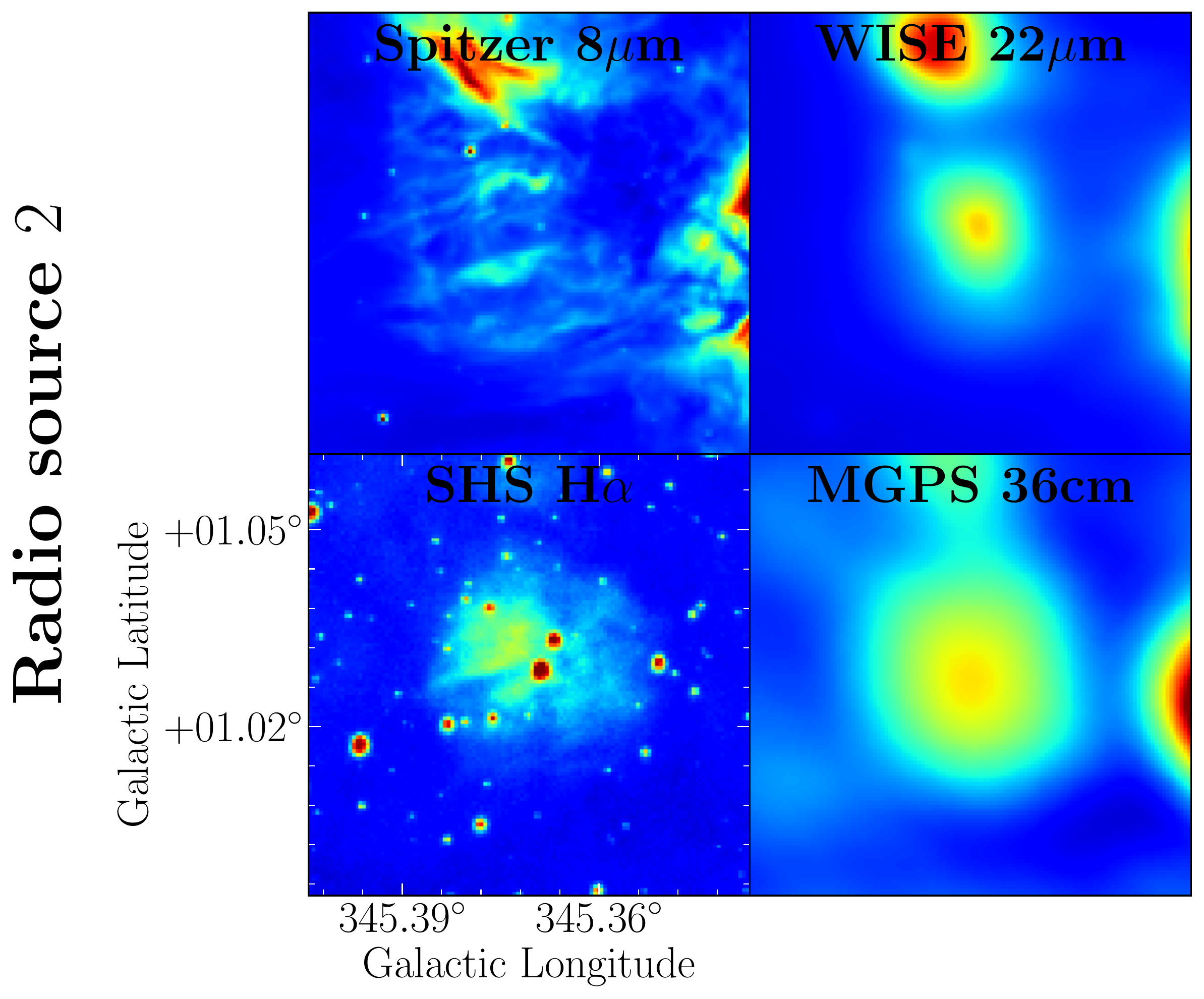}}
\subfloat{
\includegraphics[width=0.33\textwidth]{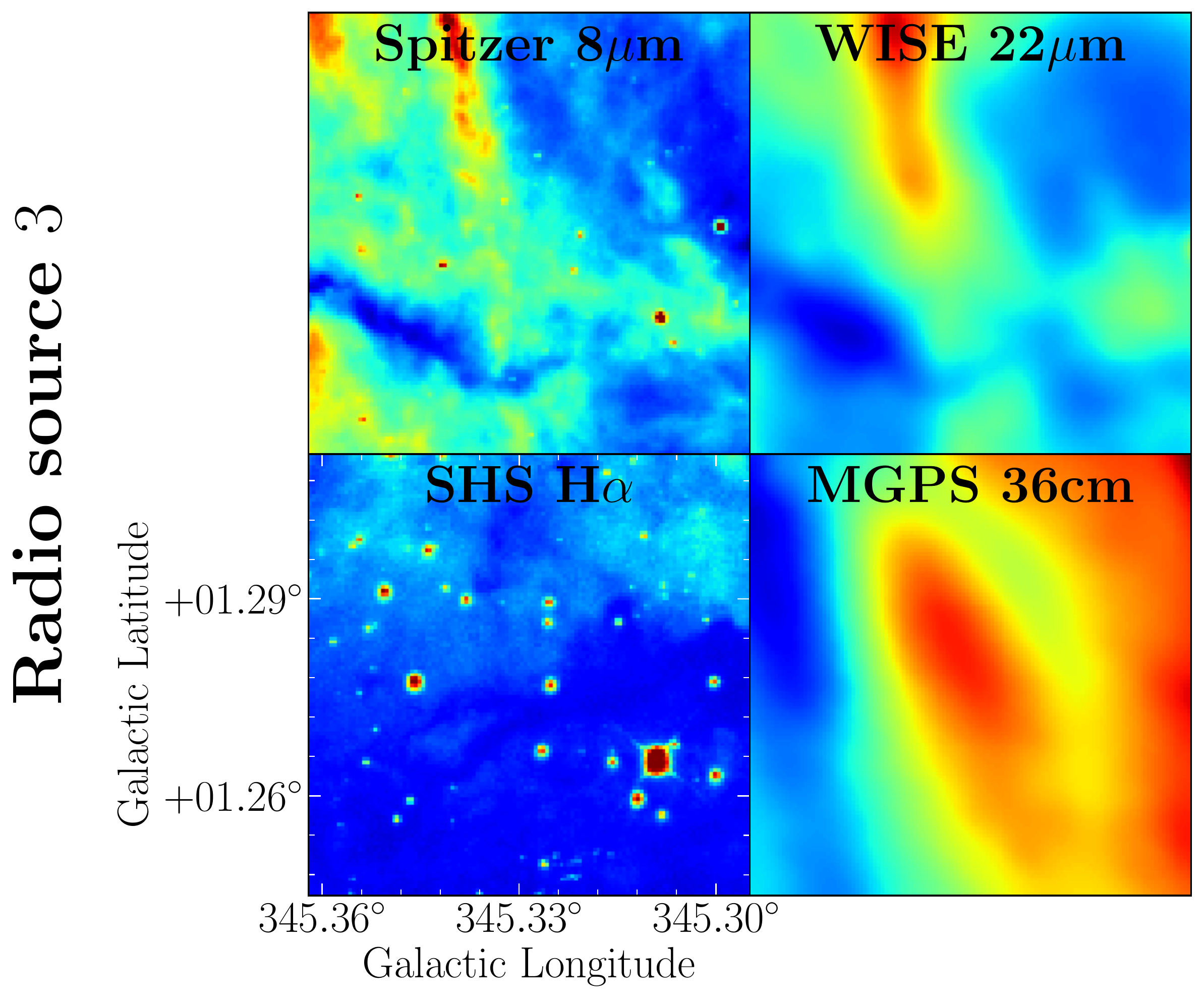}}
\end{figure*}

\begin{figure*}
\subfloat{
\includegraphics[width=0.33\textwidth]{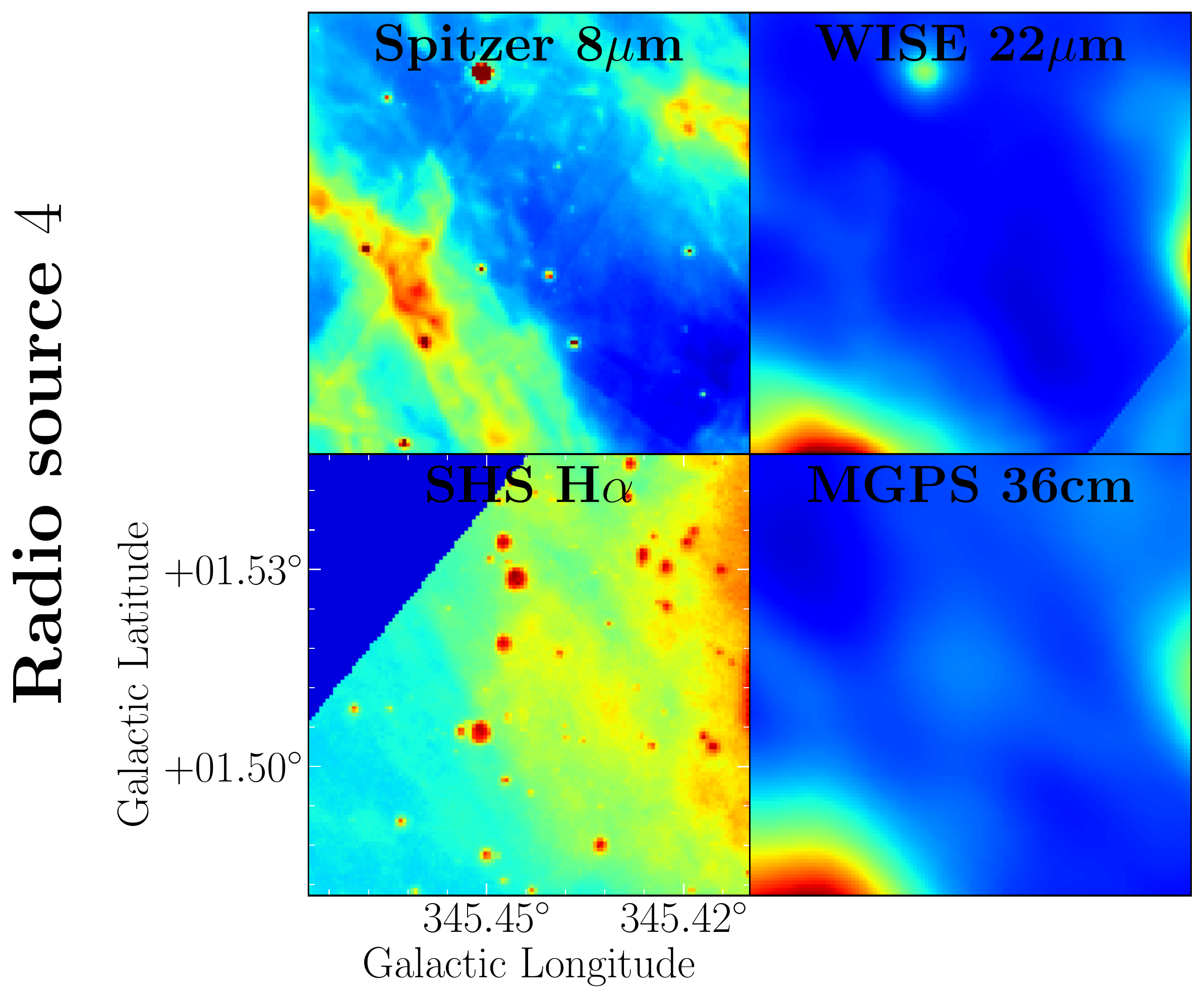}}
\subfloat{
\includegraphics[width=0.33\textwidth]{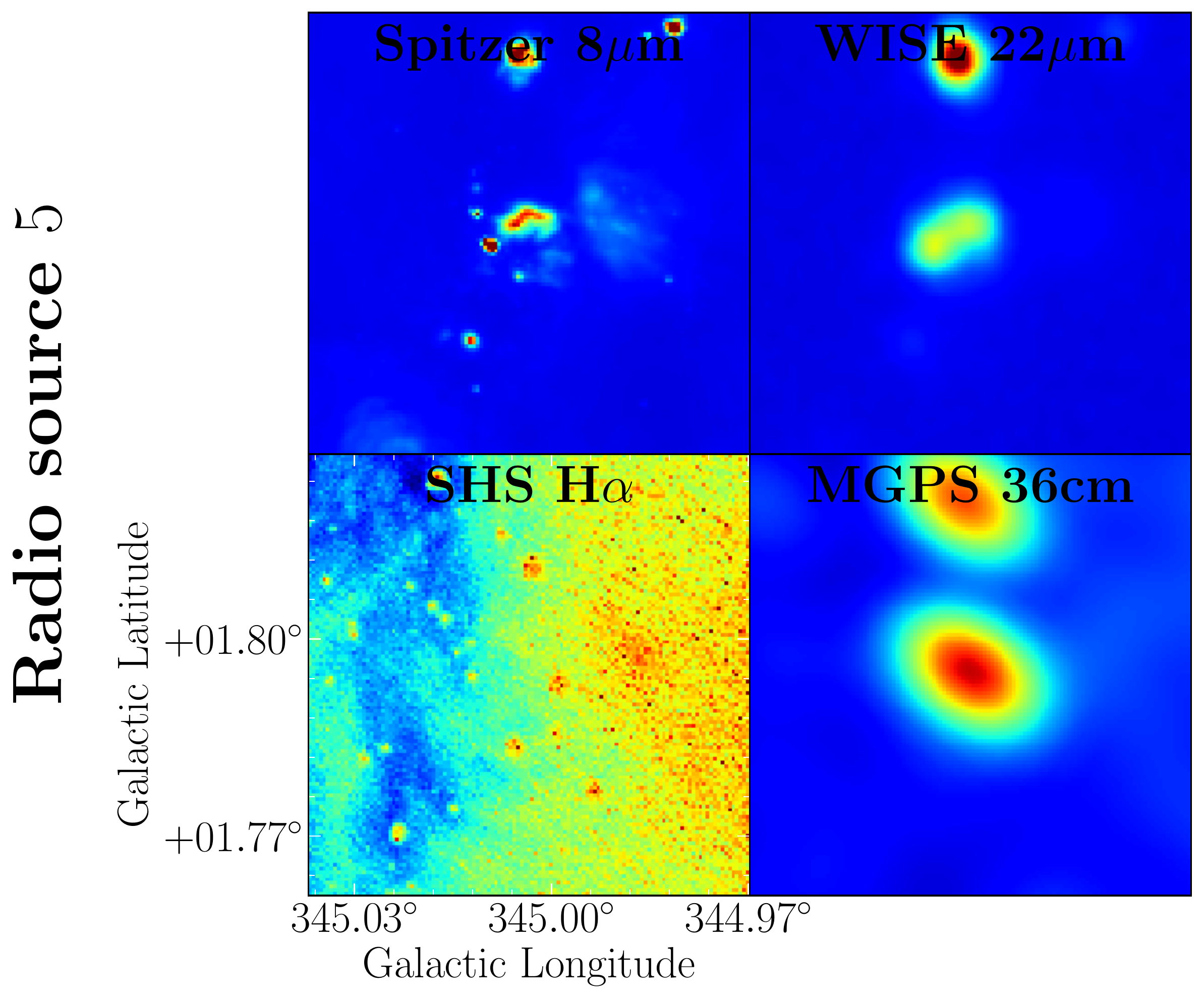}}
\subfloat{
\includegraphics[width=0.33\textwidth]{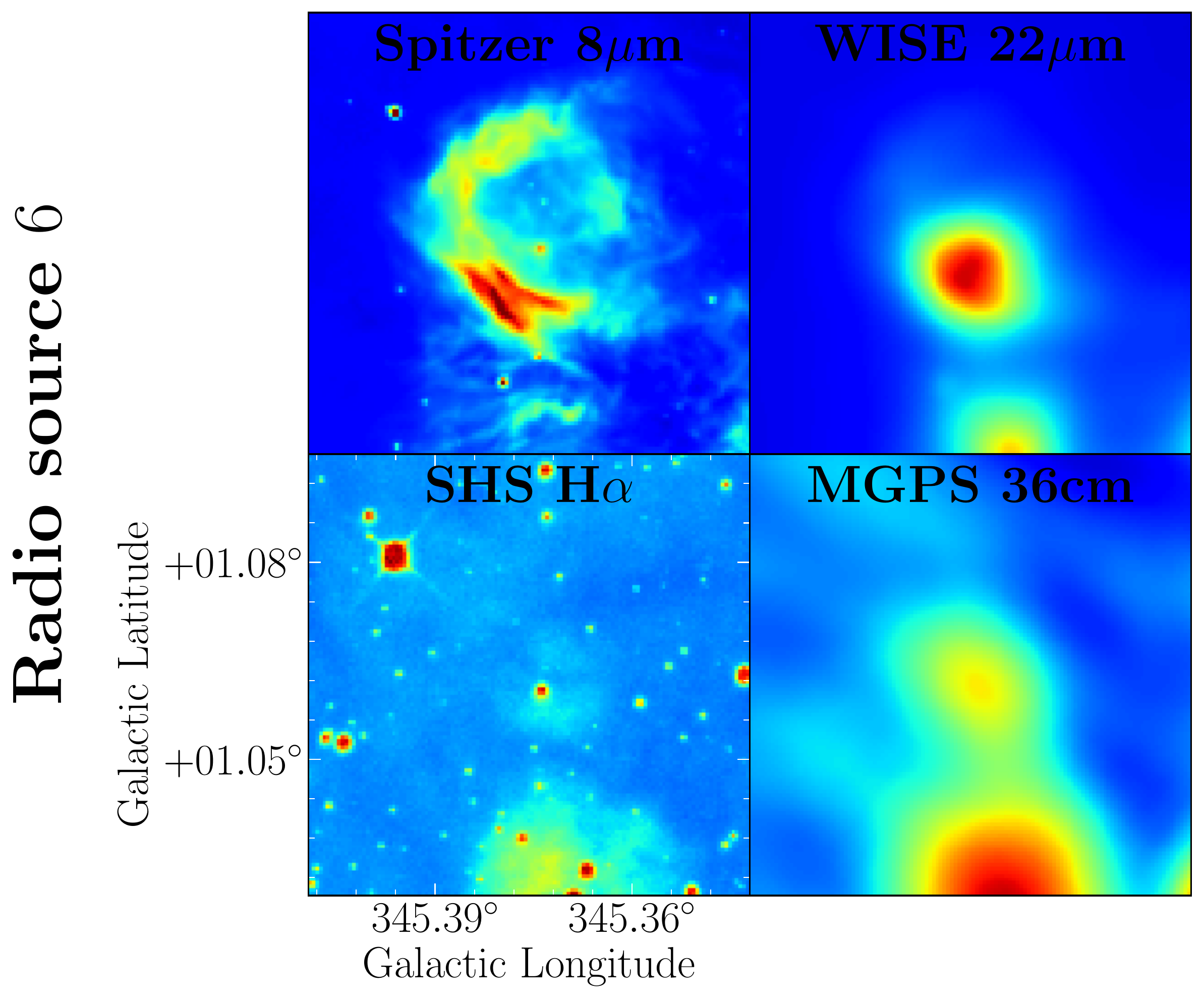}}
\end{figure*}

\begin{figure*}
\centering
\subfloat{
\includegraphics[width=0.33\textwidth]{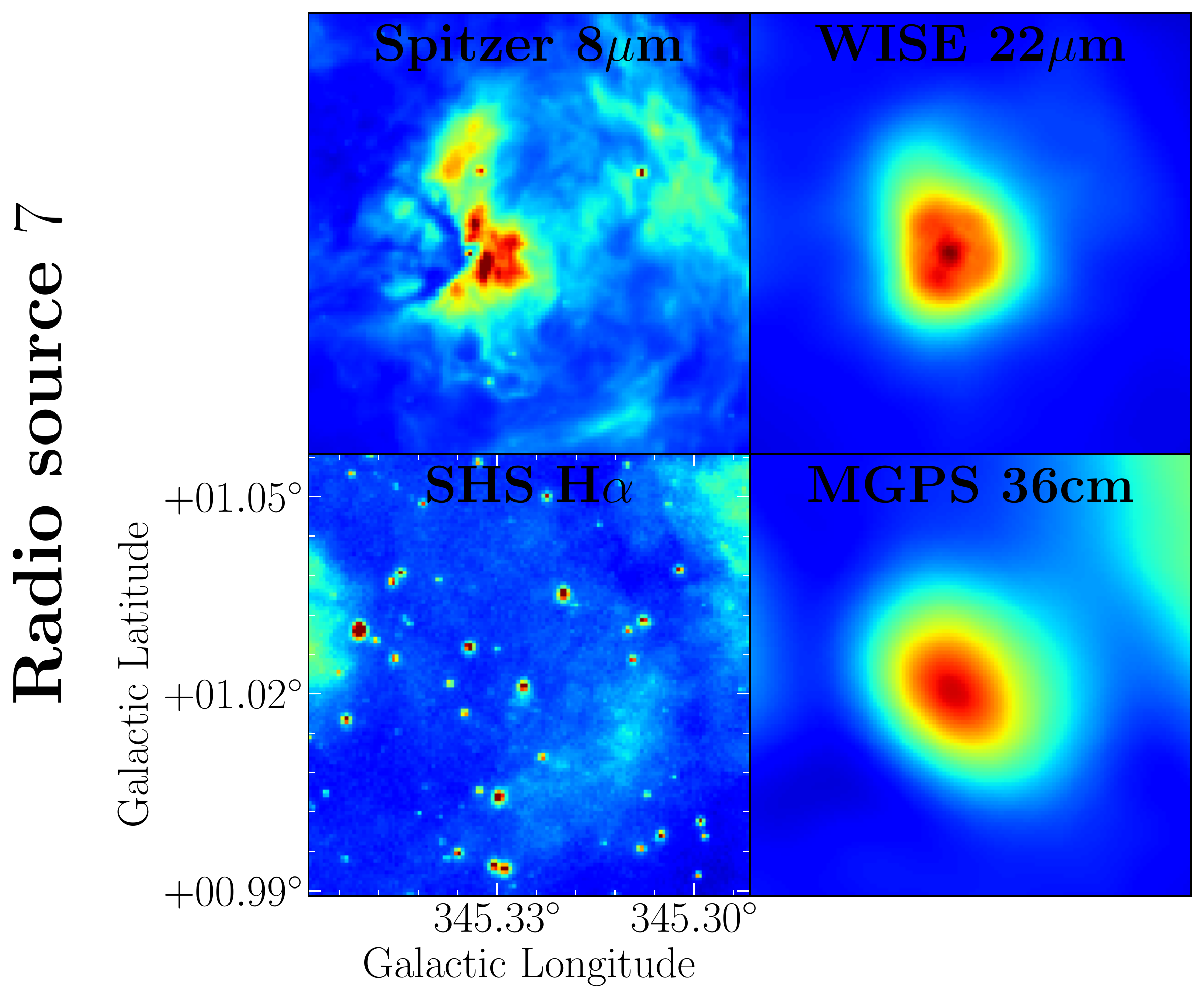}}
\subfloat{
\includegraphics[width=0.33\textwidth]{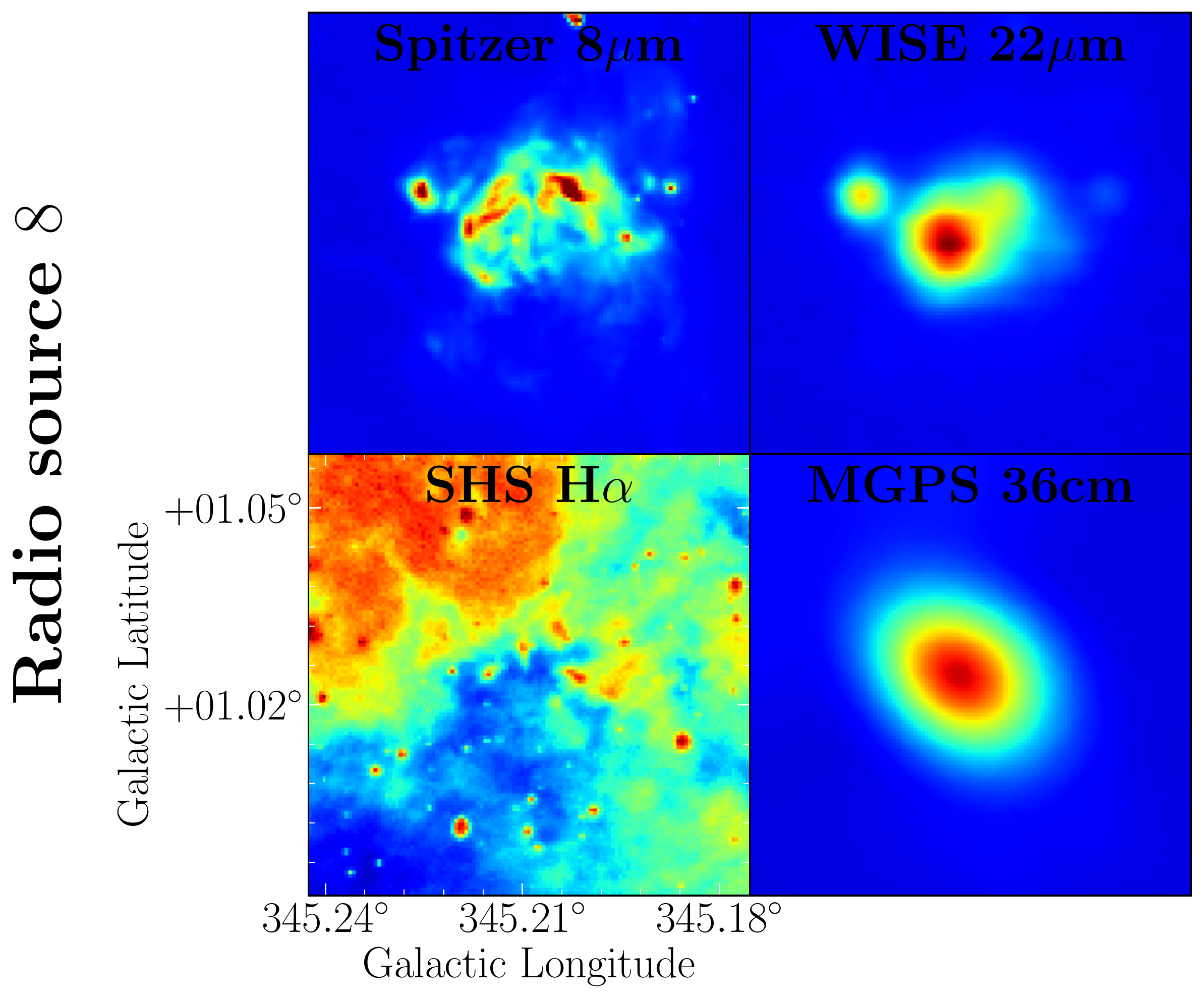}}
\subfloat{
\includegraphics[width=0.33\textwidth]{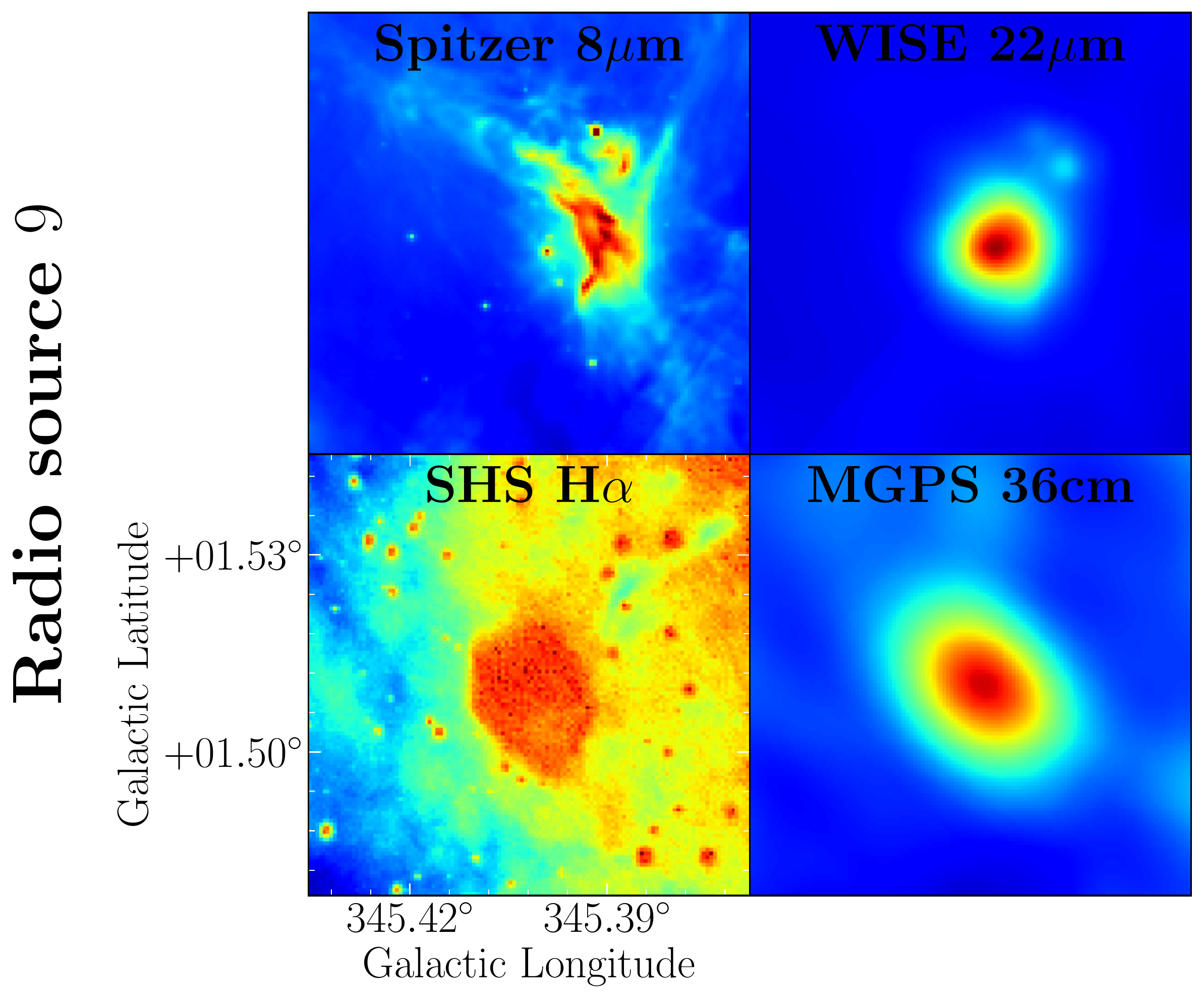}}
\end{figure*}

\begin{figure*}
\subfloat{
\includegraphics[width=0.33\textwidth]{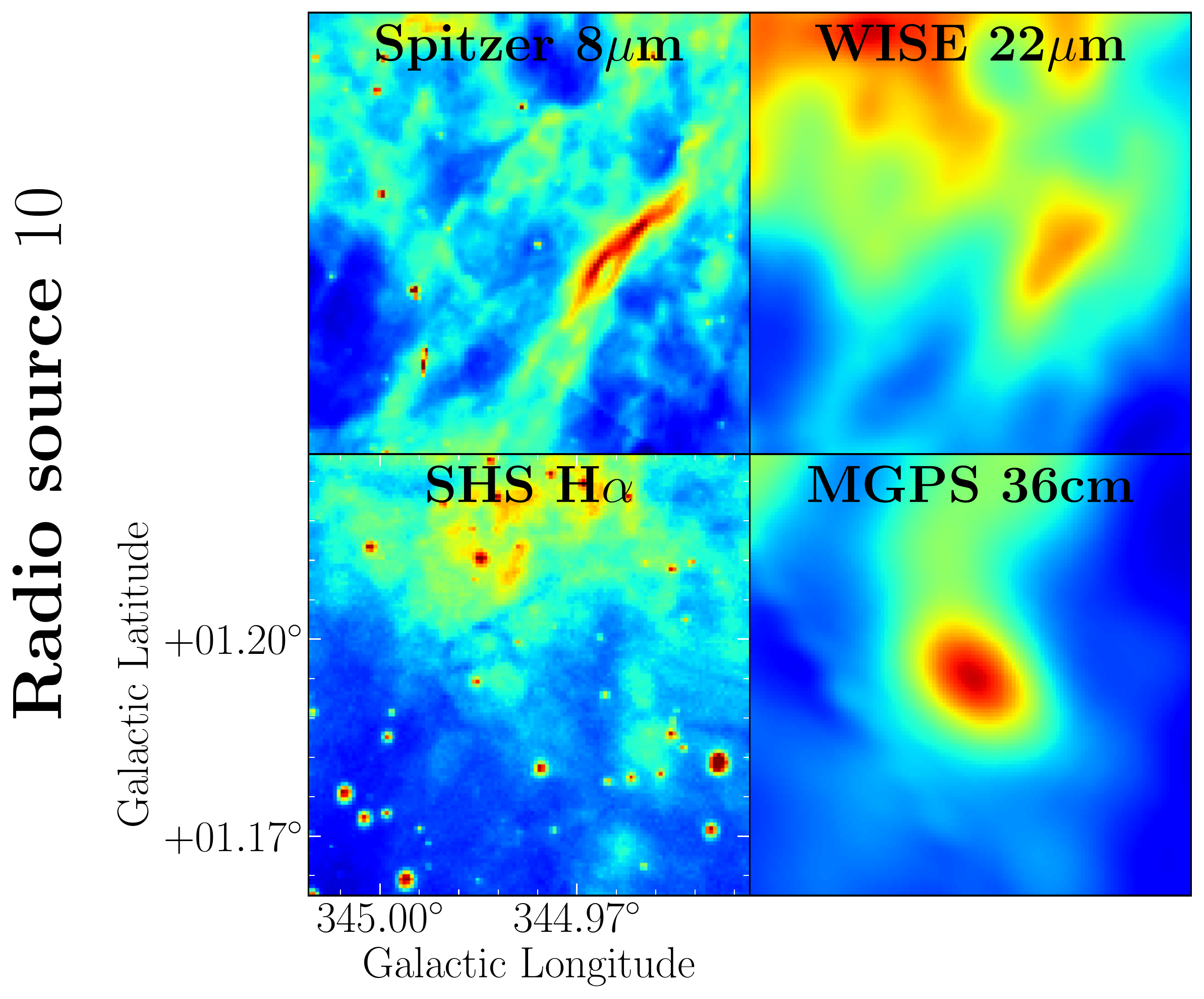}}
\subfloat{
\includegraphics[width=0.33\textwidth]{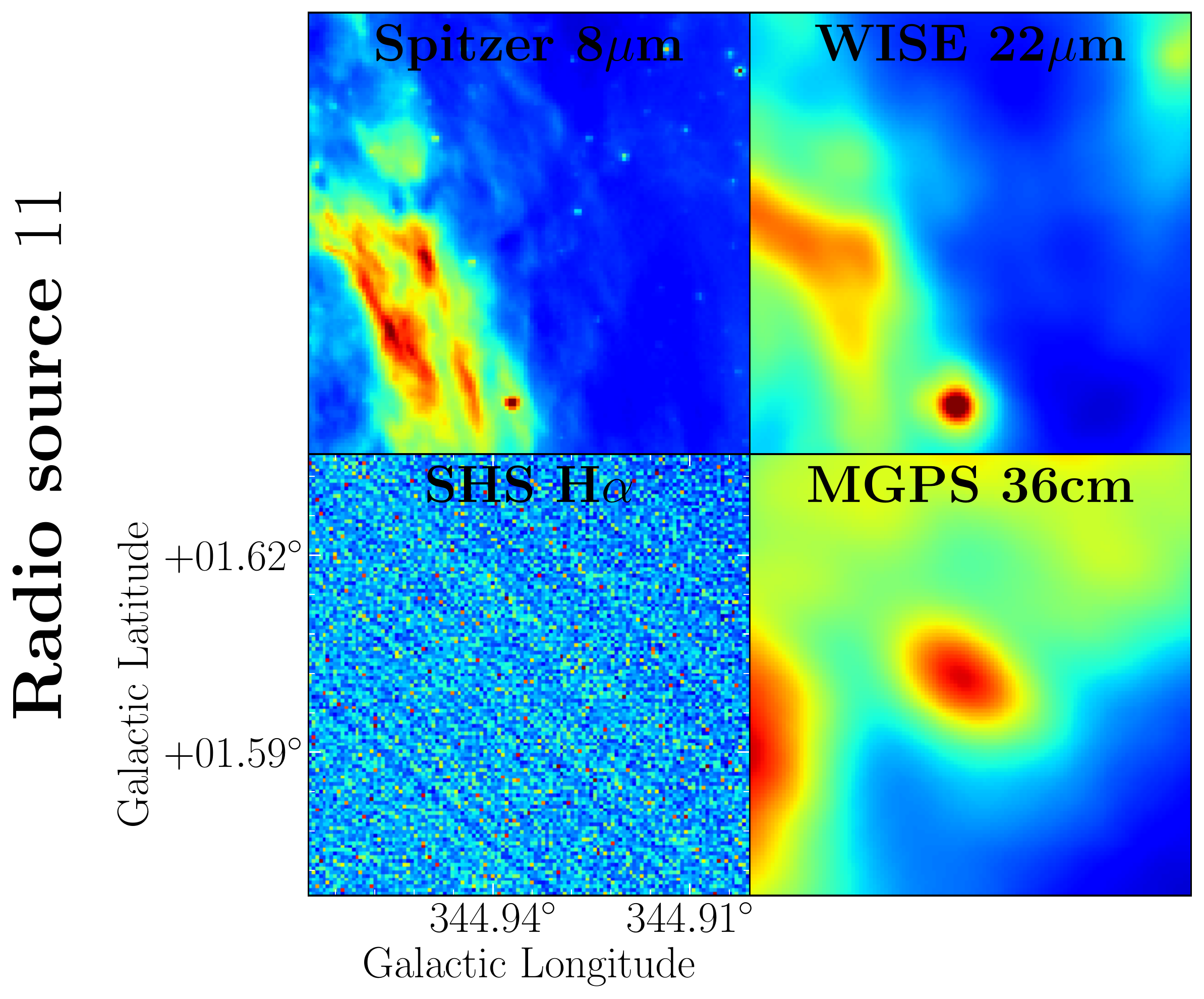}}
\subfloat{
\includegraphics[width=0.33\textwidth]{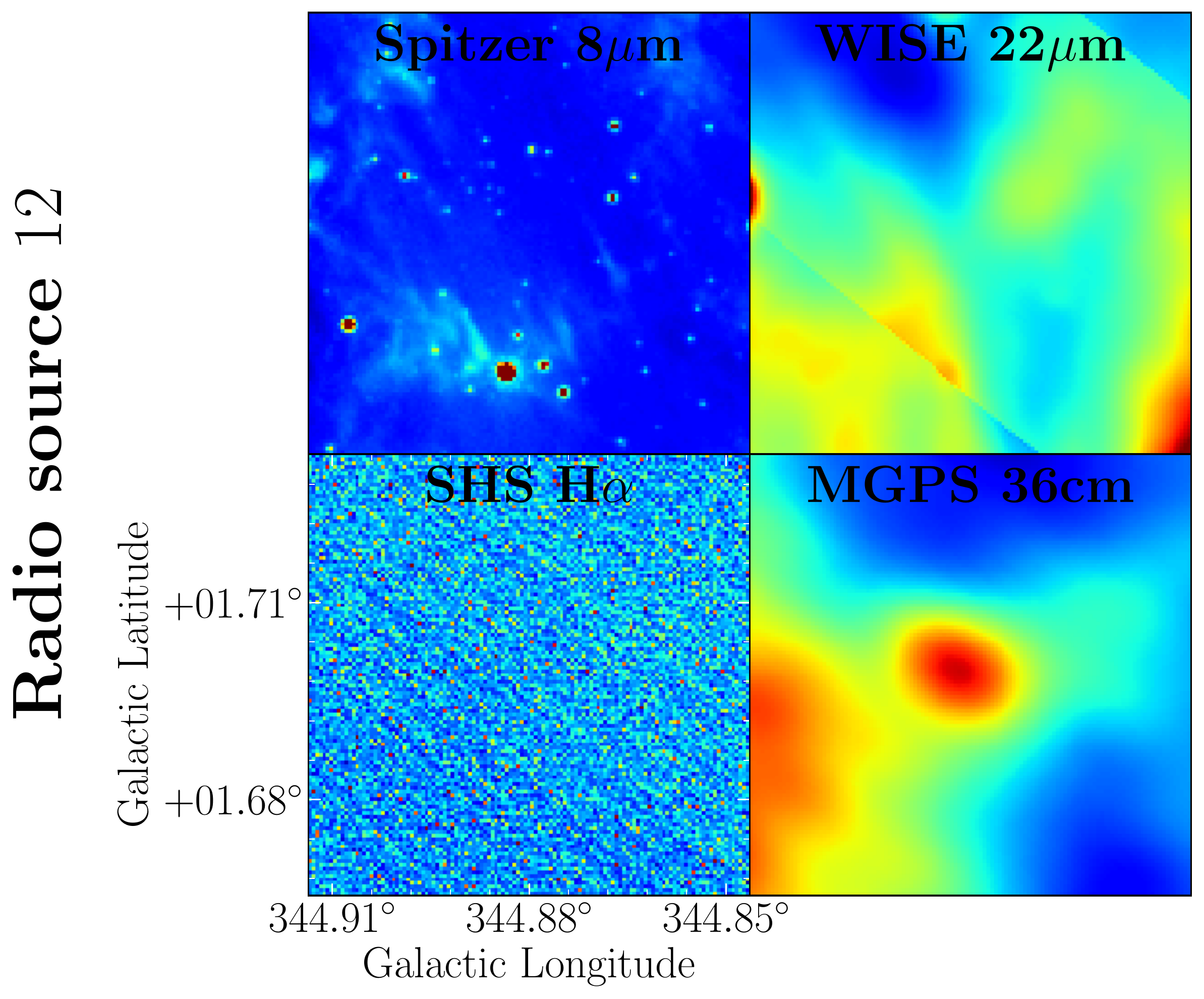}}
\end{figure*}

\begin{figure*}
\centering
\subfloat{
\includegraphics[width=0.33\textwidth]{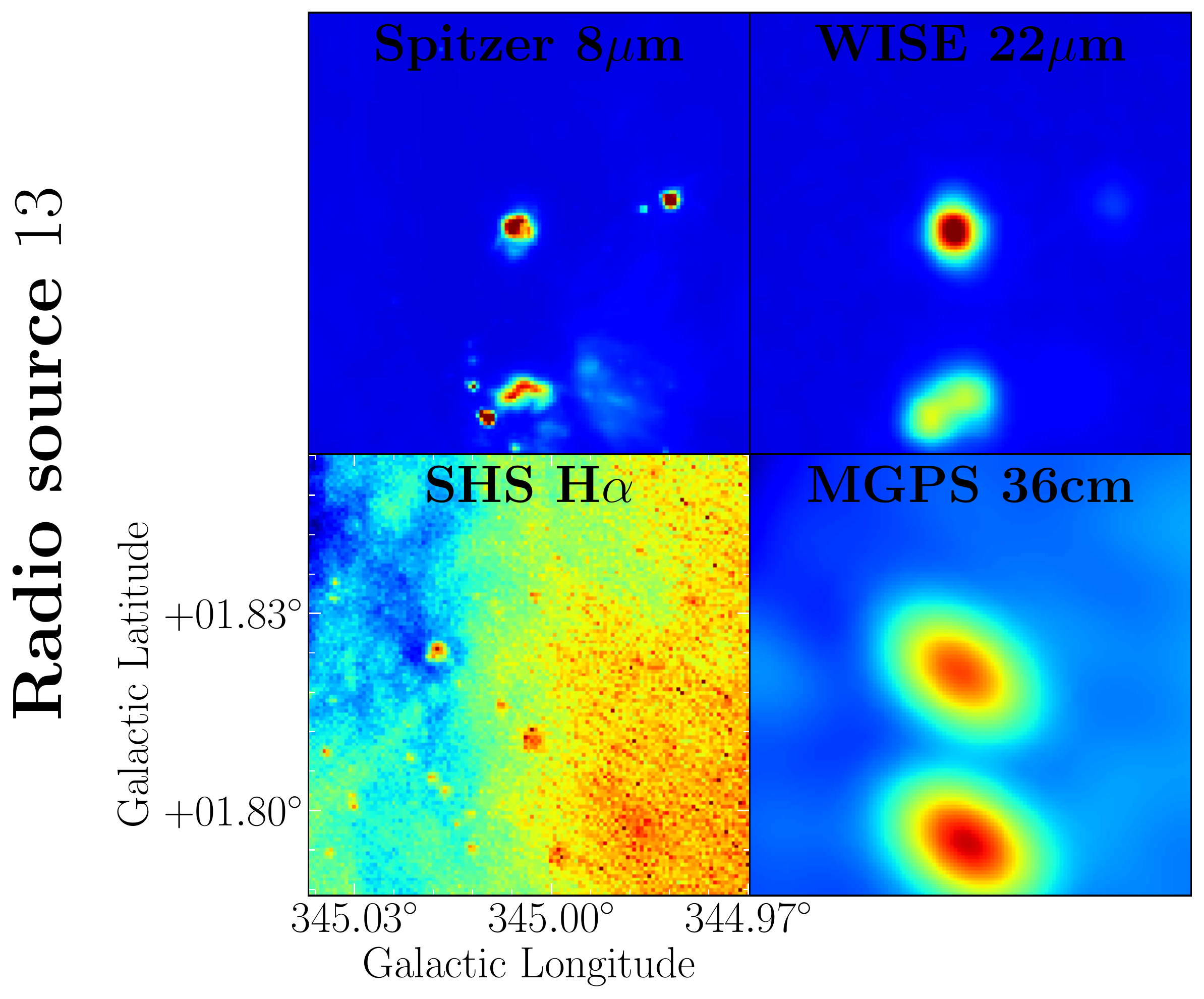}}
  \caption{Mosaic of the radio sources discussed in Sect.~\ref{subsec:radio_compact} that were observed with \spitzer 8~$\mu$m, WISE 22~$\mu$m, SHS H$\rm{\alpha,}$ and MGPS 36~cm.}
\label{fig:mosaic_radiosource}
\end{figure*}

\end{appendix}


\end{document}